\documentclass[fleqn,usenatbib]{mnras}

\usepackage{newtxtext,newtxmath}

\usepackage[T1]{fontenc}

\DeclareRobustCommand{\VAN}[3]{#2}
\let\VANthebibliography\thebibliography
\def\thebibliography{\DeclareRobustCommand{\VAN}[3]{##3}\VANthebibliography}

\usepackage{graphicx}	
\usepackage{amsmath}	

\usepackage{changes}
\usepackage{caption}
\usepackage{subcaption}
\usepackage{amsmath}
\usepackage{chngcntr}

\def \kms {km~$\rm s^{-1}$ }

\title[Characterising CO excitation in quasar host galaxies]{The Quasar Feedback Survey: characterising CO excitation in quasar host galaxies}

\author[S. J. Molyneux et al.]{
S. J. Molyneux,$^{1,2}$\thanks{Email: smolyneux.astro@gmail.com}
G. Calistro Rivera,$^{2}$ C. De Breuck,$^{2}$ C. M. Harrison,$^{3}$ V. Mainieri,$^{2}$ A. Lundgren,$^{2}$ \newauthor
D. Kakkad,$^{4}$ C. Circosta,$^{5,6}$ A. Girdhar,$^{2,7,3}$ T. Costa,$^{8}$ J. R. Mullaney,$^{9}$ P. Kharb,$^{10}$ F. Arrigoni Battaia,$^{8}$ \newauthor E. P. Farina,$^{11}$ D. M. Alexander,$^{12}$ S. R. Ward,$^{2,7}$ Silpa S.,$^{10}$ R. Smit$^{1}$
\\
$^{1}$Astrophysics Research Institute, Liverpool John Moores University, 146 Brownlow Hill, Liverpool L3 5RF, UK \\
$^{2}$European Southern Observatory, Karl-Schwarzschild-str. 2, 85748 Garching, Germany\\
$^{3}$School of Mathematics, Statistics and Physics, Newcastle University, NE1 7RU, UK\\
$^{4}$Space Telescope Science Institute, 3700 San Martin Drive, Baltimore, MD 21218, USA\\
$^{5}$European Space Agency (ESA), European Space Astronomy Centre (ESAC), Camino Bajo del Castillo s/n, 28692 Villanueva de la Cañada, Madrid, Spain\\
$^{6}$Department of Physics and Astronomy, University College London, Gower Street, London WC1E 6BT, UK\\
$^{7}$ Ludwig-Maximilians-Universit{\"a}t, Professor-Huber-Platz 2, D-80539 M{\"u}nchen, Germany\\
$^{8}$Max Planck Institut f{\"u}r Astrophysik, Karl-Schwarzschild-str. 1, D-85748, Garching bei M{\"u}nchen, Germany\\
$^{9}$Department of Physics and Astronomy, The University of Sheffield, Hounsfield Road, Sheffield, S3 7RH, UK\\
$^{10}$National Centre for Radio Astrophysics – Tata Institute of Fundamental Research, S. P. Pune University Campus, Ganeshkhind, Pune 411007, India\\
$^{11}$Gemini Observatory, NSF’s NOIRLab, 670 N A’ohoku Place, Hilo, Hawai'i 96720, USA\\
$^{12}$Centre for Extragalactic Astronomy, Department of Physics, Durham University, South Road, Durham, DH1 3LE, UK
}

\date{Accepted XXX. Received YYY; in original form ZZZ}

\pubyear{2023}

\begin{document}
\label{firstpage}
\pagerange{\pageref{firstpage}--\pageref{lastpage}}
\maketitle

\begin{abstract}
  We present a comprehensive study of the molecular gas properties of 17 Type 2 quasars at $z <$ 0.2 from the Quasar Feedback Survey (L$_{\rm [O~{\sc III}]}$ > $10^{42.1}$ $\rm ergs^{-1}$), selected by their high [O~{\sc III}] luminosities and displaying a large diversity of radio jet properties, but dominated by LIRG-like galaxies. With these data, we are able to investigate the impact of AGN and AGN feedback mechanisms on the global molecular interstellar medium. Using APEX and ALMA ACA observations, we measure the total molecular gas content using the CO(1-0) emission and homogeneously sample the CO spectral line energy distributions (SLEDs), observing CO transitions (J$_{up}$ = 1, 2, 3, 6, 7). We observe high $r_{21}$ ratios (r$_{21}$~=~L'$_{\rm CO(2-1)}$/L'$_{\rm CO(1-0)}$) with a median $r_{21}$ = 1.06, similar to local (U)LIRGs (with $r_{21}$ $\sim$ 1) and higher than normal star-forming galaxies (with $r_{21}$ $\sim$ 0.65). Despite the high $r_{21}$ values, for the 7 targets with the required data we find low excitation in CO(6-5) \& CO(7-6) ($r_{61}$ and $r_{62}$ < 0.6 in all but one target), unlike high redshift quasars in the literature, which are far more luminous and show higher line ratios. The ionised gas traced by [O~{\sc III}] exhibit systematically higher velocities than the molecular gas traced by CO. We conclude that any effects of quasar feedback (e.g. via outflows and radio jets) do not have a significant instantaneous impact on the global molecular gas content and excitation and we suggest that it only occurs on more localised scales.
\end{abstract}




\section{Introduction}
\label{sec:intro}

A fundamental outstanding question of galaxy evolution is what impact active galactic nuclei (AGN) have on the interstellar medium (ISM) and star formation in their host galaxies. AGN can release energy into their host galaxies via processes known as AGN feedback, which are required by our current models of galaxy evolution to regulate star formation \citep{Bower06, Schaye15}, and believed to be the mechanism that regulates the co-evolution of accreting black holes (BH) and their host galaxies that is observed across cosmic time \citep[e.g.][]{Madau14, Kormendy13, Cresci18}. Observational and theoretical studies have proposed both a suppression \citep[][]{Silk98, Hopkins06, Booth10, Feruglio10, Cicone14, King15, Fiore17, Costa18, Ellison21, Bertemes23} and an enhancement of star formation in AGN host galaxies \citep[][]{Ishibashi12, Silk13, Zubovas13, Lacy17, Fragile17, Gallagher19}. However, these interactions are still not fully understood and the diversity of results stress the need for multiwavelength, multi-tracer studies to characterise the interplay between the central supermassive black hole and the host galaxy.

A natural assumption may be that the most powerful and luminous AGN and quasars will have the largest impact on their host galaxy. Studies suggest that they might be able to drive kpc-scale outflows across the entire galaxy, expelling the interstellar star-forming gas \cite[e.g.][]{Cicone12, Harrison14, Feruglio15, Circosta18, Longinotti23}. However, studies at low redshift ($z$) have also shown that AGN and quasars tend to reside in gas rich, star forming galaxies and find no instantaneous depletion of total gas content \citep{Saintonge17, Shangguan20, Jarvis20, Koss21}. These findings also agree with recent simulations \citep[e.g.][]{Piotrowska22, Ward22}, therefore supporting the idea that large gas reservoirs are needed to fuel the accreting supermassive BHs in quasar and AGN hosts. It may therefore be the case that any impact from feedback is limited to a more localised scale and the global properties of the ISM are left largely unaffected. Indeed, there are works which show the possible impact of AGN feedback on the molecular gas content in central/localized part of the galaxy \citep[e.g.][]{Rosario19, Feruglio10, Ellison21, Almeida21, Audibert23}.

The molecular phase of the ISM, commonly traced by observing low transitions of carbon monoxide (CO), plays a critical role in galaxy evolution as it is this gas which is redistributed to both promote star-formation activity and fuel BH growth \citep[e.g.][]{McKee07, Carilli13, Vito14, Tacconi20}. However, no consensus has yet been reached on the impact of AGN on the overall molecular gas content in the ISM \citep{Kakkad17, Perna18, Kirkpatrick19, Rosario19, Circosta21, Morganti21}. One possible reason for this might be due to the time scales over which any impact may take place \citep{King11, Mukherjee18, Zubovas18, Ward22}. There are also complexities due to the resolution of observations, biases in the sample selection, what tracers of the gas are used, and the uniformity in the observations.

While most studies of the AGN impact on molecular gas have focused on the total gas content, much is still unknown about other molecular gas properties such as molecular gas excitation. Knowledge of the ground state CO(1-0) line is a crucial reference that is often used to not only compare to higher transitions and measure the excitation, but to also convert to the total molecular gas content of the galaxy. However, there is discussion in the community about how reliable the ground state is in doing these calculations for different objects \citep[e.g. star forming galaxies (SFGs) or (ultra) luminous infrared galaxies (U)LIRGs, see][]{Leroy22, Montoya23}. This therefore increases the importance in characterising the CO(1-0) across different samples.

Due to the expensive observations required to detect multiple emission lines for individual sources, most of our knowledge on molecular gas excitation is based on inhomogeneous coverage of few transitions, and limited to for the most luminous galaxies \citep[e.g.][]{Kakkad17, Saintonge17, Lamperti20, Circosta21, Boogaard21, Valentino21, Harrington21, Leroy22}. 
Further, studies have investigated the driving mechanism for the excitation of the molecular gas \citep[e.g.][]{Daddi15, Pozzi17, Mingozzi18, Leroy21, Esposito22}, suggesting photodissociation regions (PDRs) and X-ray dominated regions (XDRs), of diverse temperature and gas densities, are the key physical components driving CO excitation.  

Models of CO excitation suggest that AGN-related processes, such as X-ray emission \citep{Meijerink07} and shock heating induced by AGN jets and outflows \citep{Kamenetzky16}, would mainly affect the molecular gas excitation at the higher CO transitions.
It is therefore crucial to study both low and high CO transitions as the impact of feedback may only be present at higher CO excitations ($J_{up} > 5$), whereas the bulk of the molecular gas content is still traced by the ground transition. 
However, this is challenging at low redshifts ($z < 0.2$), where even using the maximum frequency limit of the ALMA bands, only $J_{up} \le 8$ can be reached. 
Due to the observational difficulty of observing at higher frequencies, there are few examples of these critical higher CO transitions observed at low-$z$ \citep[e.g.][]{vanderWerf10, Greve14, Rosenberg2015, Liu2015, Kamenetzky16, Yang2017}. 
Indeed, most $J_{up} > 7$ observations come from higher redshift ($z > 1$), highly luminous quasars, which are far more easily observed \citep{Carilli13, Wang19, Yang19, Li20, Pensabene21, Decarli22}, which in turn lack observations of low-J transitions, consequently lacking a complete characterisation of the molecular gas content and excitation.

Previous studies have suggested a close relationship between AGN feedback diagnostics and the properties of the ionised phase of the ISM. For example, a study of optically selected AGN from SDSS found that those with higher radio luminosities were more likely to have larger FWHM$_{\rm [O~{\sc III}]}$ \citep{Mullaney13}, suggesting a relation between the radio emission and the kinematics of the ionised gas. Further work on the same sample showed that the most extreme ionised outflows (FWHM$_{\rm [O~{\sc III}]}$ > 1000\kms) were found to be more common when the radio emission was compact \citep{Molyneux19}. With high resolution radio observations for a sample of 42 of these targets \citep[presented in][]{Jarvis19} from the Karl G. Jansky Very Large Array (VLA), a prevalence of small scale radio jets (in the central few kpc) was found, leading to a suggestion that they could be the driver of these ionised outflows. Alternatively star formation driven outflows and quasar winds that shock the ISM may be responsible for producing the observed radio emission and correlation with outflow properties \citep[e.g.][]{Condon13, Nims15, Zakamska16, Hwang18, Panessa19}.

With the advent of deeper radio data, increasing evidence is being found of potentially ubiquitous low-level radio emission in radio-quiet quasars \citep[e.g.][]{Mukherjee18, Jarvis19, Jarvis21, Macfarlane21}. These observations suggest that radio jets are potentially an important feedback mechanism in radio quiet quasars. Indeed radio jets have been found to have an impact on the surrounding multi-phase ISM \citep[e.g.][]{Morganti15, Oosterloo17, Jarvis19, Morganti21, Girdhar22}. However, an outstanding question is how and when these jets can couple to the ISM, and have a positive and/or negative impact on the star-forming molecular gas content \citep[e.g.][]{Silk13, Gabor14, Bieri16, Costa18}. Studying the CO excitation of quasars with known outflows/jets is therefore key to solving these outstanding questions.

The Quasar Feedback Survey (QFeedS) is a multi-wavelength survey aiming to address these open questions in order to understand the co-evolution between quasars and their host galaxy, in particular in the context of ionised outflows and radio jets. These are luminous systems at $z < 0.2$ and so it is possible to study the impact that feedback (e.g. via radio jets) has on the multi-phase ISM on both resolved and global scales, whether it be driving outflows, disturbing the gas kinematics, affecting the molecular gas excitation, or impacting on star formation \citep{Harrison15, Lansbury18, Jarvis19, Jarvis20, Jarvis21, Girdhar22, Silpa22}.

In this work we present a comprehensive study of the molecular gas properties of 17 quasars of the QFeedS sample which have multi-wavelength data. We characterise the molecular excitation in these sources, presenting the CO(1-0), CO(2-1) and CO(3-2) for the entire sample and also CO(6-5) or CO(7-6) for 7 of the 17 targets. 
With the addition of ancillary multi-wavelength data, we will explore the impact of feedback, if any, on the total molecular gas content and molecular gas excitation within the quasar host galaxies.

In Section~\ref{sec:sampleselect} we introduce the quasar sample presented in this work as part of the Quasar Feedback Survey. In Section~\ref{sec:observations} we describe the observations used and the data reduction. In Section~\ref{sec:results} we describe the analysis techniques used to study our CO data, including spectral fitting, definition of detection, flux measurements and line profile characterisation. We also introduce the comparison samples from the literature that we utilise in our analysis. In Section~\ref{sec:discussion} we present our results of the CO excitation, line profile properties and gas fractions, and at all times comparing to relevant samples from the literature. We then discuss our findings in the overall context of galaxy evolution and quasar feedback. Our final conclusions are presented in Section~\ref{sec:conclsions}.\\

We adopt $H_0=70$\,\kms\,Mpc$^{-1}$, $\Omega_M=0.3$, and $\Omega_\Lambda=0.7$ throughout.

\section{Sample selection}
\label{sec:sampleselect}

\begin{figure}
\centering
\includegraphics[width=\columnwidth]{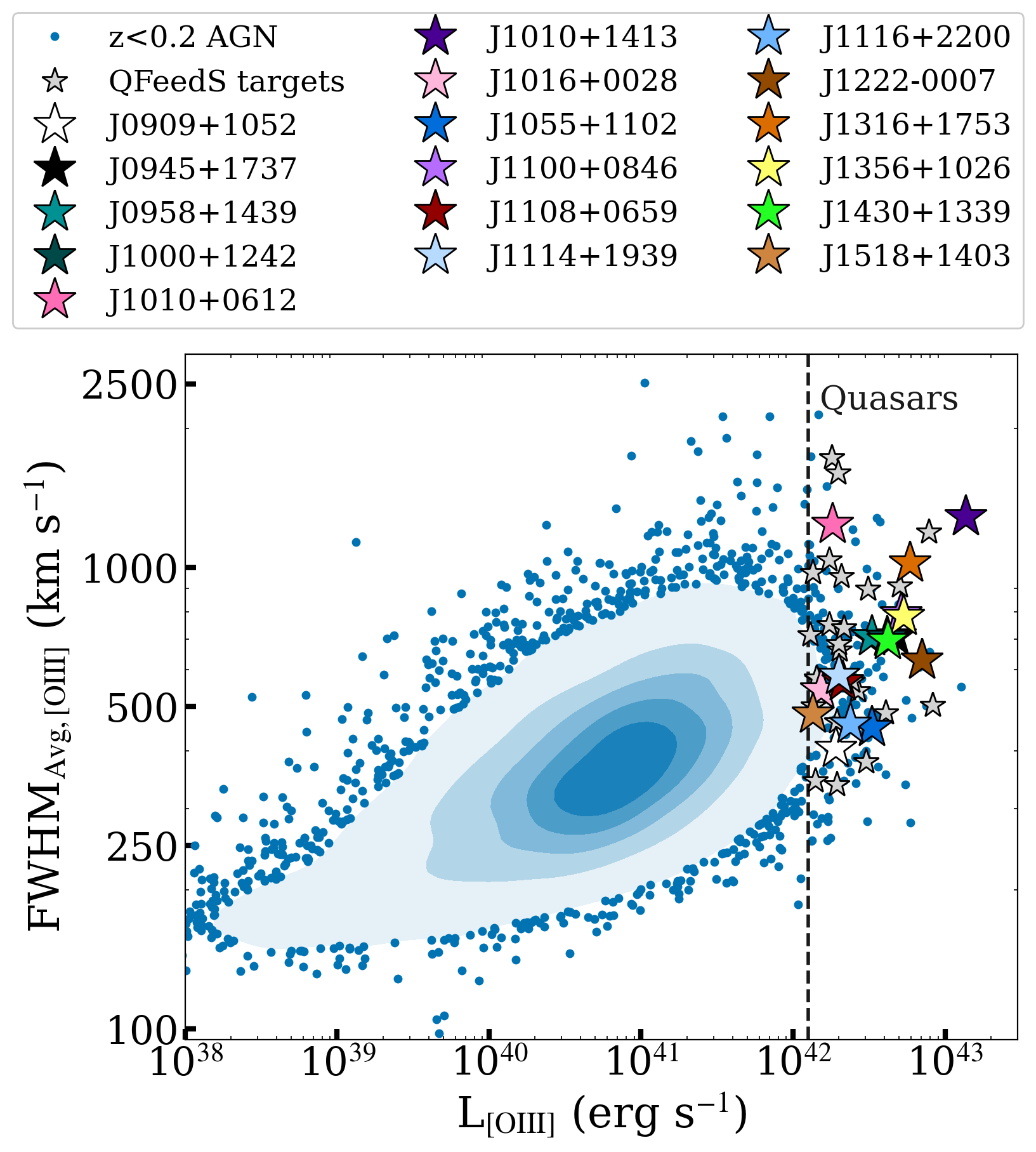}
\caption{Sample Selection: [O~{\sc III}] luminosity versus emission line widths, demonstrating how the 42 quasars in the QFeedS sample (star symbols) are selected from $z <$ 0.2 AGN sample \citep[][blue points and contours]{Mullaney13}. The 17/42 selected for this work are highlighted by the larger coloured stars and the remaining 25/42 of the QFeedS sample are shown by small grey stars. The colours for the 17 targets in this sample (shown in the legend) are carried through in figures throughout the rest of this paper. The dashed line represents the selection criteria in QFeedS of $\rm L_{[\rm O~{\sc III}]}$ > 10$^{42.1}$ $\rm ergs\, s^{-1}$.
}
\label{fig:SampleSelection}
\end{figure}

QFeedS, presented in \cite{Jarvis21}, is a multi-wavelength study of 42 quasars at $z < 0.2$. 
This main sample was selected from a parent sample of 24264 optically selected AGN from SDSS at $z < 0.4$ from \cite{Mullaney13}. These 42 quasars were selected to have L$_{\rm [O~{\sc III}]}$ > $10^{42.1}$ ergs s$^{-1}$ and to cover the full range of FWHM$_{\rm Avg [O~{\sc III}]}$ (a flux weighted average of the FWHM of the two Gaussian components present in the spectra) with velocities in the range = 339 -- 1289 \kms (see Figure~\ref{fig:SampleSelection}).

Here we introduce a study as part of QFeedS to provide a detailed characterisation of molecular gas in 17 Type 2 quasars, studying properties such as molecular gas masses, gas fractions and CO excitation. The 17 targets were selected to be Type 2 quasars which are visible from the Very Large Telescope (VLT) and the Atacama Pathfinder EXperiment telescope (APEX) and that are representative of the parent population (see Figure~\ref{fig:SampleSelection}). We selected Type 2 quasars in order to achieve a more robust characterisation of the host-galaxy stellar-emission properties \citep{Jarvis19}. These 17 targets also have available optical (Multi Unit Spectroscopic Explorer, MUSE) and radio (VLA) data allowing us to perform a full multi-wavelength analysis of the quasar and host continuum emission, in addition to a multi-tracer characterisation of the ISM (these ancillary data are discussed further in Section~\ref{sec:ancillary}). These 17 sources are representative of the survey sample as they cover the full range of QFeedS redshifts ($z$ $\sim$ 0.1 -- 0.2) as well as [O~{\sc III}] and radio luminosities, L$_{[\rm O~{\sc III}]}=$ 10$^{42.1}$ -- 10$^{43.2}$ erg s$^{-1}$ and L$_{\rm 1.4 GHz}=$ 10$^{23.5}$ -- 10$^{24.4}$ W Hz$^{-1}$.

Based on the criteria of \cite{Xu99} using the [O~{\sc III}] and radio luminosity division all 17 of our sample are defined as `radio-quiet' \citep[see also][]{Jarvis21}. From previous work \citep{Jarvis19}, we also know that at least 8 of these 17 sample galaxies are consistent with being luminous infrared galaxies (LIRGs, $10^{11}$ L$_\odot$ $\lesssim$ $L_{\rm IR, SF}$ $\lesssim$ $10^{12}$ L$_\odot$, where $L_{\rm IR, SF}$ is the far infrared luminosity associated with star formation). As only 9 of these targets have the required $L_{\rm IR, SF}$ data then the number of sources consistent with being LIRGs is likely to be higher. This is an important consideration for when we make comparisons with samples in the literature.

The source selection for this study is shown in Figure~\ref{fig:SampleSelection}. The colours for the 17 targets in this sample (shown in the legend of Figure~\ref{fig:SampleSelection}) are used in all further figures in this work. Further, basic properties of these sources can be found in Table~\ref{tab:sampleintro}.

\begin{table*}
\centering
\begin{tabular}{ |c|c|c|c|c|c|c|c| } 
 \hline
 \rule{0pt}{2ex}
 Name & RA & DEC & z &  log(L$\rm _{1.4GHz}$) & S$\rm _{1.4GHz}$ & log(L$_{\rm [O~{\sc III}]}$) & SDSS $W_{80}$ \\ 
 & (J2000) & (J2000) & & (W Hz$^{-1}$) & (mJy) & (erg s$^{-1}$) & (\kms)  \\
 (1) & (2) & (3) & (4) & (5) & (6) & (7) & (8) \\
 \hline
 \rule{0pt}{2ex}
 J0909+1052 & 09:09:35.49 & +10:52:10.5 & 0.166 & 23.6 & 6.0 $\pm$ 0.5 & 42.28 & 399 $\pm$ 24 \\ 
 \rule{0pt}{2ex}
 J0945+1737 & 09:45:21.33 & +17:37:53.2 & 0.128 & 24.3 & 45.6 $\pm$ 1.4 & 42.67 & 799 $\pm$ 26 \\ 
 \rule{0pt}{2ex}
 J0958+1439 & 09:58:16.88 & +14:39:23.7 & 0.109 & 23.5 & 10.9 $\pm$ 0.5 & 42.52 & 786 $\pm$ 10 \\ 
 \rule{0pt}{2ex}
 J1000+1242 & 10:00:13.14 & +12:42:26.2 & 0.148 & 24.3 & 34.8 $\pm$ 1.1 & 42.62 & 813 $\pm$ 5 \\ 
 \rule{0pt}{2ex}
 J1010+0612 & 10:10:43.36 & +06:12:01.4 & 0.098 & 24.3 & 92.4 $\pm$ 3.3 & 42.26 & 1462 $\pm$ 8 \\ 
 \rule{0pt}{2ex}
 J1010+1413 & 10:10:22.95 & +14:13:00.9 & 0.199 & 24.1 & 11.1 $\pm$ 0.5 & 43.14 & 1426 $\pm$ 16 \\ 
 \rule{0pt}{2ex}
 J1016+0028 & 10:16:53.82 & +00:28:57.1 & 0.116 & 23.6 & 11.8 $\pm$ 0.9 & 42.18 & 596 $\pm$ 8 \\
 \rule{0pt}{2ex}
 J1055+1102 & 10:55:55.34 & +11:02:52.2 & 0.145 & 23.5 & 5.7 $\pm$ 0.4 & 42.52 & 478 $\pm$ 7 \\
 \rule{0pt}{2ex}
 J1100+0846 & 11:00:12.38 & +08:46:16.3 & 0.100 & 24.2 & 59.8 $\pm$ 1.8 & 42.71 & 883 $\pm$ 10 \\ 
 \rule{0pt}{2ex}
 J1108+0659 & 11:08:51.03 & +06:59:01.4 & 0.181 & 24.0 & 11.1 $\pm$ 0.5 & 42.32 & 660 $\pm$ 5 \\ 
 \rule{0pt}{2ex}
 J1114+1939 & 11:14:23.81 & +19:39:15.8 & 0.199 & 24.0 & 8.4 $\pm$ 0.5 & 42.30 & 650 $\pm$ 6 \\ 
 \rule{0pt}{2ex}
 J1116+2200 & 11:16:25.34 & +22:00:49.3 & 0.143 & 23.7 & 10.5 $\pm$ 0.5 & 42.38 & 465 $\pm$ 17 \\ 
 \rule{0pt}{2ex}
 J1222-0007 & 12:22:17.85 & -00:07:43.7 & 0.173 & 23.6 & 4.5 $\pm$ 0.4 & 42.85 & 839 $\pm$ 56 \\ 
 \rule{0pt}{2ex}
 J1316+1753 & 13:16:42.90 & +17:53:32.5 & 0.150 & 23.8 & 10.3 $\pm$ 0.5 & 42.77 & 1165 $\pm$ 8 \\
 \rule{0pt}{2ex}
 J1356+1026 & 13:56:46.10 & +10:26:09.0 & 0.123 & 24.4 & 62.9 $\pm$ 1.9 & 42.73 & 871 $\pm$ 72 \\ 
 \rule{0pt}{2ex}
 J1430+1339 & 14:30:29.88 & +13:39:12.0 & 0.085 & 23.7 & 26.5 $\pm$ 0.9 & 42.62 & 772 $\pm$ 9 \\ 
 \rule{0pt}{2ex}
 J1518+1403 & 15:18:56.27 & +14:03:19.0 & 0.139 & 23.6 & 8.6 $\pm$ 0.9 & 42.13 & 520 $\pm$ 28 \\ 
 \hline
 \end{tabular}
\caption{(1) Source name; (2) -- (3) Optical RA and Dec positions from SDSS (DR7) in the format hh:mm:ss.ss for RA and dd:mm:ss.s for DEC; (4) spectroscopic redshift of the source from SDSS DR7 \citep[with an rms error on the redshift of 0.025,][]{Abazajian09}; (5) Rest-frame 1.4 GHz radio luminosities from NVSS using a spectral index of $\alpha = -0.7$ and assuming ($S_{\nu}$ $\propto$ $\nu^{\alpha}$). The typical log errors are $\sim$ 0.03; (6) 1.4 GHz flux density of the target from NVSS; (7) Total observed [O~{\sc III}]$\lambda$5007 luminosity calculated using the fluxes from \citet{Mullaney13}, the typical log errors are $\sim$ 0.01; (8) The line width ($W_{80}$) of the [O~{\sc III}]$\lambda$5007 line measured from SDSS spectra.}
\label{tab:sampleintro}
\end{table*}

\section{Observations and data reduction}
\label{sec:observations}

We use APEX and the Atacama Compact Array (ACA) to observe the carbon monoxide (CO) emission in the CO(1-0), CO(2-1), CO(3-2), CO(6-5) and CO(7-6) transitions for our sample of 17 Type 2 quasars (as detailed in Table~A1 in the supplementary material and emission line properties are provided in Tables~{A2, A3, A4, A5 and A6 in the supplementary material.} APEX is a single dish, 12 metre diameter telescope whereas the ACA is a subset of the Atacama Large Millimeter/submillimeter Array (ALMA) comprising of twelve 7 metre antennae. A description of all the observations used in this paper along with details of how the data was reduced is provided below.

\subsection{CO(1-0) observations}
\label{sec:ACAobs}

Thirteen out of our 17 targets have CO(1-0) ACA observations [proposal ID: 2019.2.00194.S, PI: Calistro-Rivera], acquired between December 2019 and March 2020. Sources from this sample without CO(1-0) data are J1010+0612, J1010+1413, J1356+1026 and J1430+1339.

The required sensitivity for the ACA observations were estimated based on two different approaches. In the case of sources with archival infrared data around the dust SED peak, a conversion was made from total IR (L$_{\rm IR}$) from SED-fitting to CO luminosities L'$_{\rm CO}$. Otherwise, conversions were estimated based on the SED-inferred stellar mass and the average gas fraction value.

We image the CO(1-0) emission using the TCLEAN function in CASA and apply natural weighting with the H\"ogbom deconvolver. Bin widths of 100 \kms were used for non-detections and 50 \kms bins were used if the S/N was high enough to see more structure in the line profile. In a few specific cases, slightly different bin sizes were used either to match to other available data or as a result of the data quality.

The beam size of the ACA observations ranged between 12 -- 14 arcsec. However, to obtain the CO(1-0) spectra we take an aperture equivalent to the APEX beam size when observing CO(2-1), which is $\sim$ 30 arcsec diameter at an observing frequency of $\sim$ 200 GHz (observation frequency of CO(2-1) at the samples median redshift of 0.14). Using this aperture consistently to extract the spectra allowed us to compare the fluxes obtained from the same regions, making calculations of line ratios and other properties more reliable. It further allowed us to investigate whether any extended diffuse gas was present, or at least detectable, when comparing to smaller apertures. The apertures used for the extraction and the contours of the CO(1-0) ACA data are shown in the Appendix (Figure~C1 in the supplementary material), plotted over rgb images from the DESI Legacy Imaging Survey in the (z, r, g) bands.
The 2 sigma contours in the highest S/N data show an extent of up to 27 arcsec (in J1108+0659). Further, those with low S/N (such as the cases of J0945+1737 and J1055+1102) show positional offsets extending out to the 30 arcsecond diameter aperture and slightly beyond. As positional uncertainty in the observations is proportional to $\frac{\rm beam~size}{\rm S/N}$, using the 30 arcsecond aperture also allows us to account for these potential offsets and ensure we accurately measure the fluxes. From the CO(1-0) data there are no signs of any companions that are spatially and spectrally aligned with our targets, such that they would impact upon the measured flux values, aside from the apparent mergers occurring in J1222-0007 and J1518+1403 (which are both treated as single systems in this work). Companions that are visible in the background rgb images do not appear in the ACA data and so either are not emitting at those frequencies, or our observations are not deep enough to observe the emission from them. Therefore we can be confident of the fluxes measured in our ACA observations and that these also represent the total CO fluxes in these galaxies. Since this is the only aperture we have control over for the flux/spectra extraction, we choose to match this to the APEX CO(2-1) aperture of 30 arcsec to be as consistent as we can be in the region we are calculating fluxes.

Recent work in the literature has shown evidence for extended, low surface brightness emission in quasars, with CO emission detected out to 100s kpc.\citep[e.g.][]{Cicone21, Li21, Scholtz23}. This provides further support to the approach taken here, where we extract spectra with an aperture diameter of $\sim$ 80 kpc at the median redshift.

\begin{figure}
\centering
\includegraphics[width=\columnwidth]{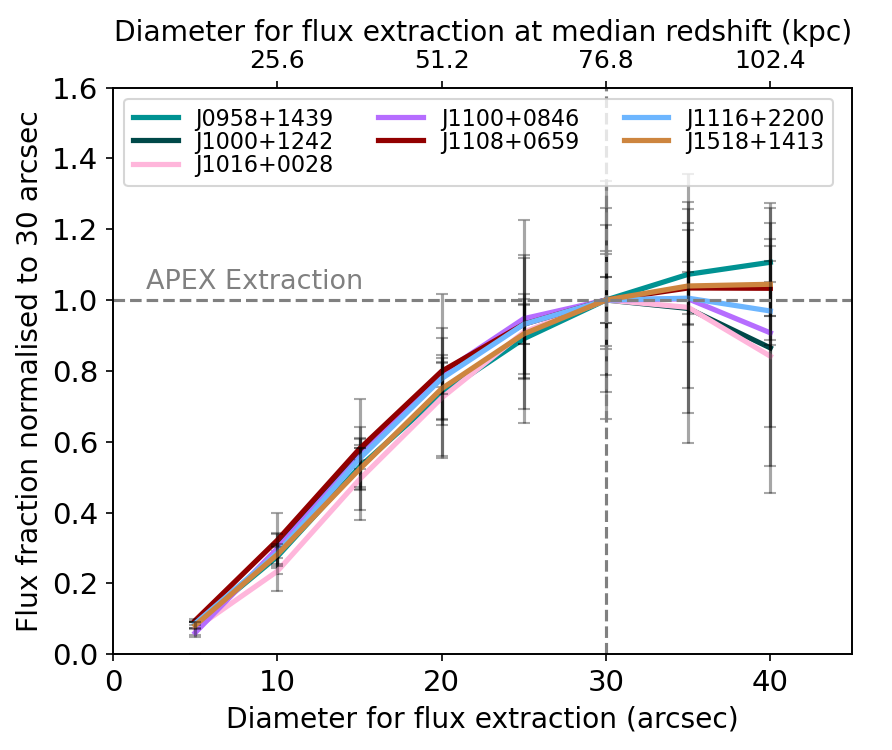}
\caption{For the 7 targets with S/N > 5 in CO(1-0) we plot the flux extracted from various diameter apertures of the ACA data up to 40 arcsec. The flux presented here is normalised to 30 arcsec which is where we extract our flux to match with APEX data (dashed lines). Uncertainties on each flux estimate are presented with grey error bars.}
\label{fig:curves_of_growth_kpc}
\end{figure}

To determine whether we obtain the total flux we plot the curves of growth of the ACA CO(1-0) (see Figure~\ref{fig:curves_of_growth_kpc}) where we indeed see that extracting the spectra at 30 arcsec is required to obtain a more accurate total flux value. Beyond 30 arcsec the flux density flattens off in almost all cases (note that in this figure we only plot those with an integrated signal-to-noise ratio (S/N) greater than 5). For a few sources we note that the curves of growth do continue to rise slightly after 30 arcsec, but by $<$ 10\% and within uncertainties. Given the larger uncertainties, we are still confident that we are consistent with obtaining the total flux. We also find that the curves of growth up to 30 arcsec follow the same trend and are consistent with each other so no conclusions can be drawn about any differences in morphology at these scales, with respect to galactic or feedback properties.

We note that if the CO(1-0) spectra were extracted using a 3$\sigma$ minimum level, we would measure on average 60\% less flux when compared to the extraction at 30 arcsec, and in one case almost 90\% less flux (values range from 25 -- 89\%). Such differences would have a significant impact on the analysis of the excitation, stressing the importance of low-resolution data for a complete census of molecular gas content.

From available multi-wavelength data \citep{Jarvis21} we also note that J1518+1403 has a secondary source located 14.7 arcsec away to the north east and J1222-0007 has a secondary source 4.7 arcsec away which are both likely to be on-going mergers. Further, J1108+0659 and J1356+1026 show evidence of hosting two nuclei. In all these cases the flux from secondary sources is likely to be included in the flux calculations for each. However, since they will also be covered by the CO(2-1) and CO(3-2) APEX observations then this is unavoidable in the analysis. No other sources have known companions that would affect the total fluxes measured.

\subsection{CO(2-1) Observations}
\label{sec:co21obs}

The CO(2-1) emission of 8/17 sources of the sample were observed with SEPIA180 on APEX between 09 December 2020 and 21 June 2021, (proposal ID: E-0105.B-0713A-2020 [PI: Calistro-Rivera]). 
The remaining 9 targets had equivalent APEX archival CO(2-1) observations presented as a pilot sample by \cite{Jarvis20} (Proposal ID: E-0100.B-0166, [PI: Jarvis]). For this work we have re-analysed the raw archival data from \cite{Jarvis20} using the new analysis techniques presented here for consistency, however we note that we find the same results reported by \cite{Jarvis20} within uncertainties.

The data was reduced using the standard procedures in the Continuum and Line Analysis Single-dish Software (CLASS) \citep{Pety05}. In all cases the reduction of the APEX data was done using a consistent strategy by modifying the template reduction script from APEX in CLASS. As with the CO(1-0) data we binned to 50 and 100 \kms where appropriate. 
The observing frequency range of 192.3 -- 212.5 GHz yields an APEX beam size in the range of 29 -- 32 arcsec, corresponding to a physical size of 68 -- 75 kpc at the median redshift of 0.14.

We fit Gaussians using standard procedures in python to obtain the integrated flux values and corresponding uncertainties for each target. We note that the results of fitting Gaussians in python match that of the Gaussian fits produced in CLASS. There is no spatial information for these data, however the beam size is large enough to cover the host galaxy, and so we are confident we measure the total flux, including any diffuse gas and do not over resolve. 

Upper limits are calculated using line widths estimated from other CO transitions for the same targets where available. If none are available, the average CO $W_{80}$ in that transition for all targets is used, for which these values are 397 \kms for CO(1-0), 477 \kms for CO(2-1) and 507 \kms for CO(3-2).
We note that this strategy is different from the calculations used by \cite{Jarvis20}, where upper limits were calculated by taking the maximum CO line width from the sample in CO(2-1). The case by case approach based on information from the other CO lines for each target, and assuming we should see similar line widths between transitions, should therefore provide a more accurate and constraining upper limit. 
The two cases for which both methods have been applied are J0958+1439 and J1356+1026. For J0958+1439, the estimated CO(2-1) upper limit estimated with our method is 30 per cent of the value reported in \cite{Jarvis20}, whereas for J1356+1026 our estimate is 93 per cent of the value reported in \cite{Jarvis20}.

An important note to make is that during the period in which CO(2-1) and CO(3-2) observations were taken with APEX, the telescope was operating at significantly different efficiencies (at most by a factor of 40\%). To account for this and to achieve accurate flux measurements, corrections have been made based on the following.
Main beam characteristics have been determined from de-convolved continuum slews across Mars, Uranus and Jupiter. Using CO(3-2), this yielded a mean beam size $\theta_{mb}$ = 17.5 $\pm$ 0.2~$\arcsec$, which we confirmed to be consistent with data on the CO line pointing sources (which are standard AGB stars used for pointing and focus calibration during observations).
To determine the main beam efficiency, we used cross scans\footnote{See http://www.apex-telescope.org/telescope/efficiency/index.php} obtained between December 2020 and December 2021 and cross-checked the result against the CO(3-2) flux of 8 line intensity monitoring sources\footnote{See https://www.apex-telescope.org/ns/apex-data/}. This analysis yielded main beam efficiencies (at 345 GHz) that depended on periods $\eta_{mb}$ = 0.63 $\pm$ 0.04 (Dec 2020), 0.55 $\pm$ 0.04 (May -- Jun 2020), and 0.67 $\pm$ 0.04 (Aug -- Dec 2020) and an antenna gain factors of Jy/K = 45 $\pm$ 4, 51 $\pm$ 4, 42 $\pm$ 4, respectively, which were converted to the science frequencies using the Ruze formula.
In the observations we used the wobbler in symmetrical mode with an amplitude of 50~$\arcsec$ and frequency of 0.5 Hz. Pointing and focus were checked regularly against sources from the APEX line pointing catalog using the CO(3-2) emission line. We estimate the overall calibration uncertainty at 10\% and that the pointing accuracy was typically within 2~$\arcsec$. Baselines were stable and we only had to fit a first order baseline to each scan before averaging them.

\subsection{CO(3-2) Observations}
\label{sec:co32obs}

Of the 17 sources in our sample, 16 were observed in CO(3-2) with SEPIA345 between 09 December 2020 and 30 December 2021 (proposal ID: E-0105.B-0713B-2020 [PI: Calistro-Rivera]). The observing frequency range of 288.4 -- 318.7 GHz yields an APEX beam size in the range of 19 -- 22 arcsec, corresponding to a physical size of 44 -- 51 kpc at the median redshift of 0.14. These data were reduced, and the flux densities were extracted in the same way as CO(2-1), as described in Section~\ref{sec:co21obs}.

For the one remaining target, J1430+1339, we utilise archival CO(3-2) ACA observations (proposal ID: 2016.1.01535.S [PI: Lansbury]) taken on 03 November 2016. These data were available on the ALMA archive with bin widths of 27 \kms. The coverage in the velocity space is not as wide as that of the equivalent APEX observations. Caution should be taken in this case as the difference in spatial resolution (here a beam size of 4.3 arcsec and a maximal recoverable scale of 23 arcsec) means that there is a possibility it is slightly over resolved and perhaps missing flux, but even with these caveats it provides a useful data point.

\subsection{CO(6-5) and CO(7-6) Observations}
\label{sec:co65obs}

We observed 7 targets in either CO(6-5) or CO(7-6) using SEPIA660 on APEX, which were selected for observation based on their brightness in the lower CO transitions. CO(6-5) and CO(7-6) were chosen to give an indication of the excitation in these higher transitions, with CO(6-5) preferred, but if it was not observable (due to the frequency range offered by SEPIA660) then we chose CO(7-6) instead. We also note that [C~{\sc I}](2-1) are also covered by our CO(7-6) observations, however, since all three have non-detections and there are no signs of detection across the obtained spectra, we do not perform any further analysis.

All observations were taken between May -- November 2022 (proposal ID: E-0109.B-0710 [PI: Molyneux]). Previous observations of 3 targets (J1010+0612, J1100+0846 \& J1430+1339) are also utilised by combining this archival data to our own (proposal id. E-0104.B-0292 [PI: Harrison]). These data were reduced in the same way as CO(2-1) (see Section~\ref{sec:co21obs} for details). 

From the range of frequencies 613 -- 708 GHz, the corresponding beam size was 9 -- 10 arcsec, which relates to a physical size of 21 -- 23 kpc at the median redshift of 0.14. This beam size should still allow us to retrieve the full flux values for two reasons. Firstly, for targets in this sample that are observed at higher spatial resolution (0.2 arcsec) and presented by \cite{Almeida21}, the moment maps and position velocity diagrams show that the CO emission is confined within the APEX beam size. Furthermore, we would expect the CO(6-5) and CO(7-6) to be more compact than the emission of the lower transitions and we are therefore confident that we are measuring the total flux in these data. One caveat would be that if any extended, diffuse emission exists in these higher CO transitions we would potentially be resolving out some of the flux, but we consider this unlikely due to the reasons outlined above.

\subsection{Ancillary Multiwavelength Data}
\label{sec:ancillary}
To achieve a detailed characterisation of the AGN feedback processes in our sample, the 17 quasars studied in this work have ancillary radio and optical data from the VLA and MUSE on the VLT, respectively. Here we describe these data and all values used in this paper can be found in Table~\ref{tab:sampleintro}. 

VLA radio data are available for all 17 quasars in the sample at 1--6 GHz and at a resolution of 0.3 -- 1 arcsec. For a full review and analysis of the radio data see \cite{Jarvis19, Jarvis21}. In this work we utilise knowledge of the 1.4 GHz radio data to aid in the interpretation of our findings. 
The quasars show a range of moderate radio luminosities of $\rm \log (L_{\rm 1.4\,GHz} / W Hz^{-1})=$ 23.5 -- 24.4. Crucially, although the quasars in our sample are `radio-quiet', according to widely-used radio-loudness definitions \citep{Xu99}, the bulk of them exhibit extended radio structures on 0.2 -- 34 kpc scales, with evidence of jets and/or shocked winds being the dominant cause of the extended radio structures \citep{Jarvis19, Jarvis21}.

There is also strong evidence for these radio jets interacting with the ionised gas and driving outflows \citep{Jarvis19, Girdhar22}. We therefore know that within this sample we are observing a diversity of radio AGN emission and ionised outflows in quasars. Further, from these ancillary data we know that this sample is dominated by AGN which are driving outflows, host jets/winds and show interactions with the ISM. The question remains however, as to how these feedback mechanisms impact the molecular gas properties, which we aim to address in this work.

We have also obtained MUSE VLT observations for these 17 quasars (proposal ID: 0103.B-0071 [PI: Harrison]). In this work we use the MUSE data to extract spectra of the [O~{\sc III}]$\lambda$5007 emission line where possible, and otherwise [O~{\sc III}]$\lambda$4959 (1 case) or $\rm H\beta$ (2 cases) if no [O~{\sc III}]$\lambda$5007 line were available. We use these lines as tracers of the ionised gas kinematics and to compare to the molecular CO gas presented here.
The spectra were all extracted using the same aperture (diameter $\sim$ 30 arcsec) as our APEX data (details in Section~\ref{sec:observations}) to make a comparison of ionised gas on the same scales. Specifically we use the [O~{\sc III}] $W_{80}$ and the properties of the line profile to analyse the differences between the impact of feedback on the ionised and molecular gas properties.

\section{Analysis and results}
\label{sec:results}

In this section we present the main analysis and results of this work. Firstly, in Section~\ref{sec:comparisons} we introduce the comparison samples that are used to put our results into context of the overall population of both AGN and non-AGN. We then present the analysis undertaken of the observed CO transitions in Section~\ref{sec:spectra}.
We present the calculations and results of the molecular gas masses and gas fractions in Section~\ref{sec:gasmass}. Finally, we present our findings on the CO excitation via the use of CO Spectral Line Energy Distributions (CO SLEDs) and CO line ratios (Section~\ref{sec:coexcite}).
Further, an example of the spectra obtained can be found in Figure~\ref{fig:J1100_Spectra_ex} and the remaining spectra, alongside tables of the line properties, are contained within the supplementary material.

\subsection{Comparison samples}
\label{sec:comparisons}

Throughout Section~\ref{sec:results} we present comparison samples from the literature to put our work into context and aid in the interpretation of our analysis. These comparison samples are described below:

We first utilise non-AGN and AGN from \cite{Tacconi18} to put the gas fractions of our sources in context and show that they are consistent with both AGN and non-AGN. The comparison sample is a compilation of data from xCOLD GASS \citep{Saintonge17}, EGNOG \citep{Bauermeister13} and GOALS \citep{Armus09} surveys as well as from the sample presented in \cite{Combes11}. 
We matched this sample to be within $z$ $\pm$ 0.05 of the full range of redshifts spanned by our sample. AGN hosts for the sample were identified using BPT-based AGN classifications. The galaxies in this comparison sample also span the full range of stellar mass, sSFR, and $\Delta$MS found for our sample \citep[see Figure 6 in][]{Jarvis20} meaning that the dependency on the specific star-formation rate has been removed and we are focusing on any possible impact of having an active BH rather than the star-formation efficiency of the given galaxy. To ensure consistency in the comparison, the molecular gas masses presented for our sample are calculated using the same method as shown in \cite{Tacconi18}. This comparison is presented in Figure~\ref{fig:Miranda_Tacconi}. For further information on this comparison sample also see \cite{Jarvis20}.

In our CO SLEDs (Section~\ref{sec:COSLEDs}) we utilise the compilations by \cite{Valentino21} and \cite{Carilli13} to compare to our CO SLEDs. From \cite{Valentino20, Valentino21} we present the CO SLEDs (both Figures~\ref{fig:CO_SLED} and \ref{fig:CO_SLED_hightrans}) of starburst and main sequence galaxies in the redshift range $z \sim$ 1 -- 2 with L$_{IR}$ > 10$^{12}L\odot$. These luminosities are similar to those in our sample (see Section~\ref{sec:ancillary}) and therefore provide a useful comparison.
From \cite{Carilli13} we utilise the compilation of high-$z$ quasars ($z$ $\sim$ 1 -- 6), shown as "C\&W13" in Figures~\ref{fig:CO_SLED} and \ref{fig:CO_SLED_hightrans}, to see how our low-z quasar sample compare to these more distant and more luminous objects ($\rm L_{bol}$ > 10$^{47}$ erg$s^{-1}$, compared to QFeedS with $\rm L_{bol}$ < 10$^{46.5}$ erg$s^{-1}$). 
We also show these high-$z$ quasars in Figure~\ref{fig:lineratio} as individual points to show how they compare in the lower transitions to the range of line ratios in our sample and other comparison samples listed below. 
Finally, in Figure~\ref{fig:LineRatio_BolLum} we present the line ratios of our high J$_{\rm CO}$ transitions compared to the high-$z$ quasar sample, as a function of bolometric luminosities.

As mentioned above, in Figure~\ref{fig:lineratio} we present the range of line ratios found in our sample compared to others in the literature in the form of violin plots. \cite{Montoya23} analysed a sample 40 local (U)LIRGs ($L_{\rm IR, SF}$ $\gtrsim$ $10^{12}$ L$_\odot$) in the same redshift range as our sample ($z < 0.2$). The targets were selected based on OH absorption and not on the presence of radio jets, however this does not exclude radio jets being present. This sample also shows a range of AGN fractions, from 0 -- 0.92, with 50 \% having an AGN fraction greater than 0.5. They find no correlation between AGN fraction or AGN luminosity within the sample of (U)LIRGs. Since 8 sources out of 9 in the QFeedS sample with the required measurements are known to be LIRGs, \cite{Montoya23} provides a useful comparison to determine whether the presence of radio jets or shocked winds and ionised outflows found in our sample makes a significant difference to the observed line ratios.

In Figure~\ref{fig:lineratio} we also perform a similar comparison to a sample of (U)LIRGs at $z$ $\leq$ 0.1 from \cite{Greve14}.
This sample of (U)LIRGs was selected against AGN, all with an AGN contribution of $\lesssim$ 0.3. With this, alongside the sample in \cite{Montoya23}, we have comparisons samples with similar IR luminosities and a range of AGN contribution. Since the QFeedS sample is compiled of quasars with LIRG-like infrared luminosities, 
but with additional known radio jets and ionised outflows, any differences in the excitation of CO could potentially be attributed to the jet and outflow properties of our sample.

In order to also investigate how local AGN (with a median $z$ $\sim$ 0.05) with lower luminosities (median $\rm L_{bol}$ $\sim$ $10^{44.8}$) compare with our sample, we utilise the sample by \cite{Lamperti20}. We can therefore test how our more luminous quasars are different in CO excitation. This comparison sample comprises of X-ray selected AGN, for which further information can also be found in \cite{Ricci17,Koss21}. These data are also used in Figure~\ref{fig:lineratio}. Finally, we also used a compilation of local star forming galaxies as a comparison \cite{Leroy22}, which includes data from HERACLES \citep{Leroy09}, the James Clerk Maxwell Telescope (JCMT) Nearby Galaxy Legacy Survey \citep[NGLS][]{Wilson12}, the CO Multiline Imaging of Nearby Galaxies (COMING) survey \citep{Sorai19}, PHANGS ALMA \citep{Leroy21a}, IRAM 30m CO (2–1) observations, and Large APEX Sub-Millimetre Array (LASMA) CO (3–2) observations.

\subsection{Spectral properties}
\label{sec:spectra}

We analyse the spectral properties across different CO transitions for all targets in our sample to investigate the integrated fluxes, line profiles, including line widths, velocity offsets and features within the lines (e.g. potential outflow components). These can then be compared to properties of the host galaxies and the line profiles of the ionised gas to search for any influence of AGN activity.

Spectra are mostly plotted for each target with the same bin widths across all transitions so that the line profiles of each transition can be easily compared (see Figure~\ref{fig:J1100_Spectra_ex} and the remaining spectra in the supplementary material, Figures B1~--~B17).
In some exceptional cases, where we had enough S/N in some transitions to investigate the line profile in more detail, but not enough S/N in other transitions, we choose the bin widths accordingly.
Central frequencies (where $v$ = 0 \kms) have been defined using the SDSS redshifts quoted in Table~\ref{tab:sampleintro}. We choose the SDSS redshifts as this has been used throughout the QFeedS survey work, and since we are only comparing CO and ionised gas lines within the same target, the specific reference velocity/redshift is not important. The aperture from which all the spectra are taken is consistent between CO(1-0) and CO(2-1) for each source, and for higher transitions will be slightly smaller due to the APEX beam size reducing as observing frequency increases (as mentioned in Section~\ref{sec:observations}). However, our analysis is done in a way that is consistent between transitions and therefore any differences found are potential indications of the impact from feedback mechanisms being different on the different CO transitions.

As an example we present the spectra of all CO transitions (CO(1-0), CO(2-1), CO(3-2), and CO(6-5)) and the MUSE [O~{\sc III}] emission for one target in the main body of the paper (J1100+0846 shown in Figure~\ref{fig:J1100_Spectra_ex}) and all remaining spectra are then presented in the supplementary material. For all sources, [O~{\sc III}] line profiles are plotted in a separate panel below the CO spectra for comparison (in some cases the [O~{\sc III}] line was not available so the $\rm H\beta$ emission line was used instead) extracted from MUSE data using the same aperture as that of the CO data. For some sources of our sample (J1010+0612, J1100+0846, J1356+1026 and J1430+1339), ALMA observations of the CO(2-1) at 0.2 arcsec resolution presented by \citet{Almeida21} are available. We show these ALMA spectra plotted in orange over our APEX CO(2-1) spectra to compare. Making these comparisons required a velocity shift to the \cite{Almeida21} data to match the zero velocity used here, which we determined using the SDSS redshift, as opposed to the approach taken in \cite{Almeida21}. Specifically, they used the SDSS redshift as the initial $v$~=~0~\kms and then applied a small shift to make the peak (or centre of two peaks) at $v$~=~0~\kms. Despite this small difference, we find consistent flux values and line profiles within errors when comparing to \cite{Almeida21}. We also note the availability of CO(1-0) and CO(3-2) data for J1356+1026 from \cite{Sun14}, however due to the differences in resolution (at 1.3 and 0.6 arcsec for CO(1-0) and CO(3-2) respectively) and therefore high chance of over-resolving the total CO emission compared to the QFeedS observations, these are excluded from any analysis.

From a first look at our CO spectra it is immediately clear that there is a large variety of line profiles, luminosities and detections for the different transitions within our sample. 
We also find a large diversity of broad line profiles, double peaked profiles as well as a blue wing and offsets from $v$ = 0 \kms. A detailed comparison of the molecular and ionised gas line profiles is discussed in Section~\ref{sec:CO_vs_OIII}.

To analyse the line profiles of both the CO transitions and MUSE [O~{\sc III}] spectra, we fit either one, two, or three Gaussian components to the spectra where appropriate. 
In the CO data, since we have relatively low S/N, we find only two targets with more than one Gaussian component. 
However, in the MUSE data we see a wide range of line profiles, including many with multiple components. From the data we analyse the central velocity ($V_{50}$) and line width over which 80 percent of the flux is contained ($W_{80}$) which can either be done using the fits to the data or to the data itself. We choose to present the values calculated on the fits to the data for the following reasons.

There are 21 cases for which we are measuring $V_{50}$ and $W_{80}$ of spectra with S/N > 5 (used in future analysis) and of those 21, 17 of the $W_{80}$ values from the data are within the uncertainties of the $W_{80}$ from the fit, whilst 4 are outside the fit uncertainties. Three sources have higher S/N data published by \cite{Almeida21}; of these, two have line widths closer to the $W_{80}$ values from the fit, while one is closer to the $W_{80}$ of the data. Finally, of these 21 cases there are 16 where the $W_{80}$ of the fit is larger, while there are five where the $W_{80}$ of the data is larger. Given that both measures are mostly consistent with each other, that in two out of three cases the $W_{80}$ of the fit is closer to higher S/N data in the literature and that in most cases $W_{80}$ of the fit is larger, we decide that the $W_{80}$ of the fits is more appropriate for this work. Since we are looking at relatively low S/N data across the sample, either measure used will give large uncertainties and so using the larger of the two means we are closer to the maximum value from the data, which is considered later in the analysis of the line profiles and further discussion in Section~\ref{sec:CO_vs_OIII}.

Therefore, in all cases we use the $V_{50}$ and $W_{80}$ of the fits to the data to analyse as the description of the line profiles.
In the case of spectra which show a single Gaussian profile, the uncertainties on $V_{50}$ and $W_{80}$ are the errors on the fit. The uncertainties for the total flux is the uncertainty on the fit plus the uncertainty on the telescope efficiency.
For those with multiple components, the uncertainty on $V_{50}$ presented is the uncertainty on the narrowest components peak velocity. The uncertainty on $W_{80}$ is the uncertainty on the $W_{80}$ of the broadest component. The uncertainty on the flux for those with multiple components is the uncertainties on each component added in quadrature plus the uncertainty on the telescope efficiencies ($\sim$ 5 -- 10\% for all observations).
The noise for each channel for APEX data was taken from the results of the CLASS reduction scripts. For ACA observations the rms was calculated by taking the median flux from the remaining spectra, whilst masking the line (where present).

\begin{figure}
\centering
\includegraphics[width=0.9\columnwidth]{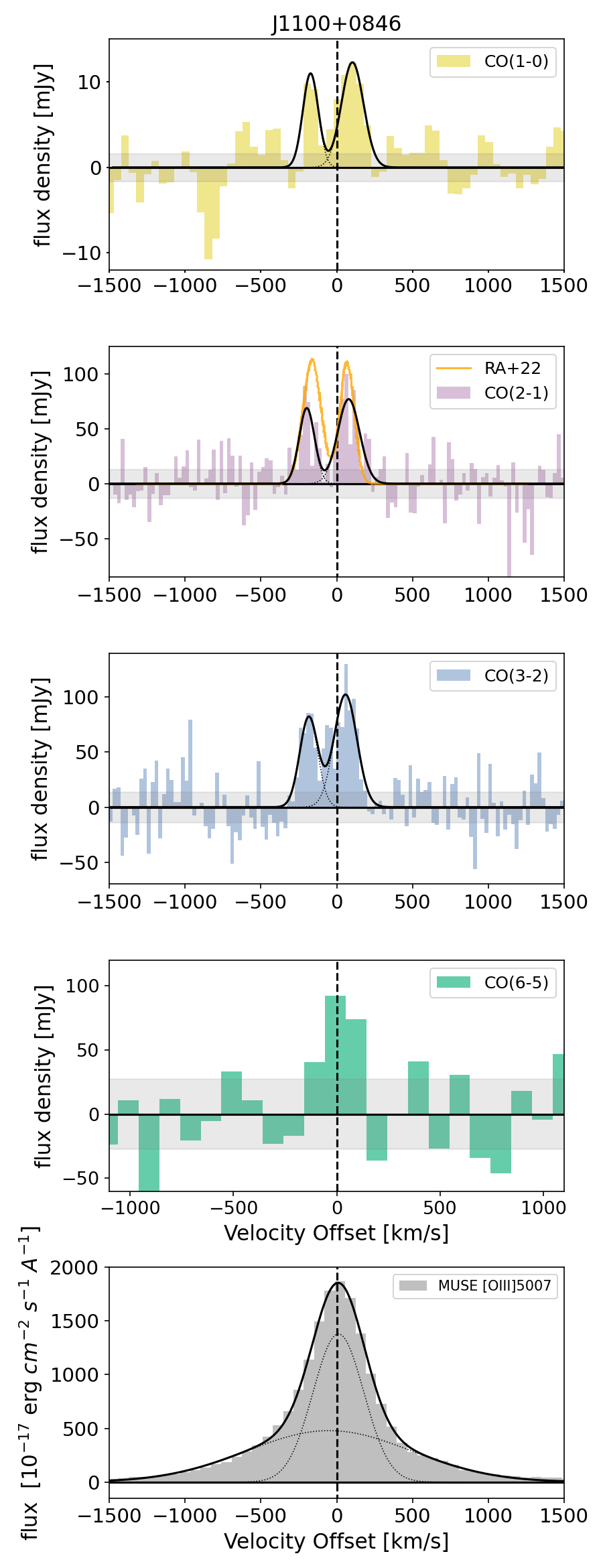}
\caption{Example of multiple CO spectra and MUSE data obtained, here showing J1100+0846. Top panel: CO(1-0) from ACA. Second panel CO(2-1) from APEX \citep[in this case the data was from][]{Jarvis20}. Third panel: CO(3-2) data from APEX. Fourth panel: CO(6-5) data from APEX. Fifth panel: MUSE spectra for the [O III] line extracted from a 30 arcsec diameter aperture. In all cases, solid black lines denote fits to the data. Here there are multiple components and as such, the dotted black lines denote the different components that make the total fit to the spectra. In the CO(2-1) spectra we also show the fit from higher resolution ALMA observations \citep[solid orange line labelled RA+22,][]{Almeida21}. Shaded grey regions represent the 1$\sigma$ level. Spectra for all targets are shown in the appendix, using the same presentation methods.}
\label{fig:J1100_Spectra_ex}
\end{figure}

In order to determine which lines and fits are robust and can be used for further analysis we measure the data quality based on the S/N of the lines within our sample.
We then define different levels of detection: "detections", "low S/N detections" and "non-detections", which are classified in the following way:

\begin{itemize}
    \item "Detections" are defined as spectra which show lines with an integrated S/N $\geq$ 5.\\
    \item "Low S/N detections" are defined as those lines with 3 $\leq$ integrated S/N $<$ 5.\\
    \item Anything with no clear line or a line with an integrated S/N $<$ 3 is defined as a "non-detection".
\end{itemize}

When referring to "detections" from this point onward we refer to both "detections" and "low S/N detections" as defined here unless otherwise stated.

For those with low S/N detections we present the fits to the spectra with a dashed black line to differentiate from detections shown by solid black lines. The integrated S/N values are also shown in the tables presenting CO data. For a list of the classifications of all detections, low S/N detections and non-detections see Table~A1 in the supplementary material.

There are three sources which show non-detections in all CO transitions which are J0909+1052, J1222-0007 and J1356+1026. However, we note that J1356+1026 does have a CO(2-1) detection in the deeper ALMA data from \cite{Almeida21} which we utilise in later analysis. Several other targets also show non-detections in at least one transition. To test whether the non-detections might show any low-brightness signal when combined, we stack these data by bringing the observations from different targets to the same velocity reference. However, from stacking non-detections we don't find any underlying flux and thus we can make no more conclusions based on these data. 

From these spectra we can calculate total fluxes, line widths ($W_{80}$), velocity offsets ($V_{50}$) and line luminosities (all data are presented in the appendix in Tables: A2, A3, A4, A5, A6 and A7.
For those with non detections we choose to present the 3$\sigma$ upper limits for CO flux and luminosity. Further observations would be required to confirm any of these detections. Line luminosities (L$'_{CO}$) are calculated using the following equation from \cite{Solomon97} \citep[also used in the analysis of previous QFeedS work][]{Jarvis20}: 

\begin{equation}
\label{eq:linelum}
\rm L'_{CO} [{\rm K~ km~ s^{-1} pc^2}] = \frac{3.25 \times 10^7}{\nu ^2_{\rm CO,rest}}\left(\frac{D^2_L}{1+z}\right)f
\end{equation}
where $\nu_{\rm CO,rest}$ is the rest frequency of the CO line, $\rm D_L$ is the luminosity distance, $z$ is the redshift and $f$ is the velocity integrated line flux density measured in Jy \kms.

\subsection{Molecular gas masses}
\label{sec:gasmass}

Studying the molecular gas masses in this sample will allow us to determine whether the presence of a quasar has an impact on the total gas fraction.
We calculate the CO(1-0) molecular gas masses using the mass–metallicity relation used by \citealt{Tacconi18} \citep[see also][]{Genzel15}, along with the following equations~\ref{eq:aco} and \ref{eq:tacc}: 

\begin{equation}
\label{eq:mco}
\rm M_{\rm CO} = \alpha_{\rm CO} \times L'_{\rm CO}(1-0)
\end{equation}
where $\rm L'_{\rm CO}(1-0)$ is the CO(1-0) luminosity and $\alpha_{\rm CO}$ is the conversion factor calculated as a function of metallicity (following the $\alpha_{\rm CO}$ calculation from \citealt{Tacconi18}, taking the geometric mean of the metallicity-dependent $\alpha_{\rm CO}$ recipes of \citealt{Genzel12} and \citealt{Bolatto13}): 

\begin{multline}
\label{eq:aco}
    \alpha_{\rm CO} = 4.36 \times \sqrt{
    \begin{aligned}
    0.67 \times \exp (0.36 \times 10^{-1 \times(12+ \log(O/H)-8.67)},\\
    \times 10^{-1.27\times(12+\log (O/H)-8.67)}
    \end{aligned}
    }
\end{multline}
where $\rm \alpha_{CO}$ has units $\rm M_{\odot}$ ( K \kms pc$^2$)$^{-1}$. Also following
\cite{Tacconi18} we use the following mass metallicity relation from \cite{Genzel15}:

\begin{equation}
\label{eq:tacc}
12 + log(O/H) = a - 0.087 \times (\log M_{\star} - b)^2
\end{equation}
where, $a = 8.74$, $b = 10.4 + 4.46 \times \log(1+z) - 1.78 \times (\log (1+z))^2$ and $\rm M_{\star}$ is the stellar mass obtained from SED fitting \citep{Jarvis20}. \\

For those without CO(1-0) detections we use the CO(2-1) luminosity where possible and convert to CO(1-0). Conversions made from $\rm L'_{CO}(2-1)$ use the median line ratios observed within this sample (presented in Table~\ref{tab:LineRatios}) as conversion factors (median line ratio of 1.06). Here the sources J1010+0612, J1010+1413, J1356+1753 and J1430+1339 are calculated using $\rm L'_{CO}(2-1)$. For those with non-detections across all CO transitions we provide 3$\sigma$ upper limits of M$_{\rm CO}$ based on the upper limits of the CO(1-0) flux.

The calculated values of $\rm \alpha_{CO}$ from our sample are within the range 4.0 -- 4.2 (shown in Table~\ref{tab:gasfrac}). These values are consistent with typical high redshift, high star-forming, quasar host galaxies \citep{Bolatto13, Carilli13}. However, $\rm \alpha_{CO}$ may also be significantly lower in LIRGs, submillimeter galaxies, mergers, starbursts and AGN, with values as low as $\sim$ 0.6 -- 1 \citep[e.g.,][]{Bolatto13, Sargent14, Gaby18}. Therefore there are uncertainties that arise in these values and the calculated gas masses. There can also be dependencies on the metallicity and SFR \citep[see e.g.][and references therein]{Bolatto13, Sandstrom13}, but for most galaxies a value of $\sim$ 4 is found, as is identified in our targets using the same method of calculation as the comparison sample in \cite{Tacconi18} (which we follow for consistency).

\begin{figure}
\centering
    \includegraphics[width=\columnwidth]{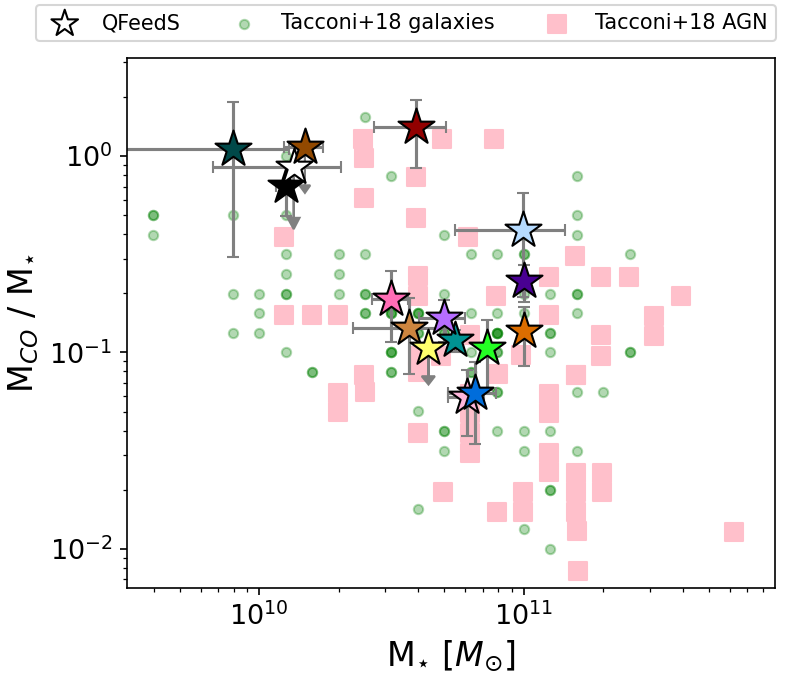}
    \caption{For all targets with data for both stellar mass (M$\rm _{\star}$) and CO gas masses (M$\rm _{CO}$) we present the stellar mass versus the gas fraction ($\rm M_{CO}$/M$_{\star}$). Large coloured stars show targets from this work, with colours as in Figure~\ref{fig:SampleSelection}. Background small green circles and pink squares show a compiled literature sample of AGN and non-AGN from \citealt{Tacconi18}. Here 16/17 targets are presented, with J1116+2200 missing due to a lack of stellar mass information. }
    \label{fig:Miranda_Tacconi}
\end{figure}

The molecular gas masses in our sample range from 0.36~--~5.50~$\times 10^{10}$ M$_\odot$ and all molecular gas masses can be found in Table~\ref{tab:gasfrac} along with stellar mass estimates obtained from SED fitting \citep[see][for a description of these calculations]{Jarvis20}. There are cases with large uncertainties which result from poor constraints on the SED fitting. These uncertainties also follow through into Figure~\ref{fig:Miranda_Tacconi} where we present the stellar and CO gas masses. Gas fractions are also presented in Table~\ref{tab:gasfrac} calculated as $\rm M_{CO}/M_{\star}$. We find values in the range 0.06 -- 1.4. These are used in Figure~\ref{fig:Miranda_Tacconi} to compare to AGN and non-AGN from the literature \citep{Tacconi18}. See section~\ref{sec:comparisons} for further information about this comparison sample. We find that our sources are consistent with the comparison sample of non-AGN and AGN. There are a few cases where we see higher gas fractions, in particular J0945+1737, J1108+0659 and J0958+1439 but these are still consistent with both AGN and non-AGN from the literature (see Figure~\ref{fig:Miranda_Tacconi}). In fact, AGN have been found to have similar, or higher gas fractions than non-AGN \citep[e.g.][]{Rosario18, Kirkpatrick19, Jarvis20, Shangguan20, Koss21, Zhuang21, Salvestrini22}. Further, in AGN CO gas can also be detected in a warm phase \citep[e.g.][]{Rosario19} so the molecular gas levels here could also be considered a lower limit of the total molecular gas content. However, it is important to note that even with the presence of high gas fractions, it does not exclude the possibility of AGN feedback (including cases with the presence of outflows) as shown in comparisons of simulations \citep{Ward22}.

\cite{Jarvis20} discuss gas fractions as well as analyse the star formation rate, specific star formation rate and distance from the main sequence for a subsample of these data. We note however, that there are differences in this work when comparing to the same targets in \cite{Jarvis20}, which used $r_{21}$ of 0.8 to convert from CO(2-1) to CO(1-0). Given that we have the CO(1-0) measurements here we simply use those, and as discussed before we find higher values of $r_{21}$ with a median of 1.06 within our sample. As such, the results of the total gas masses and gas fractions are also different here compared to \cite{Jarvis20}, with this work finding lower total gas masses, but within uncertainties and therefore still consistent.

\begin{table}
\begin{tabular}{ |c|c|c|c|c| } 
 \hline
 Name & $\rm \alpha_{CO}$ & M$\rm _{CO}$ &  M$_\star$  & M$\rm _{CO}$/M$_\star$ \\ 
  & & M$_\odot$ ($\times 10^{10}$)  & M$_\odot$ ($\times 10^{10}$) & \\ 
 \hline
 \rule{0pt}{2ex}
 J0909+1052 & 4.28 & < 1.20 $^a$ & 1.35 $\pm$ 0.68 & < 0.88$^a$ \\
 J0945+1737 & 4.24 & 0.89 $\pm$ 0.25$^a$  & 1.25 $\pm$ 0.10 & 0.71 $\pm$ 0.21$^a$ \\
 J0958+1439 & 4.24 & 0.64 $\pm$ 0.08$^a$  & 5.50 $\pm$ 0.10 & 0.12 $\pm$ 0.01$^a$ \\
 J1000+1242 & 4.34 & 0.87 $\pm$ 0.30$^a$  & 0.79 $\pm$ 0.50 & 1.09 $\pm$ 0.79$^a$ \\
 J1010+0612 & 4.20 & 0.59 $\pm$ 0.21 $^b$ & 10.00 $\pm$ 0.50 & 0.19 $\pm$ 0.07$^b$ \\
 J1010+1413 & 4.18 & 2.30 $\pm$ 0.50 $^b$ & 3.16 $\pm$ 0.10 & 0.23 $\pm$ 0.05$^b$ \\
 J1016+0028 & 4.05 & 0.36 $\pm$ 0.12$^a$  & 6.12 $\pm$ 0.95 & 0.06 $\pm$ 0.02$^a$ \\
 J1055+1102 & 4.05 & 0.40 $\pm$ 0.16$^a$  & 6.54 $\pm$ 1.35 & 0.06 $\pm$ 0.03$^a$ \\
 J1100+0846 & 4.16 & 0.75 $\pm$ 0.11 $^a$ & 5.01 $\pm$ 1.00 & 0.15 $\pm$ 0.04$^a$ \\
 J1108+0659 & 4.10 & 5.50 $\pm$ 1.27 $^a$ & 3.91 $\pm$ 1.18 & 1.41 $\pm$ 0.54$^a$ \\
 J1114+1939 & 4.05 & 4.17 $\pm$ 1.30 $^b$ & 9.94 $\pm$ 4.43 & 0.42 $\pm$ 0.23$^b$ \\
 J1116+2200 & 4.19$^c$ & 3.07 $^c$ & -- & -- \\
 J1222-0007 & 4.27 & < 1.65 $^a$  & 1.49 $\pm$ 0.25 & < 1.11$^a$ \\
 J1316+1753 & 4.27 & 1.28 $\pm$ 0.42 $^b$ & 10 $\pm$ 0.25 & 0.12 $\pm$ 0.04$^b$ \\
 J1356+1026 & 4.27 & < 0.46 $^b$ & 4.37 $\pm$ 0.10 & < 0.11$^b$ \\
 J1430+1339 & 4.25 & 0.76 $\pm$ 0.30 $^b$ & 0.79 $\pm$ 0.05 & 0.11 $\pm$ 0.04$^b$ \\
 J1518+1403 & 4.08 & 0.49 $\pm$ 0.08$^a$  & 3.68 $\pm$ 1.41 & 0.13 $\pm$ 0.06$^a$ \\

 \hline
\end{tabular}
\caption{We present $\alpha_{CO}$ (calculated using equation~\ref{eq:aco}), M$\rm _{CO}$ (calculated using equations~\ref{eq:mco}, \ref{eq:aco} and \ref{eq:tacc}), M$\rm _{\star}$ and gas fractions ( M$\rm _{CO}$/M$_{\star}$) for each target in the sample. For J1116+2200 the stellar mass is unconstrained by the SED and so there is no value of M$_{\star}$ (and therefore also $\rm \alpha_{CO}$). $^a$~Calculated from CO(1-0) line luminosity. $^b$~Converted to CO(1-0) from CO(2-1) using the median line ratio of those in the rest of the sample with detections. $^c$~Median $\rm \alpha_{CO}$ value of 4.19, and M$\rm _{CO}$ calculated from this median.}
\label{tab:gasfrac}
\end{table}

\subsection{CO excitation}
\label{sec:coexcite}

To measure and analyse the excitation of the molecular gas we analyse both the observed shape of the CO SLEDs (Section~\ref{sec:COSLEDs}) as well as the line ratios of the CO transitions (Section~\ref{sec:lineratios}).
Studying the excitation of the gas and comparing to literature samples of both AGN and non-AGN will again allow us to determine whether the excitation in the quasar host galaxies is systematically different to that of other relevant galaxy samples.

\subsubsection{CO SLEDs
\label{sec:COSLEDs}}

\begin{figure*}
     \centering
     \begin{subfigure}[b]{0.49\textwidth}
         \centering
         \includegraphics[width=\textwidth]{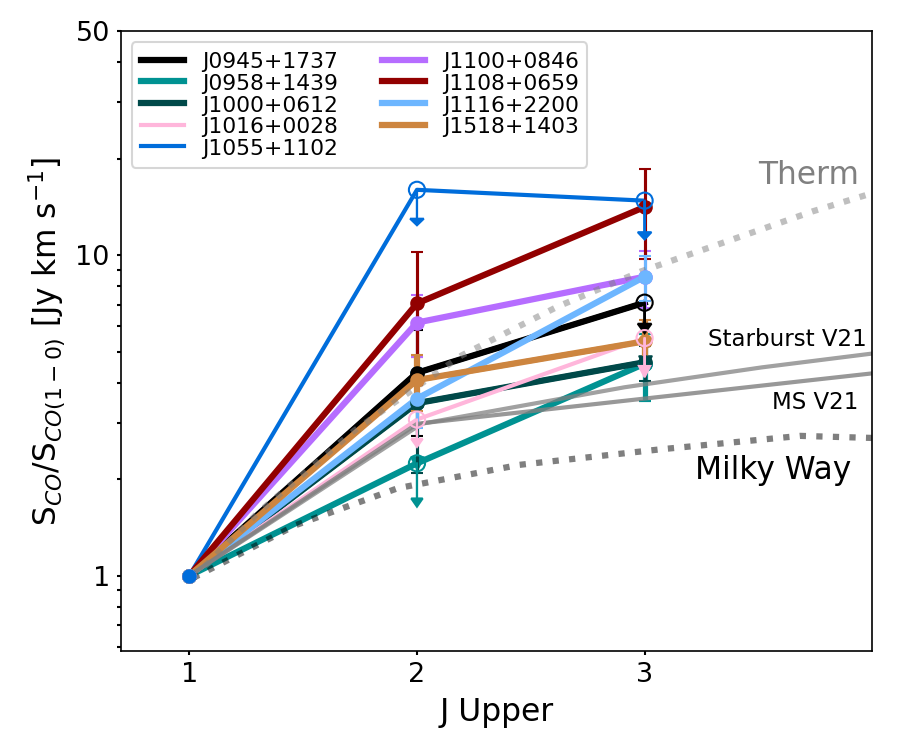}
         \caption{For the 9 targets which have detections in CO(1-0) we present the CO SLEDs, normalised to CO(1-0) with Milky Way and thermalised relations for comparison \citep{Carilli13}. Also shown are $z$ = 1 -- 2 Starburt and main sequence (MS) CO SLEDs from \cite{Valentino21} (V21). The CO SLEDs are colour coded by target (as in Figure~\ref{fig:SampleSelection}). 3 sigma upper limits are indicated by empty circles with downward arrows. No uncertainties are shown for CO(1-0) as these are incorporated into the uncertainties on CO(2-1) and CO(3-2) to best represent the uncertainty on the excitation level.}
         \label{fig:CO_SLED}
     \end{subfigure}
     \hfill
     \begin{subfigure}[b]{0.49\textwidth}
         \centering
         \includegraphics[width=\textwidth]{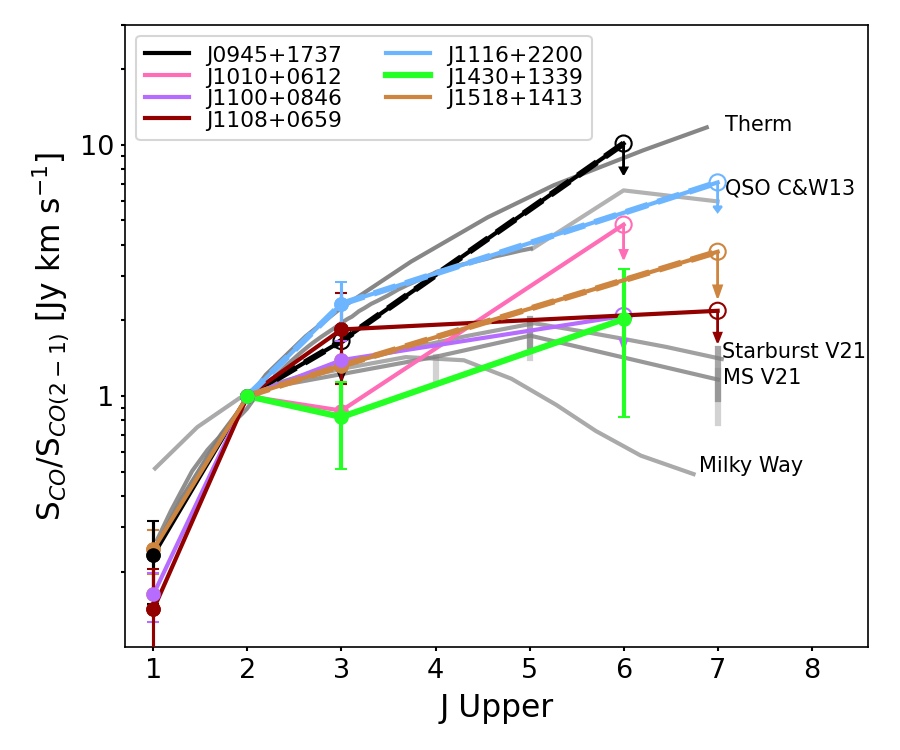}
         \caption{For the 7 targets which have either CO(6-5) or CO(7-6) data we present the CO SLEDs up to these higher transitions. Here we are normalising to the CO(2-1). Literature comparison data showing high-$z$ quasars \cite{Carilli13} (C\&W13) and $z$ = 1 -- 2 Starburst and main sequence (MS) \cite{Valentino21} (V21). 3 sigma upper limits are indicated by empty circles with downward arrows. No uncertainties are shown for CO(2-1) as these are incorporated into the uncertainties on CO(1-0), CO(3-2) and CO(6-5) to best represent the uncertainty on the excitation level.}
         \label{fig:CO_SLED_hightrans}
     \end{subfigure}
     \caption{CO SLEDs}        
\end{figure*}

We present the CO SLEDs of our sources in two different sets. Firstly, in Figure~\ref{fig:CO_SLED} we show CO SLEDs from the ground state up to CO(3-2), making comparisons to literature samples. We plot only those with detections in CO(1-0) (including low S/N detections), so that we have a reliable normalisation to the ground state. For those without a detection in CO(1-0) deeper observations would be required to provide a reliable CO SLED. The reference SLEDs shown here are the Milky Way and thermalised SLEDs, shown by dotted lines \citep{Carilli13}. We further make comparison to starburst and main sequence galaxies at $z \sim$ 1 -- 1.7 taken from \cite{Valentino21} (see Section~\ref{sec:comparisons} for further information on these comparison samples).

We report that the CO SLEDs of 2 out of 9 sources are consistent with being above the thermalised relation at the CO(2-1) level (excluding the upper limit of J1055+1102) and 2 out of 9 at the CO(3-2) level (see Figure~\ref{fig:CO_SLED}).

However, including the uncertainties on the CO(1-0) fluxes in addition to the CO(2-1) or CO(3-2) they are all still consistent with being thermalised, with the exceptions of $r_{21}$ for J1100+0846 and $r_{32}$ for J1010+1413 (see Table~\ref{tab:LineRatios} for individual values).

For sources within our sample with available observations in higher CO transitions (J$_{up}$ = 6, 7) we plot the CO SLEDs extending to these higher transitions and normalise the SLED to CO(2-1) instead of CO(1-0) (see Figure~\ref{fig:CO_SLED_hightrans}). 
Normalisation to the CO(2-1) transition was done for two reasons. Firstly, two of our targets with high CO transition data have not been observed in CO(1-0) and so to make comparisons within our own sample the next transition available with detections for all targets was CO(2-1). Secondly, based on our observations of several super-thermal SLEDs at CO(2-1) and the ongoing discussion in the community about the optical thickness of CO(1-0) (see Section~\ref{sec:high_r21}), we argue that CO(2-1) might be more reliable transition for normalisation. 
Again, in Figure~\ref{fig:CO_SLED_hightrans} we make comparisons to other samples from the literature. In addition to the literature samples mentioned previously we also include high-$z$ quasars ($z \sim$ 1 -- 6) compiled by \cite{Carilli13}, to make comparisons to the low-$z$ counterparts in our sample.

Out of the 7 sources observed, 6 show non-detections and only 1 source, J1430+1339, has a detection in either CO(6-5) or CO(7-6), based on the same criteria stated in Section~\ref{sec:spectra}.
We find that this detection is relatively low in the CO SLED, suggesting that the peak of the SLED may be at J$_{CO}$ < 6 (see Figure~\ref{fig:CO_SLED_hightrans}). It appears to be more similar to main sequence and starburst galaxies as opposed to high-$z$ quasars \citep{Carilli13}. 

For the remaining CO SLEDs in Figure~\ref{fig:CO_SLED_hightrans} we only have upper limits, however two of these are very constraining namely J1100+0846 and J1108+0659, which are at the same excitation level as the already detected J1430+1339). 5 targets are less excited that the high-$z$ quasar CO SLED, with 3 of these showing consistent excitation with starburst and main sequence galaxies. J1100+0846 shows signs of a detection, with a S/N of 2.2, but the 3$\sigma$ upper limit is at a similar excitation to our detected J1430+1339. J1108+0659 shows no sign of any detection at a very constraining level in the CO SLED, placing it already at a similar excitation to starburst and main sequence galaxies. Likewise, J1518+1403 shows no signs of a detection but is between the excitation of the high-$z$ and starburst CO SLED. J1010+0612 shows a very tentative signs of a line with a S/N of 1.1 at a higher excitation than starburst/MS galaxies, but still less than the high-$z$ QSO CO SLED. Further observations would be required to confirm any detection and place a proper constraint on the excitation. The remaining two targets (J0945+1737, J1116+2200) can't be analysed in much detail since the upper limits are not very constraining. However, the fact that these upper limits are already at the level of the thermalised relation and that of the high-$z$ quasars \citep[see Figure~\ref{fig:CO_SLED_hightrans} and][]{Carilli13} shows we are likely observing systems with a significantly lower excitation. This along with the other targets clearly showing lower excitations shows a clear difference between our sample of quasars at $z < 0.2$ and those at $z$ $\sim$ 1 -- 6.

\subsubsection{Line ratios}
\label{sec:lineratios}

We can also investigate the excitation of the gas by calculating the ratios of line luminosities of different CO transitions. We calculate these via the following equation:
\begin{equation}
\label{eq:lineratio}
\hspace{0.7cm} r_{xy} = \frac{\rm L'_{CO(x-(x-1))}}{\rm L'_{CO(y-(y-1))}}, \hspace{0.3cm} e.g., \hspace{0.1cm} r_{21} = \frac{\rm L'_{CO(2-1)}}{\rm L'_{CO(1-0)}}
\end{equation}
where $\rm L'_{CO}$ is the CO luminosity of a given CO transition line.\\

\begin{figure*}
\centering
\includegraphics[width=1.95\columnwidth]{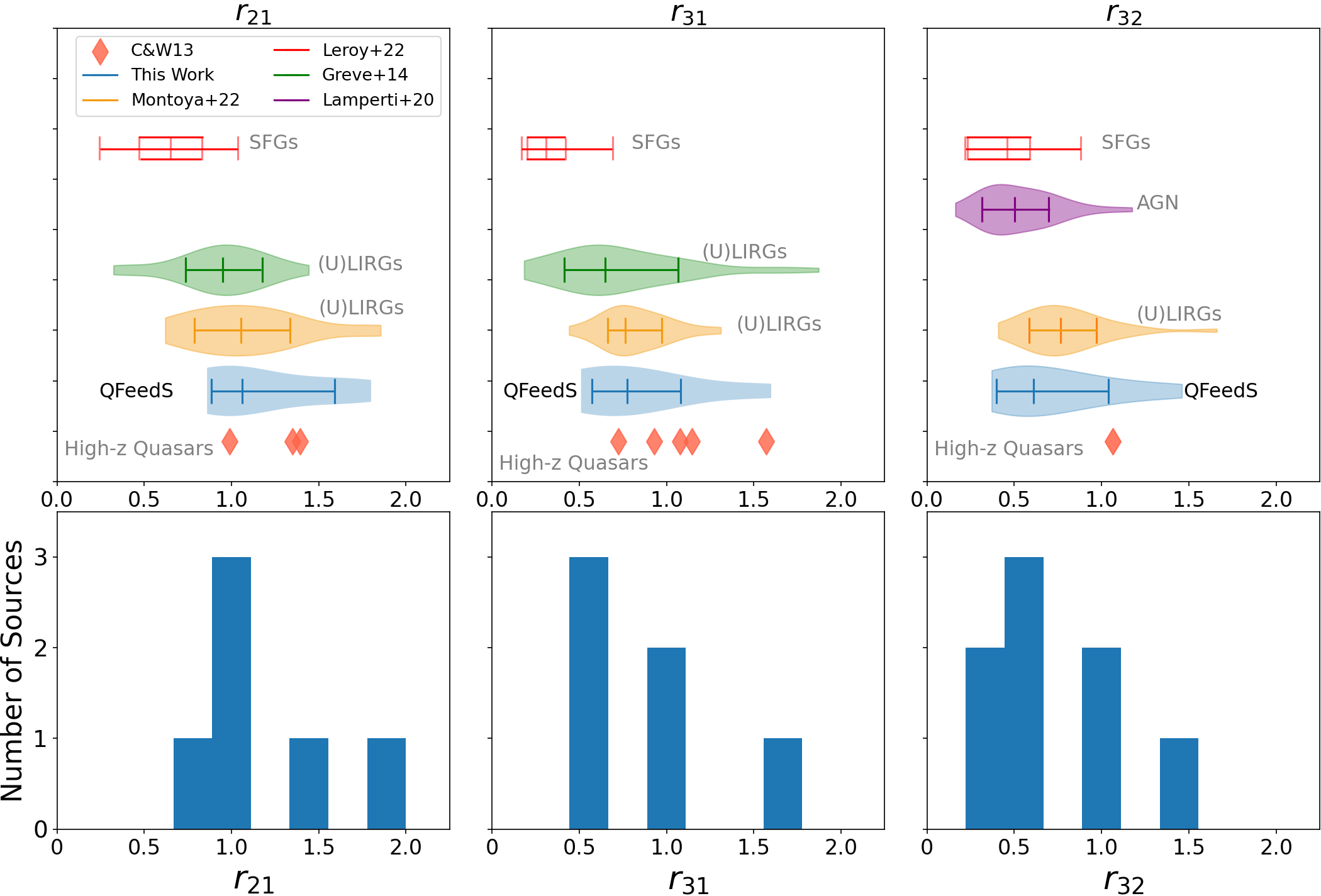}
\caption{Bottom panels: Histograms to show the distribution of line ratios within our sample. Top panels: Violin plots showing the distribution of line ratio values from our sample and selected comparison samples including the median, 16th and 84th quartiles of data (ignoring upper limits). The violin plots shown in blue are those for our sample and are shown in the same colour in the corresponding histograms below. Individual high-$z$ quasars from \citealt{Carilli13} are shown by red diamond markers. Other literature samples of SFGs and (U)LIRGS are also shown for comparison \citep[][more details of the selected literature comparisons can be found in section~\ref{sec:coexcite}]{Greve14, Lamperti20, Leroy22, Montoya23}. 
}
\label{fig:lineratio}
\end{figure*}

All measured line ratios for each individual target are shown in Table~\ref{tab:LineRatios}. We present the observed line ratios found in our sample of quasars and compare to literature values in Figure~\ref{fig:lineratio}. In the lower panel of Figure~\ref{fig:lineratio} we present histograms of the line ratios from our sample, as well as violin plots for different reference samples in the upper panels. We choose violin plots as these show more information of the distribution of the data, with a wider section showing a larger number of data, as well as the maximum and minimum values in the range of the data plus a defined median and 16/84th quartiles.
For targets within our sample which show multiple components in their CO spectra, we calculate the line ratios for these individual components and these are also presented in Table~\ref{tab:LineRatios}.

The overall line ratios observed in this sample (only including those targets with detections, and ignoring non-detections) are as follows: 

For those in our sample with detections in both CO(1-0) and CO(2-1) (6 sources), we calculate the line ratios using equation~\ref{eq:lineratio} and find a median $r_{21}$ of 1.06$_{-0.18}^{+0.53}$, where the negative and positive uncertainties represent the 16th and 84th quartiles respectively. For $r_{31}$ we find a median of 0.77$_{-0.20}^{+0.31}$ from 6 sources with detections in both CO(3-2) and CO(1-0). We find a median $r_{32}$ of 0.61$_{-0.21}^{+0.43}$ from 8 sources with detections in both CO(3-2) and CO(2-1). We find that in those targets with multiple components, the excitation of the different, individual components and of the total emission across the entire spectral line from the galaxy are consistent (within uncertainties). 
Therefore, with the data we have available, we cannot measure any difference between the excitation levels in these different components. The median line ratios in our sample along with literature comparisons can be found in Table~\ref{tab:LineRatiosTotal}.

\begin{table*}
\centering
\begin{tabular}{ |c|c|c|c|c|c|c|c| } 
 \hline
 Name & $r_{21}$ & $r_{31}$ & $r_{32}$ & $r_{61}$ ($r_{71}$) & $r_{62}$ ($r_{72}$) & $r_{63}$ ($r_{73}$)  \\ 
 \hline
 \rule{0pt}{2ex}
 J0909+1052 & -- & -- & -- & -- & -- & -- \\
 J0945+1737 & 1.10 $\pm$ 0.40 & < 0.80 & < 0.73 & < 1.20 & < 1.10 &  \\
 J0958+1439 & < 0.62 & 0.59 $\pm$ 0.14 & > 0.96 & -- & -- & --  \\
 J1000+1242 & 0.86 $\pm$ 0.34 & 0.51 $\pm$ 0.25 & 0.60 $\pm$ 0.24 & -- & -- & -- \\
 J1010+0612 & -- & -- & 0.37 $\pm$ 0.18 & -- & < 0.50 & < 1.40  \\
 J1010+1413 & -- & -- & 1.46 $\pm$ 0.33 & -- & -- & --  \\
 J1016+0028 & < 1.20 & < 1.60 & -- & -- & -- & -- \\
 J1055+1102 & < 4.20 & < 1.70 & -- & -- & -- & --  \\
 J1100+0846 & 1.54 $\pm$ 0.34 & 0.95 $\pm$ 0.19 & 0.62 $\pm$ 0.12 & < 0.43  & < 0.28  & < 0.45 \\
 J1100+0846 (red) & 1.60 $\pm$ 0.40 & 0.96 $\pm$ 0.22 & 0.61 $\pm$ 0.14 & -- & -- & -- \\
 J1100+0846 (blue) & 1.60 $\pm$ 0.60 & 0.97 $\pm$ 0.31 & 0.62 $\pm$ 0.20 & -- & -- & -- \\
 J1108+0659 & 1.80 $\pm$ 0.80 & 1.60 $\pm$ 0.50 & 0.89 $\pm$ 0.35 & (< 0.30) & (< 0.20) & (< 0.20)  \\
 J1108+0659 (core) & 1.60 $\pm$ 0.80 & 1.70 $\pm$ 0.60 & 1.10 $\pm$ 0.50 & -- & -- & --  \\
 J1108+0659 (blue wing) & 2.00 $\pm$ 1.20 & 1.40 $\pm$ 0.70 & 0.70 $\pm$ 0.40 & -- & -- & -- \\
 J1114+1939 & < 1.60 & -- & > 1.00 & -- & -- & -- \\
 J1116+2200 & 0.89 $\pm$ 0.17 & 0.95 $\pm$ 0.15 & 1.06 $\pm$ 0.24 & (< 0.50) & (< 0.60) & (< 0.50)  \\
 J1222-0007 & -- & -- & --  & -- & -- & -- \\
 J1316+1753 & > 1.30 & -- & < 0.47 & -- & -- & -- \\
 J1356+1026 & -- & -- & -- & -- & -- & -- \\
 J1430+1339 & -- & -- & 0.37 $\pm$ 0.14 & -- & 0.22 $\pm$ 0.13 & 0.61 $\pm$ 0.27 \\
 J1518+1403 & 1.00 $\pm$ 0.20 & 0.60 $\pm$ 0.10 & 0.60 $\pm$ 0.20 & (< 0.43) & (< 0.42) & (< 0.71) \\
 \hline
\end{tabular}
\caption{Table of line ratios along with uncertainties for all targets. Those with no data shown do not have the required data to present. Sources with multiple components are split into the line ratios for the individual components (components mentioned in brackets) as well as the total values. }
\label{tab:LineRatios}
\end{table*}

\begin{table}
\centering
\begin{tabular}{ |c|c|c| } 
 \hline
 \rule{0pt}{2ex}
 Line Ratio/sample & $z$ range & Median  \\ 
 \hline
 \rule{0pt}{2ex}
 $r_{21}$ & &  \\
 \rule{0pt}{2ex}
 QFeedS (This work) & 0.1 -- 0.2 & 1.06 $^{+0.53}_{-0.18}$ \\[0.2cm]
 High-$z$ Quasars (C\&W+13) & 1 -- 6 & 1.35 \\[0.2cm]
 SFGs (Leroy+22) & 0 & 0.65 $^{+0.18}_{-0.15}$ \\[0.2cm]
 (U)LIRGs (Montoya+22) & < 0.2 & 1.05 $^{+0.32}_{-0.30}$ \\[0.2cm]
 (U)LIRGs (Greve+14) & & 0.95 $^{+0.23}_{-0.21}$  \\[0.2cm]
 \hline
 \rule{0pt}{2ex}
 $r_{31}$ & &  \\
 \rule{0pt}{2ex}
 QFeedS (This work) & 0.1 -- 0.2 & 0.77 $^{+0.31}_{-0.20}$  \\[0.2cm]
 High-$z$ Quasars (C\&W+13) & 1 -- 6 & 1.08 \\[0.2cm]
 SFGs (Leroy+22) & 0 & 0.31 $^{+0.11}_{-0.11}$  \\[0.2cm]
 (U)LIRGs (Montoya+22) & < 0.2 & 0.76 $^{+0.22}_{-0.10}$ \\[0.2cm]
 (U)LIRGs (Greve+14) & & 0.65 $^{+0.42}_{-0.23}$ \\[0.2cm]
 \hline
 \rule{0pt}{2ex}
 $r_{32}$ & & \\
 \rule{0pt}{2ex}
 QFeedS (This work) & 0.1 -- 0.2 & 0.61 $^{+0.43}_{-0.21}$  \\[0.2cm]
 High-$z$ Quasars (C\&W+13) & 1 -- 6 & 1.06  \\[0.2cm]
 SFGs (Leroy+22) & 0 & 0.46 $^{+0.13}_{-0.20}$ \\[0.2cm]
 (U)LIRGs (Montoya+22) & < 0.2 & 0.76 $^{+0.23}_{-0.17}$  \\[0.2cm]
 AGN (BASS, Lamperti+20) & median $\sim$ 0.05 & 0.50 $^{+0.20}_{-0.19}$ \\[0.2cm]
 \hline
\end{tabular}
\caption{Table of median line ratios along with 16th and 84th quartiles (indicated by plus and minus values) for all targets and comparisons to the literature used in Figure~\ref{fig:lineratio} along with the $z$ range for each sample. \citealt{Montoya23}, \citealt{Leroy22}, \citealt{Greve14}, \citealt{Lamperti20}, \citealt{Carilli13}.}
\label{tab:LineRatiosTotal}
\end{table}

Making comparisons to lower redshift, less luminous samples of AGN and star forming galaxies however, we find higher line ratios (see Figure~\ref{fig:lineratio}). For example, \cite{Lamperti20} present a study of 36 Hard X-ray selected AGN at z = 0.002 -- 0.04, conducted as part of the BASS AGN sample. They present twelve targets with the requisite data to calculate the $r_{21}$ values and find a median of 0.72$_{-0.17}^{+0.17}$. These are lower than those measured in our sample. Further, the $r_{32}$ of the same sample also shows lower excitation with a median of 0.50$_{-0.19}^{+0.20}$.

As expected the line ratios of our sample are higher than for normal, star forming galaxies \citep[e.g.][]{Saintonge17, Yajima21, denBrok21, Leroy22}. For example, a compilation of low-z samples found a median $r_{21}$ of 0.65 \cite{Leroy22}.

\begin{figure}
\centering
\includegraphics[width=\columnwidth]{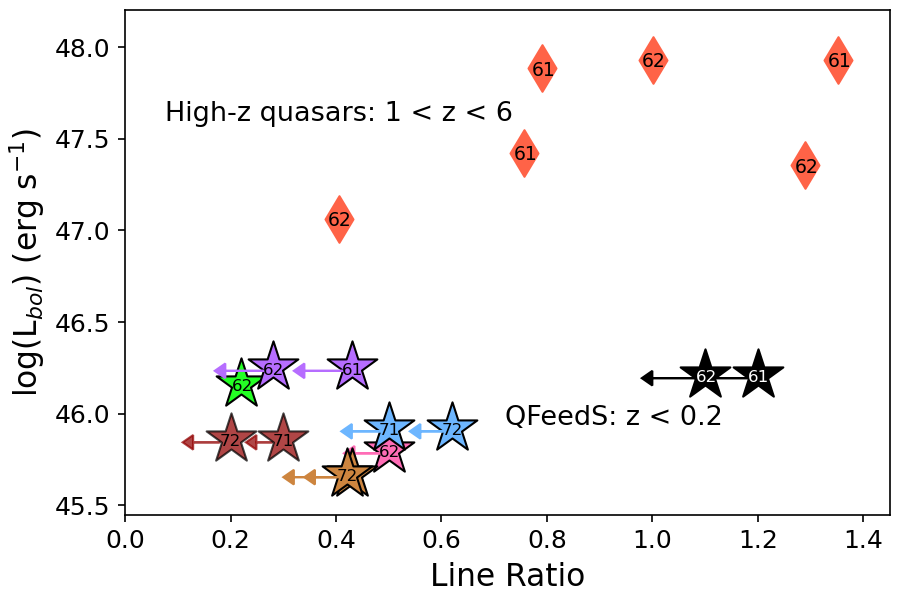}
\caption{For the 7 targets which have either CO(6-5) or CO(7-6) data we compare the bolometric luminosity and line ratios of a sample of high-$z$ quasars \citep[from][shown by red diamond markers]{Carilli13} to our sample of low-z quasars (shown by star markers and arrows indicating those that are upper limits). The numbers in each marker indicates the line ratio being plotted, e.g. "61" indicates r$_{61}$. Markers of the QFeedS targets are also coloured in the same way as in all previous plots (see Figure~\ref{fig:SampleSelection}).}
\label{fig:LineRatio_BolLum}
\end{figure}

Further, we analyse the higher CO line ratios, which are mostly upper limits for our sample, and utilise data of high-$z$ quasars which have CO line fluxes in at least one higher transition \citep[taken from the compilation by ][]{Carilli13} as well as corresponding bolometric luminosities \citep[taken from][]{Trentham95, Lewis98, Lutz07, Aravena08, Bradford09, Wang10}. We also note that in comparing bolometric luminosities we convert our [O~{\sc III}] luminosities to bolometric luminosities via the following equation: $\rm L_{bol}/L_{\rm [O~{\sc III}]}$ = 3500 \citep[from][]{Heckman04}. We present these data in Figure~\ref{fig:LineRatio_BolLum} by also comparing the bolometric luminosities of the targets. \\

\subsection{Gas temperature and density}
\label{sec:temp_dens}

\begin{figure}
\centering
\includegraphics[width=\columnwidth]{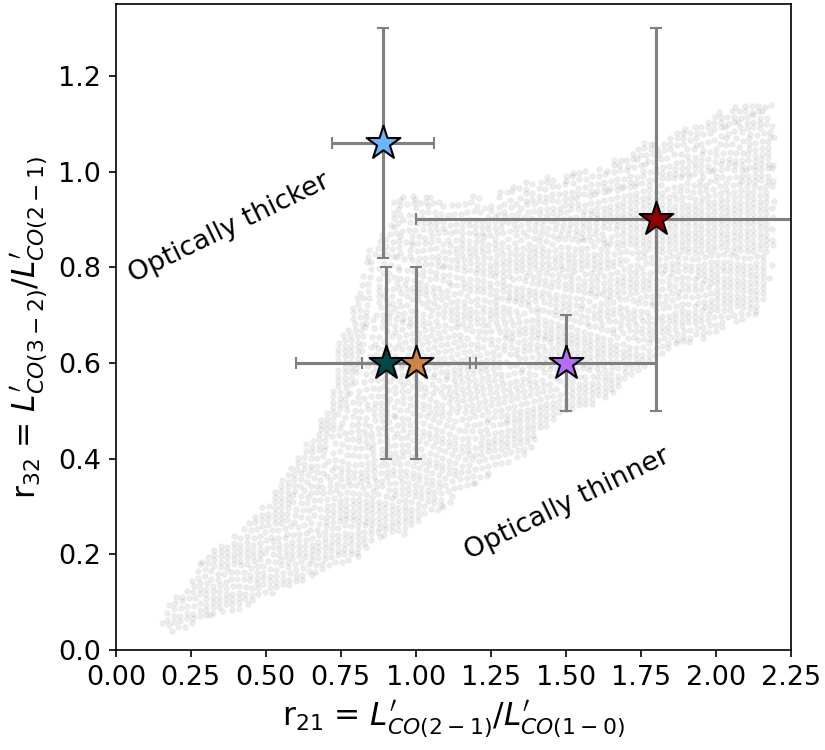}
\caption{For the 5 targets which have detections in all three transitions r$_{21}$ plotted vs r$_{32}$ to give an indication of temperature, density and optical thickness. Background grey points taken from \citealt{Leroy22}.}
\label{fig:temp_dens}
\end{figure}

We can utilise the measured line ratios to help understand the physical conditions of the molecular gas in individual targets. In particular, comparing the measured line ratios can give us an indication of the gas temperature and density within these quasar host galaxies \citep{Penaloza17, Leroy22}. Specifically, the ratio of $r_{32}$ with $r_{21}$ gives hints as to these properties (see Figure~\ref{fig:temp_dens}). We present these data plotted over simulations of the expected parameter space covered for these line ratios based on variable temperature, densities and optical depths \citep{Leroy22}. Figure~\ref{fig:temp_dens} shows this effect of temperature, density, and optical depth on the observed line ratios \citep[grey points from][]{Leroy22}. The expected line ratios for starbursts and AGN would be close to 1 \citep[e.g.][]{Mao10, Lamperti20, Yajima21}. These values are consistent with having both higher densities and hotter gas \citep{Leroy22}.

To further investigate the temperature and density of gas in these quasars we analysed the 5 targets with detections in all of the first three transitions (J1000+1242, J1100+0846, J1108+0659, J1116+2200 and J1518+1403), which provide best opportunity to test these properties. Using the Dense Gas Toolbox \citep[][]{vandtak07, Leroy17, Johannes21}, the density of the gas and the dense gas fraction (fraction of gas with density $>$ 10$^5$ cm$^{-3}$) were calculated by fixing the temperatures in 5 degree increments in the range of 10 -- 50 K. 
In 4 out of 5 sources (except J1000+1242) a temperature of less than 35K lead to the highest density provided in the model (10$^5$ cm$^{-3}$) and dense gas fractions greater than 90\%. From 40K -- 50K these four targets give densities in the range 10$^5$ -- 5000 cm$^{-3}$. J1000+1242 is the only target showing different properties of the temperature and density. Only at 20K does the model give the highest density with a dense gas fraction greater than 90\%. From 50 -- 25K a gas density ranging between 600 -- 4000 cm$^{-3}$ was calculated. This is significant as J1000+1242 has the lowest line ratios amongst the 5 and is the only one with $r_{21} <$ 1.
This therefore shows that within our sample, those with high line ratios would require higher temperatures ($>$ 35K) and densities. For those in this sample with lower line ratios, temperatures as low as 25K can still provide realistic scenarios.

From analysing the line ratios in Figure~\ref{fig:temp_dens} we find that the molecular gas in J1116+2200 seems consistent with being optically thicker and J1100+0846 seems consistent with being optically thinner. J1000+1242 and J1518+1403 are most likely somewhere in the middle and J1108+0659 is more difficult to determine due to the larger uncertainties. Further studies with higher S/N observations across the entire sample would be required to make any further conclusions. However, from the large variety identified in these limited data we can say that there is not a particular tendency in optical depth valid for all our sources. 

\subsection{Comparing CO and ionised gas line profiles}
\label{sec:lineprofiles}

\begin{figure}
\centering
\includegraphics[width=\columnwidth]{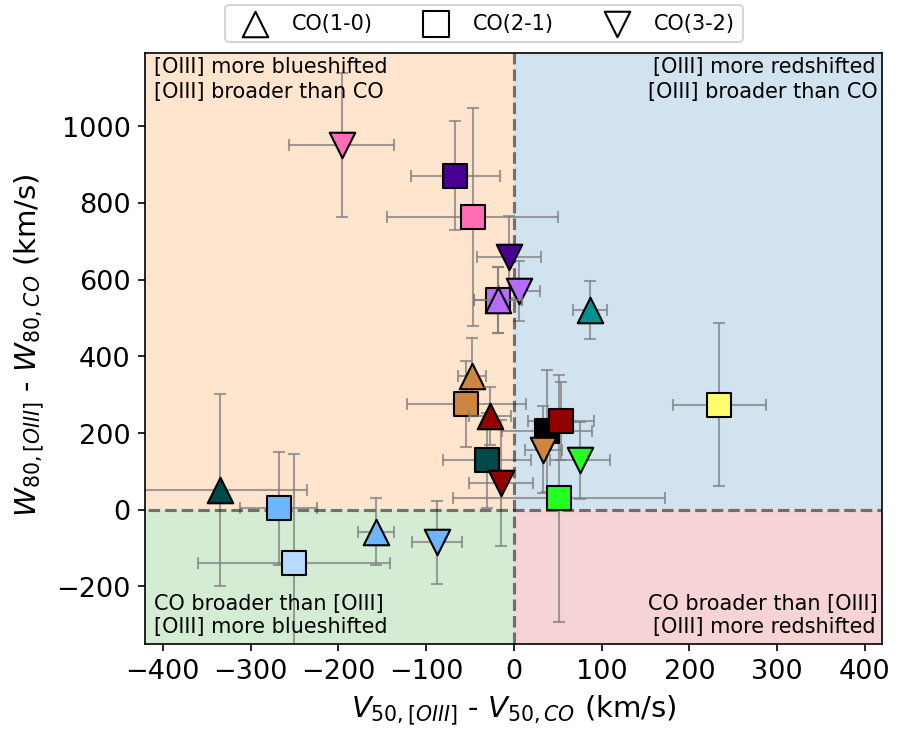}
\caption{For the 11 targets which have lines with S/N > 5 we present an analysis of the velocity offsets ($V_{50}$) and line widths ($W_{80}$) of the CO and [O~{\sc III}] spectra. Colours of markers for the individual targets are the same as throughout this work (colours shown in the legend of Figure~\ref{tab:sampleintro}). Some targets have multiple transitions shown, which are differentiated by: Upward triangles indicate CO(1-0), squares are CO(2-1) and downward triangles are CO(3-2) data (see legend). Each region of the plot is shaded to highlight the corresponding properties and the differences between the [O~{\sc III}] and the CO. Note: J1108+0659 is $\rm H\beta$ not [O~{\sc III}]. Also note that the CO data for J1356+1026 (yellow square) is taken from \citealt{Almeida21}.}
\label{fig:CO_OIII_Profiles}
\end{figure}

To investigate any differences between the molecular and ionised phase of the ISM we analyse the CO and [O~{\sc III}] line profiles. We do this by comparing both the velocity offsets ($V_{50}$) and the line width ($W_{80}$) for the CO lines compared to MUSE observations of the ionised gas (in most cases traced by [O~{\sc III}] and for two cases $\rm H\beta$, where the [O~{\sc III}] was not available). We present these results in Figure~\ref{fig:CO_OIII_Profiles}.
We only perform this analysis on those with an integrated velocity integrated S/N greater than 5 so that the line profiles we compare are more reliable. This threshold will ensure that there is a lower uncertainty on the CO line profile, allowing us to make more reliable comparisons between the ionised and molecular phase. From these results, in all cases we observe consistent or broader [O~{\sc III}] line profiles than in the CO transitions (as measured by $W_{80}$, shown in Figure~\ref{fig:CO_OIII_Profiles}).

As well as the line widths, we can also analyse the velocity offsets of CO compared to [O~{\sc III}] (see Figure~\ref{fig:CO_OIII_Profiles}). From this we find that a number of targets are consistent in terms of the $V_{50}$ measurements, whilst others do show significant differences, both positive and negative. 
While sources which show more redshifted [O~{\sc III}] lines than CO also exhibit broader [O~{\sc III}] line widths, the few cases with more blueshifted [O~{\sc III}] lines present equivalent 
[O~{\sc III}] line widths to those observed in CO, which are among the most similar line profiles that we observe. 

For our sample overall we also find more complex line profiles in the ionised gas than in the molecular phase. Specifically we identify 2 out of 17 sources with CO spectra that show multiple components in CO whereas in the MUSE ionised gas data we find 13 out of 17 sources with multiple components. This is limited by S/N in our CO data but interestingly, in the 2 cases where we find multiple components in the CO (J1100+0846 \& J1108+0659) we identify the following:

\textbf{J1100+0846} shows a clear double peaked line profile in the first three CO transitions. This double peaked profile was also identified in \cite{Almeida21} and we confirm the similarity by plotting them together onto our CO(2-1) spectra. This double peak however is not present in the [O~{\sc III}] where instead a broad line is identified.

Another target within our sample (\textbf{J1010+0612}) also shows a double peak CO(2-1) profile \citep[identified in higher resolution ALMA data by][]{Almeida21}. In our APEX spectra we also see tentative signs for a double peak profile in both our CO(2-1) and CO(3-2) data (see spectra in Figure~B5 in the supplementary material). However, like J1100+0846, J1010+0612 shows a broad line in the MUSE [O~{\sc III}] data, without signs of two narrower components as identified in CO.

\textbf{J1108+0659} shows a prominent blue wing component, identified in the first three CO transitions. The spatial resolution in our CO(1-0) data is not enough to spatially locate where this potential outflow component is located. Interestingly, the $\rm H\beta$ line profile also shows the same blue wing component, but with a more obviously present blue wing in $\rm H\beta$ than the CO. The peaks of the two components are also at very similar velocities across the three CO transitions and the $\rm H\beta$ data.

\begin{figure*}
     \centering
     \begin{subfigure}[b]{0.49\textwidth}
         \centering
         \includegraphics[width=\textwidth]{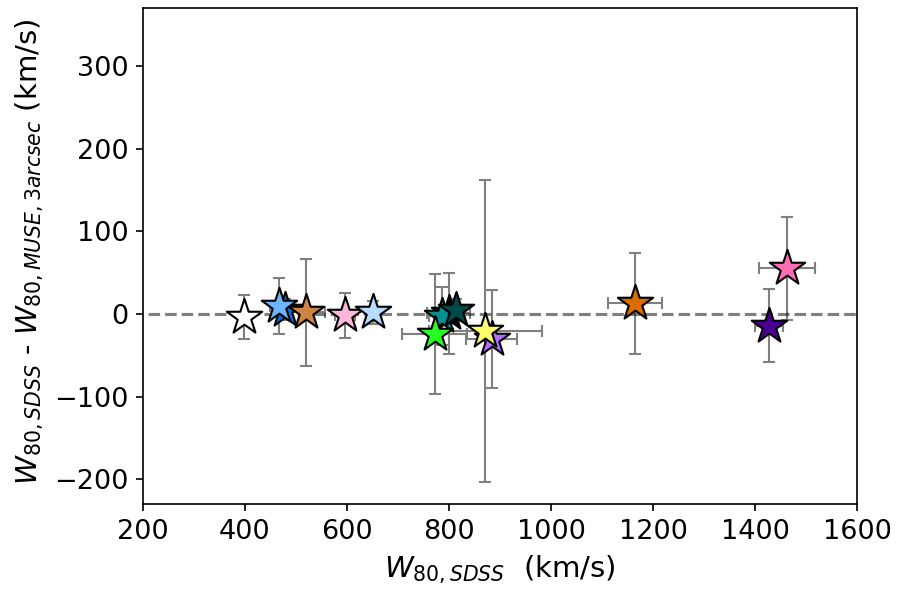}
         \caption{Difference between SDSS [O~{\sc III}] $W_{80}$ and [O~{\sc III}] $W_{80}$ extracted from MUSE at 3 arcsec, as a function of [O~{\sc III}] $W_{80}$ from SDSS.}
         \label{fig:muse3_vs_sdss}
     \end{subfigure}
     \hfill
     \begin{subfigure}[b]{0.49\textwidth}
         \centering
         \includegraphics[width=\textwidth]{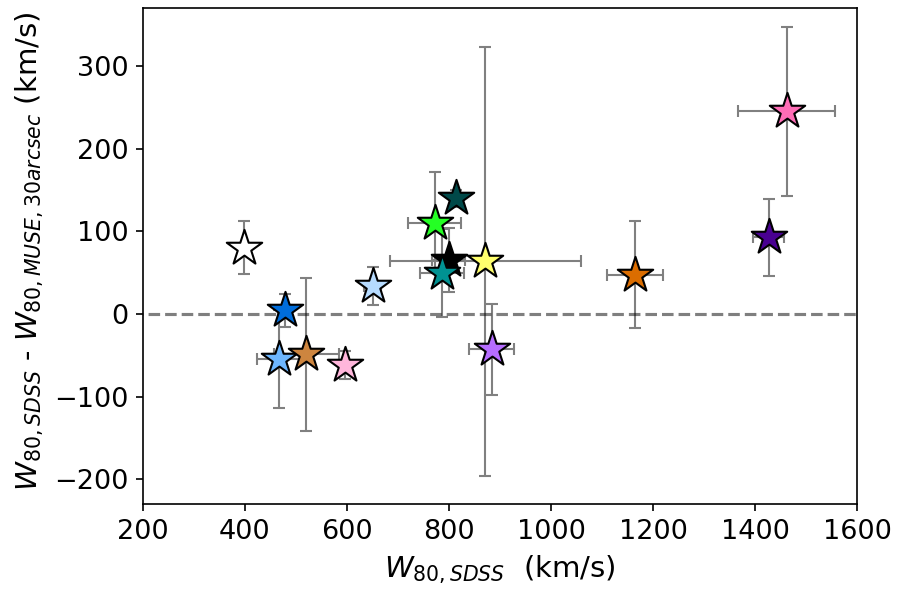}
         \caption{Difference between SDSS [O~{\sc III}] $W_{80}$ and [O~{\sc III}] $W_{80}$ extracted from MUSE at 30 arcsec, as a function of [O~{\sc III}] $W_{80}$ from SDSS.}
         \label{fig:muse30_vs_sdss}
     \end{subfigure}     
    \caption{Comparing line widths of the [O~{\sc III}] spectra extracted from SDSS and MUSE data cubes at two different extraction radii.}
\end{figure*}

Looking at the ionised gas in isolation, we extracted [O~{\sc III}] spectra from the MUSE cubes at both 30 arcsec (to match our CO observations) as well as at 3 arcsec (to match the SDSS observations). We find that the observed [O~{\sc III}] line profiles in SDSS and extracted from 3 arcsec apertures from MUSE are consistent within uncertainties (Figure~\ref{fig:muse3_vs_sdss}). However, we find a larger scatter when comparing SDSS line profiles to those extracted from 30 arcsec apertures in MUSE (Figure~\ref{fig:muse30_vs_sdss}). There are 8 cases for which they are not consistent, 7 where the SDSS lines are broader, and 1 case where MUSE at 30 arcsec is broader. Cases where SDSS are broader suggest a larger impact on velocities of ionised gas close to the core but further work would be needed to confirm this. Interestingly, the only case which is broader in MUSE at 30 arcsec than SDSS is J1016+0028, with the largest radio size, known radio lobes extending to distances of $\sim$ 15 arcsec, suggesting that impact on velocity of ionised gas is present out to these larger distances. There is no clear overall picture or trends from these results and so any differences likely depend on a case by case basis, but these observations do hint at potential differences in velocities of ionised gas in the halos, and that the integrated velocity is dominated by the kinematics in the core, since 7 out of 8 are broader at 3 arcsec. However, further and more focused studies, are needed to investigate these effects in more detail.

\section{Discussion}
\label{sec:discussion}

Here we discuss the findings and an interpretation of the results in relation to the literature.
In Section~\ref{sec:noimpact} we discuss the observed CO excitation within our sample using CO SLEDs and line ratios, including comparisons to the literature.
In Section~\ref{sec:CO_vs_OIII} we discuss the differences in the line profiles seen in our CO data compared to ionised gas observations from MUSE. Finally, in section~\ref{sec:temp_dens} we discuss the implications of our observed line ratios on the molecular gas temperature and density in our sources, based on theoretical models.

\subsection{No observed impact on total CO excitation}
\label{sec:noimpact}

Here we will discuss the findings of the CO excitation, in the context of the selected comparison samples from the literature (presented in Section~\ref{sec:comparisons}).
Overall, these sources which have evidence for AGN-driven ionised outflows and jets/winds do not have exceptional global CO content or excitation, beyond that seen in relevant comparison samples of (U)LIRGS and less luminous AGN.
However, we discuss further that a local impact on CO excitation is still a possible scenario, and if present is likely to be driven by AGN feedback processes such as via radio jets. 

\subsubsection{Consistent CO excitation to local (U)LIRGs}
\label{sec:high_r21}

Previous studies into the molecular excitation have found no consensus on whether AGN have systematically different r$_{21}$ line ratios \citep[see][]{Flaquer10,Papadopoulos12, Xia12, Husemann17, Shangguan20}, or the r$_{31}$ line ratios \citep{Sharon16}.
With this work we looked for any correlations with these line ratios and AGN or galactic properties.

Our $r_{21}$ values (median $r_{21}$ of 1.06$_{-0.18}^{+0.53}$) are consistent with those of recent studies of (U)LIRGs \citep{Greve14, Montoya23} with median values of 0.95 and 1.05 respectively. The $r_{31}$ and $r_{32}$ values are also found to be consistent with these (U)LIRG studies (as shown in Figure~\ref{fig:lineratio} and Table~\ref{tab:LineRatiosTotal}).
As mentioned previously, 9 quasars in our sample of 17 have L$_{\rm IR, SF}$ data, of which 8 are consistent with being LIRGs \citep[see Section~\ref{sec:sampleselect} and][]{Jarvis19}. The ninth source, J1430+1339, is just below the threshold with log[L$_{\rm IR, SF}$] = {44.32}$^{+0.06}_{-0.07}$ erg$s^{-1}$. The fact that we see similar line ratios to these studies of local (U)LIRGs is an indication that the presence of a quasar and the observed radio jets and ionised outflows has no significant impact on the excitation of the global molecular gas content at the mentioned transitions. 

As mentioned above, we find slightly higher excitation in our sample than found in lower redshift AGN \citep{Lamperti20}, with the distribution of QFeedS line ratios extending to higher values.
These targets have a lower luminosity compared to our sample with a median $\rm L_{bol}$ $\sim$ $10^{44.8}$ which may explain the small differences observed in the CO excitation, despite the presence of AGN. However, it is important to note that the differences are not large (particularly when considering the smaller sample size in QFeedS) and so there is no clear impact from the presence of AGN-driven ionised outflows and jets/winds.

With the diversity of lines ratios observed for different populations, depending on the type of galaxy and properties, we note that extreme care should be taken when converting from higher J$_{CO}$ lines to the ground state to make sure that correct line ratio is used based on these factors.

\subsubsection{Optical depth effects}
\label{sec:optical_depth}

As presented in Section~\ref{sec:COSLEDs}, we find cases where the CO SLEDs are above the thermalised relation in CO(2-1) and CO(3-2). The most straight-forward explanation might be that some of the CO(1-0) emission may be resolved out given the ACA spatial resolution, particularly if there is extended diffuse emission beyond the beam size of the observations, therefore obtaining a lower flux value than the true value. The larger uncertainties on the CO(1-0) flux values may also be a factor in this. Interestingly two out of the four non-detections in the ACA CO(1-0) data are known to be ongoing mergers/dual AGN \citep[J1222-0007 and J1316+1753, see][]{Jarvis21}. However, apart from this there are no special properties of these sources that would differentiate them from the rest of the sample, and as mentioned in Section~\ref{sec:ACAobs}, the sensitivity calculations for the observations were done using the same method across the sample.

For those that are above the thermalised level in the CO SLEDs, and assuming the thermalised level should be the maximum flux, then this would imply that between 7 and 50 per cent of the flux would be in extended emission at scales greater than 188kpc (median recoverable scale of the observations). Since this would be an unrealistic scenario, we favour a physical interpretation rather than an observational one.
In addition, we can compare to \cite{Montoya23} which contains more nearby objects (at a median $z =$ 0.09 compared to a median $z =$ 0.14 for the QFeedS sample) and would therefore more subject to over-resolution effects with ACA CO(1-0). Despite this, they find very similar line ratios to ours (Figure~\ref{fig:lineratio}), again lending credence to the interpretation that this is a physical and not an observational effect.

Although we cannot exclude the influence of the above mentioned observational effects we note that other investigations on ULIRG samples have reported similar trends, indicating that these super-thermal $r_{21}$ ratios may have a physical explanation. 
For example, the high $r_{21}$ values in our sample could be due to optical depth effects, with highly excited gas in combination with a low opacity in the CO(1-0) transition \citep{Zschaechner18}. It has been argued that low opacities can be driven by large velocity gradients and would require the presence of turbulent or outflowing gas, perhaps also in a diffuse, warm phase \citep{Cicone18, Montoya23}. In the case of our sample this could mean kinematic disturbances as a result of quasar feedback \citep[e.g.][]{Jarvis19, Almeida21, Girdhar22}. However, in the data presented here we only see tentative indications of such outflows, but this may be the result of limited S/N and further studies with deeper observations at a higher spatial resolution are required to investigate this further.

It therefore may well be the case that the CO(2-1) is a more reliable line, especially when investigating the excitation through CO SLEDs. For example \cite{Montoya23} claimed that one cannot determine the CO excitation from the low-J CO lines, most likely because the optical depth effects add too much noise to the calculation. They also identified that weak correlations are found only from ratios involving the CO(1-0) line and $\rm L_{IR}$ and SFR. This therefore supports the idea that CO(1-0) is the only line affected by optical depth effects and not the CO(2-1) or CO(3-2).

\subsubsection{Low excitation at CO J$_{up}$ = 6, 7}
\label{sec:low_r61}

As shown in Figure~\ref{fig:CO_SLED_hightrans}, we observe relatively low excitation in the CO(6-5) and CO(7-6) transitions, for at least 4 out of 7 targets with the appropriate data. We note that the differing beam sizes between the observations at CO(2-1) compared to CO(6-5) and CO(7-6) may play a role in the low excitation observed, especially if flux is resolved out in the smaller beam for observations in higher CO transitions. However, similarly low excitation is found when compared to the CO(1-0) ground state which has similar beam sizes to the higher CO transition APEX observation. From previous work into the CO(2-1) kinematics for some of these targets \citep{Almeida21} and the fact that higher CO transitions are expected to be more centrally located we argue that these line ratios are reliable and still very constraining. 

One explanation (as discussed previously in relation to the $r_{21}$ values) could be due to a difference in bolometric luminosity. The high-$z$ quasars that show much higher line ratios do indeed have significantly higher bolometric luminosities (see Figure~\ref{fig:LineRatio_BolLum}), with luminosities of 10$^{47}$ -- 10$^{48}$ ergs$^{-1}$ compared to our sample which are all L$\rm _{bol}$ < 10$^{46.5}$ ergs$^{-1}$. The high excitation levels observed in quasar host galaxies with extreme luminosities, like those presented in \cite{Carilli13}, and in particular unobscured sources \citep{Banerji18, Wang19, Li20, Bischetti21}, can often be associated with depleted gas reservoirs \citep[e.g.][]{Brusa15, Perna18, Schreiber19, Circosta21}. These depleted gas reservoirs, potentially as a result of quasar activity, could therefore lead to lower observed CO fluxes at J$_{CO} < 3$, and consequently high excitation at J$_{CO} >$ 5. Since we observe gas-rich systems in our quasar host galaxies, the lower excitations that we measure might therefore be expected. Further studies of both low and high-$z$ quasars covering much of the CO SLED are required to test this further \citep[e.g.][]{Novak19, Pensabene21}.

It is also possible that the effect of AGN may only be detected at J$_{\rm CO}$ > 10. CO is most commonly excited by photodissociation regions (PDRs) from the UV photons emitted from young stars. However, excitation at these higher CO transitions requires shocks and/or X-ray emission (through X-ray-dominated region models, XDR), both of which can be powered by AGN or jets \citep[][]{Pereira-Santaella13, Carniani19}. The effects of both XDR and PDR regions play an important role but observations with parsec scale resolution are required to disentangle and analyse these regions \citep{Wolfire22}. Therefore, observations at even higher CO transitions and high resolution may be needed to detect the XDR-dominated CO lines and provide a more complete constraint on the influence of AGN \citep[see e.g.][]{vanderWerf10, Mashian15, Carniani19}.

Differences in the overall star formation and ISM conditions at low and high redshifts could also contribute to the differences we observe in CO excitation between our sample and high-$z$ quasars. 
Estimates of the cosmic molecular gas density indeed suggest that the molecular gas fractions peak at redshifts of $z$~=~1~--~3 \citep[see][for a review]{peroux20}, roughly mirroring the cosmic star formation rate density \citep{Madau14} and black hole accretion density \citep[e.g., ][]{aird15}.
As a result, we might therefore expect higher gas temperatures and densities at nuclear scales in quasar host galaxies at redshifts 1~--~3 (higher than our sample), excited by stronger radiation fields from star formation and the AGN.
However, there are cases where high-$z$ quasars have also been found to have lower than expected CO excitation. For example, a recent study of nine $z \sim$ 3 quasars \citep{Elgueta22} finds that their CO SLEDs peak in the range J$_{\rm up}$ = 5 -- 7 compared to the expected J$_{\rm up}>$  6 -- 8. Further studies with coverage across a large range of the CO SLED is required for both low and high-$z$ quasars to investigate these findings further.

Another factor could be the potential impact of obscuration and line-of-sight effects, when comparing Type 1, Type 2, and red quasars. For example, the CO emission in unobscured quasars might be dominated by gas within the ionising cone through which we observe Type 1 quasars \citep[e.g.][]{Vayner21, Stacey22}. Despite this, no significant differences in the molecular gas content and star formation efficiencies have been reported between obscured and unobscured AGN in the context of AGN unification and line-of-sight effects \citep[e.g.][]{Perna18}. However, the increased incidence of high-velocity [O~{\sc III}] outflows and radio emission in red quasars has been associated with higher nuclear dust reddening at high redshifts \citep[e.g., ][]{perrotta19, Klindt19, Fawcett20, CR21, Andonie22}. This connection might suggest an increased amount of obscuring material (dust and gas) at nuclear scales for luminous quasars with ongoing outflows and jets or winds.

\subsubsection{Line ratios as tracers of temperature and density}
In Section~\ref{sec:temp_dens} we presented the results of using the line ratios $r_{21}$ and $r_{32}$ to give an indication of the temperature and density of gas within the QFeedS sample. We are limited by the lack of detections across the CO SLED, with only 5 sources having detections in all of the first three J$_{CO}$ transitions, and by the large uncertainties in the line ratios (see Figure~\ref{fig:temp_dens}). Further, the dynamic range of the line ratios observed in AGN, quasars, (U)LIRGs and star forming galaxies is not large and therefore, more accurate, high S/N observations are required to analyse these to a high degree (i.e. placing accurately on Figure~\ref{fig:temp_dens}).

Despite this, we can place constraints and upper/lower limits on the temperature and density as shown in Section~\ref{sec:temp_dens}. Indeed, we have shown a difference between those with line ratios $\geq$ 1 and those $<$ 1. The 4 out of 5 sources with higher line ratios $\geq$ 1 in our sample required temperatures $>$ 35K to show dense gas fractions less than 90\%. 
On the other hand, for J1000+1242 with both $r_{21}$ and $r_{32}$ $<$ 1, this was reduced to 25K.

Observations to constrain the CO properties of other targets in QFeedS would be an important next step, as well as observing other lines such as CI, HCN or HCO+ which would also help further constrain the the gas temperature and densities.

\subsection{A localised impact on CO?}
\label{sec:spearman_corr}

Although our sample consist of luminous quasars with known ionised outflows, radio jets and/or large radio structures, we report no signs of enhanced CO excitation on the global scales in which we are measuring when compared to local (U)LIRGs of similar luminosities.

There has previously been found a positive relation with $\rm L_{IR}$ and the line ratios $r_{21}$ and $r_{31}$ \citep{Montoya23}, as well as a positive relation between $r_{31}$ and SFR. However, as mentioned in Section~\ref{sec:optical_depth} they claim that one cannot determine the CO excitation from the low-$J_{CO}$ lines because of optical depth effects.

We performed some basic analysis on the line ratios compared with galactic and AGN feedback diagnostics such as SFR, radio luminosities, FWHM$_{\rm [O~{\sc III}]}$, finding no correlations within our sample. However, when working with such small sample (less than 10 sources for which we have the requisite data) and small parameter space, it is hard to take strong conclusions from these investigations.

Despite this, a lack of any correlation may not be surprising. For example, a study on local Seyferts found no clear evidence for a systematic reduction in the molecular gas reservoir at galactic scales with respect to SFGs \citep{Salvestrini22}. Previous studies have also found weak or no correlation with properties such as stellar mass, AGN fraction and SFR offset to the main sequence \citep{Liu21}. Recent work has also found no correlation between the cold molecular gas properties and AGN properties \citep{Molina23}. Additionally, studies at both low and high redshifts have found no differences in low-$J_{CO}$ excitation between samples of star forming galaxies \citep[taken form the xCOLD GASS Survey][]{Saintonge11} and AGN host galaxies \citep{Sharon16, Lamperti20}.  
This finding again indicates a lack of influence from the AGN on the total molecular gas content. 
Further support for a lack of impact on global scales is the evidence of extended CO(1-0) emission as shown in Figure~C1 in the supplementary material These data may be an indication that for those with detections we still observe an extended molecular gas reservoir. On the other hand, those with non-detections may be an indication of disruption on these global scales, but with the data that is available we cannot make any conclusions about why a few of the targets have gone undetected in CO(1-0).

The typical theoretical prediction is that AGN outflows do not efficiently disrupt disc systems, because the outflow is deflected into the halo \citep[][]{Costa20}, therefore supporting the hypothesis of a lack of impact on the total molecular gas content. Some simulations predict that an outflow will carve out a small cavity (on scales of $\sim$ 1 kpc) in the galactic nucleus. As we are far from resolving on these scales here then it is not surprising that we do not see a significant impact on galactic scales. This would also mean that any immediate impact on star formation is also likely to be modest \citep[see also][]{Gabor14, Ward22, Piotrowska22}. However, this does not rule out any long-term impact through the effect of outflows on halo gas.

An explanation for our findings could therefore be that the impact of feedback and/or galactic properties on the excitation of the molecular gas may occur on a more localised scale, and once looking at the total molecular gas content, this effect is no longer observed. Indeed, higher spatial resolution studies have found differences in the excitation on scales < 1 kpc \citep[e.g.][]{Dasyra16, Oosterloo17, Rosario19, Zhang19, Ellison21, Audibert23}, in some cases also localised next to radio jets. 

Further supporting this idea, two of this sample have been studied in a resolved way in which the CO velocity dispersion was observed to be effected perpendicular to radio jets \citep[J1316+1737][]{Girdhar22} and CO temperature ratios were enhanced perpendicular to the radio jet \citep[J1430+1339][]{Audibert23}. This provides further motivation for resolved studies of the CO excitation and kinematics around small scale radio-jets. One interesting scenario would also be to determine any dependence on the inclination of any radio jet with respect to galaxy plane \citep{Mukherjee18, Venturi21, Girdhar22, Meenakshi22, Audibert23}, something which needs to be studied further within this survey. 

It therefore may be the case that any impact of AGN feedback on the excitation of the molecular ISM seems to only occur on localised scales \citep{Morganti21}, but the impact does not take effect over the whole galaxy. Therefore, since we are observing the total molecular gas content in the host galaxies, these smaller scale effects are likely to be lost in the full picture. 

One reason for this limited impact may be due to the power of the radio jets, being too weak to penetrate throughout the entire galaxy and they are deflected by interactions with the ISM and are contained within the central region of the galaxy. Another potential for these small scale jets is the potential time scales involved, and that what we observe are young jets which have not yet made their way to have an influence over the whole galaxy \citep[e.g.][]{ODea91, Morganti17, Bicknell18}.
An alternative scenario may be quasar driven winds that drive ionised outflows and simultaneously shock the ISM to produce radio emission in the same region of the galaxy \citep[e.g.][]{Wagner13, Zakamska14, Nims15, Zakamska16, Hwang18}.

\subsection{Comparing CO and ionised gas line profiles}
\label{sec:CO_vs_OIII}

Across our sample we see differences between the line profiles of the ionised gas and the CO (see individual spectra in the appendix), suggesting potential differences in the impact of AGN feedback on the different gas phases.

As shown in Section~\ref{sec:lineprofiles} we identify broader [O~{\sc III}] line profiles than in the CO transitions indicating a larger impact from feedback (e.g. via known radio jets) on the ionised gas kinematics than in the molecular gas. As discussed in Section~\ref{sec:spectra}, the CO $W_{80}$ measure based on the fit that can be considered as an upper limit as it is mostly higher than that measured on the data.

Despite this, we still see that the CO widths are less than, or consistent with, the [O~{\sc III}] line widths. This difference in molecular and ionised gas velocities could be attributed to the different densities, with the denser molecular gas being naturally more difficult to drive to higher velocities \citep{Nayakshin12, Mukherjee16, Mukherjee18a, Girdhar22}.

The presence of double-peaked CO lines (e.g. in the case of J1100+0846 and J1010+0612) can be indicative of jet-gas interactions whereby jets are pushing the gas in opposite directions \citep{Kharb21}. Alternatively, these profiles can indicate that the gas is in a disk, or that binary black holes with individual broad and narrow line regions are present. Binary black holes could have resulted from galaxy mergers.

However, as mentioned the lack of S/N in our CO data means that interpreting these multiple components is limited. On the other hand, we should nonetheless be sensitive to the overall gas kinematics, and so our finding of broader total [O~{\sc III}] line profiles compared to CO is reliable. It is worth mentioning that another explanation for this is that what we now observe as ionised gas was originally molecular gas that became ionised and heated in an outflow. In this case, the fact that [O~{\sc III}] is broader might reflect a shorter survival time of cold, dense gas in the outflow \citep[see e.g.][]{Costa15, Costa18}, rather than the impact of AGN feedback.

Differences in the kinematics of CO compared to the ionised gas could also indicate that the molecular gas is not mixed in the outflowing ionised medium. This may be as a result of cold gas clouds being unable to survive in hot winds \citep[e.g.][]{Farber22}.

\section{Conclusions}
\label{sec:conclsions}

We present a molecular gas excitation survey, observing a range of CO transitions (J = 1, 2, 3, 6, 7) for a sample of 17 quasars at $z <$ 0.2.
Our goal is to measure the molecular gas properties such as molecular gas masses, fractions, and CO excitation, as well as gas kinematics in order to identify any impact due to the presence of radio jets and ionised outflows on a global scale.\\

-- From all the evidence presented here we suggest that the presence of ionised outflows and radio jets in these LIRG type systems does not significantly impact the CO excitation on a global scale, but that given evidence from the literature, localised effects are likely, and do not extend to the scales of the entire galaxy.\\

-- We find no differences between the molecular gas fractions of our sample of quasars as compared to non-AGN in the literature (see Figure~\ref{fig:Miranda_Tacconi}), in agreement with previous works.\\

-- We observe median $r_{21}$, $r_{31}$ and $r_{32}$ ratios of 1.06$^{+0.53}_{-0.18}$, 0.77$^{+0.31}_{-0.20}$ and 0.61$^{+0.43}_{-0.21}$ respectively, which are consistent with those reported for (U)LIRGs of similar redshift (see Figure~\ref{fig:lineratio}).\\

-- We suggest that optical depth effects may contribute to the high line ratios involving CO(1-0) that are observed, in agreement with previous studies.\\

-- From analysing the CO SLEDs in 7 targets of our sample (see Figures~\ref{fig:CO_SLED} and \ref{fig:CO_SLED_hightrans}), we observe lower excitation in CO(6-5) and CO(7-6) as compared to a sample of quasars at higher redshift ($z$ = 1 -- 6). We suggest this difference is due to higher bolometric luminosities in the higher redshift quasars (see Figure~\ref{fig:LineRatio_BolLum}). We conclude that we detect no evidence of impact of AGN feedback on the CO SLEDs up to J $\leq 7$ for our quasar sample, despite the strong feedback signatures that characterise them (i.e., a sample with prevalent radio jets and/or shocked winds and ionised outflows).\\

-- We observe differences between the CO and [O~{\sc III}] line profiles, both in the line widths and velocity offsets, finding systematically broader [O~{\sc III}] line profiles than CO. The median difference in $W_{80}$ between [O~{\sc III}] and CO is $\sim$ 200 \kms, with a maximum difference of $\sim$ 650 \kms. This suggests a larger impact of feedback on the ionised gas than on the molecular gas (see Figure~\ref{fig:CO_OIII_Profiles}). Alternatively this can indicate cold gas clouds are unable to survive in hot winds.\\

-- We identify consistent [O~{\sc III}] line profiles in SDSS data compared to MUSE data extracted at a 3 arcsec aperture. However, differences in the line profiles are identified when extracted from MUSE data at a larger aperture of 30 arcsec. This suggests that the effects of feedback processes (such as outflows, radio jets or winds) are likely more dominant at smaller scales, closer to the central AGN/quasar (see Figures~\ref{fig:muse3_vs_sdss} and \ref{fig:muse30_vs_sdss}.\\

Overall we conclude that in these sample of quasars at $z < 0.2$ the impact of these quasars on the total molecular gas content, both in excitation and velocities, is likely to be minimal. On a global scale we see no real divergences from ULIRGs. This work therefore adds to the growing body of evidence that on global scales there is a minimal impact on CO excitation and total gas content, even in the extreme cases of luminous quasars with ionised outflows and extended radio structures.
However, we note that on smaller scales an increased velocity dispersion \citep{Girdhar22} and increased line ratios \citep{Audibert23} for two targets in our sample plus displaced molecular gas in another two targets (Girdhar et al, submitted), have been previously identified with a spatial relation to the observed radio jets. The question remains as to whether this impact is seen across the entire sample and further resolved studies will shed light on the the impact on the multi-phase ISM, in particular further investigation into targets with well characterised radio emission.

\section*{Acknowledgements}
We thank the referee for constructive comments that helped us to improve the clarity of the manuscript.
SJM acknowledges support from an STFC studentship, jointly supported
by the Faculty of Engineering and Technology at the Astrophysics
Research Institute at Liverpool John Moores University as well as an ESO studentship.
GCR acknowledges the ESO Fellowship Program and a Gruber Foundation Fellowship grant sponsored by the Gruber Foundation and the International Astronomical Union. 
CMH acknowledges funding from an United Kingdom Research and Innovation grant (code: MR/V022830/1).
EPF is supported by the international Gemini Observatory, a program of NSF’s NOIRLab, which is managed by the Association of Universities for Research in Astronomy (AURA) under a cooperative agreement with the National Science Foundation, on behalf of the Gemini partnership of Argentina, Brazil, Canada, Chile, the Republic of Korea, and the United States of America.
SS acknowledges the support of the Department of Atomic Energy, Government of India, under the project 12-R\&D-TFR-5.02-0700.
RS acknowledges support from an STFC Ernest Rutherford Fellowship (ST/S004831/1)
For the purpose of open access, the authors have applied a Creative Commons Attribution (CC-BY) license to any author accepted version arising.

We are grateful to all the APEX and ESO staff involved in obtaining the data used for this work. This publication is based on data acquired with the Atacama Pathfinder Experiment (APEX) under programme IDs 0100.B-0166B-2017, 0105.B-0713A-2020, 0105.B-0713B-2020 and 0109.B-0710A-2022. APEX is a collaboration between the Max Planck Institut f{\"u}r fur Radioastronomie, the European Southern Observatory, and the Onsala Space Observatory. This paper makes use of the following ALMA data: ADS/JAO.ALMA\#2016.1.01535.S, ADS/JAO.ALMA\#2019.2.00194.S ALMA is a partnership of ESO (representing its member states), NSF (USA) and NINS (Japan), together with NRC (Canada), MOST and ASIAA (Taiwan), and KASI (Republic of Korea), in cooperation with the Republic of Chile. The Joint ALMA Observatory is operated by ESO, AUI/NRAO and NAOJ. This work is also based on data obtained under ESO Proposal ID: 0103.B-0071.

\section*{Data Availability}


The data used in this work are available from the ESO Science Archive Facility (\url{https://archive.eso.org/}). The reduced data underlying this paper will be shared on reasonable request to the corresponding author. Data is also available at Newcastle University’s data repository (\url{https://doi.org/10.25405/data.ncl.24312502}) and can also be accessed through the Quasar Feedback Survey website. Specifically, all tables in digital format and .fits files of the spectra are available.

Further, all spectra and tables of data listing the CO properties are available in the supplementary material.



\bibliographystyle{mnras}
\bibliography{Molyneux23} 





\clearpage

\section*{Appendix}

\noindent Here we present the appendices to the main paper (The Quasar Feedback Survey: characterising CO excitation in quasar host galaxies). Firstly we present the tables of data used in the paper in appendix~\ref{sec:tables}. We then present all the spectra in appendix~\ref{sec:spectra}, for which the line profiles and flux estimates have been determined. Finally, in appendix~\ref{sec:CO_extent} we show contour plots showing the extent of CO(1-0) in the ACA data for the 9 targets with detections, plotted over optical images of the galaxies.

\appendix

\section{Tables of data}\label{sec:tables}

We first present details of the CO observations including project id, beam sizes, rms and observation times in Table~\ref{tab:sample_obvs2}. We also present tables showing information about the spectra and line fits including the S/N of lines, integrated fluxes, line widths, velocity offsets and luminosities in Tables~\ref{fig:tableco10} -- \ref{fig:tableco76}. We also present the line profile information from the fits to the ionised gas observations in Table~\ref{fig:table_muse}.

\begin{table*}
\centering
\begin{tabular}{ |c|c|c|c|c|c| } 
 \hline
 \rule{0pt}{2ex}
 Line & Source Name & Beam Size & rms, $\Delta v$ = 100 \kms & $\rm t_{obs}$  & Detection? \\
  &  & (arcsec) & (mJy) & (minutes) &   \\
 \hline
 \rule{0pt}{2ex}
   CO(1-0) Atacama Compact Array & J0909+1052 & 12.7 & 1.8 & 43.8 & N   \\
   project id: 2019.2.00194.S  & J0945+1737 & 11.9 & 1.7 &  31.2 & L   \\
   & J0958+1439 & 12.7 & 1.1 &  85.7 &  D   \\
   & J1000+1242 & 12.8 & 1.3 & 92.7 &  D   \\
   & J1016+0028 & 12.2 & 1.2 & 42.3 &  L  \\
   & J1055+1102 & 11.7 & 1.1 & 44.3 &  L   \\
   & J1100+0846 & 12.3 & 2.5 & 10.2 &  D   \\
   & J1108+0659 & 12.8 & 1.1 & 35.3 &  D   \\
   & J1114+1939 & 14.1 & 1.7 & 63.5  &  N   \\
   & J1116+2200 & 13.1 & 1.7 & 51.4 &  D   \\
   & J1222-0007 & 12.7 & 1.3 & 30.2 &  N   \\
   & J1316+1753 & 12.9 & 1.6 & 21.2 &  N   \\
   & J1518+1403 & 12.5 & 1.9 & 12.1 & D    \\
  \hline 
  \rule{0pt}{2ex}
  CO(2-1) APEX id: E-0105.B-0713A-2020  & J0909+1052 & 31.6 & 26.5 & 44.7 &  N  \\
    & J1016+0028 & 30.2 & 7.4 & 51.1 &  N  \\
   & J1055+1102 & 31.0 & 15.6 & 19.1 &  N   \\
   & J1108+0659 & 32.0 & 18.0 & 12.8 &  D   \\
   & J1114+1939 & 32.5 & 9.9 & 31.3 &  D   \\
   & J1116+2200 & 30.9 & 14.4 & 19.2 &  D   \\
   & J1222-0007 & 31.7 & 18.7 & 19.2 &  N   \\
   & J1430+1939 & 29.4 & 10.6 & 58.3 &  D   \\
   & J1518+1403 & 30.8 & 6.8 & 128.8 & D   \\
   \hline 
  \rule{0pt}{2ex}
  CO(2-1) APEX id: E-0100.B-0166B-2017
 & J0945+1737 & 30.5 & 7.6 & 42.0 &  D   \\
   & J0958+1439 & 30.0 & 6.0 & 71.9 & N    \\
   & J1000+1242 & 31.1 & 3.2 & 239.1 & D    \\
   & J1010+0612 & 29.8 & 12.0 & 23.0 &  D \\
   & J1010+1413 & 32.4 & 6.8 & 60.0 &  D \\
   & J1100+0846 & 29.8 & 18.9 & 27.5 &  D   \\
   & J1316+1753 & 31.2 & 11.0 & 49.6 &  L   \\
   & J1356+1026 & 30.4 & 24.0 & 35.8 &  N   \\ 
  \hline 
  \rule{0pt}{2ex}
  CO(3-2) APEX id: E-0105.B-0713B-2020 & J0909+1052 & 21.0 & 17.9 & 64.0 & N   \\
    & J0945+1737 & 20.4 & 14.1 & 46.5 &  N   \\
   & J0958+1439 & 20.0 & 18.1 & 248.3 &  L  \\
   & J1000+1242 & 20.7 & 10.6 & 60.1 &  L   \\
   & J1010+0612 & 19.8 & 10.0 & 76.2 & D  \\
   & J1010+1413 & 21.6 & 9.6 & 53.8 & D  \\
   & J1016+0028 & 20.1 & 9.5 & 148.8 &  N  \\
   & J1055+1102 & 20.7 & 15.4 & 15.9 &  N   \\
   & J1100+0846 & 19.9 & 14.2 & 57.3 &  D   \\
   & J1108+0659 & 21.3 & 24.2 & 32.0 &  D   \\
   & J1114+1939 & 21.6 & 25.9 & 18.5 &  N   \\
   & J1116+2200 & 20.6 & 17.8 & 63.5 &  D   \\
   & J1222-0007 & 21.2 & 21.4 & 12.8 &  N   \\
   & J1316+1753 & 20.8 & 16.0 & 64.9 &  N   \\
   & J1356+1026 & 20.3 & 11.3 & 95.5 &  N   \\
   & J1518+1403 & 20.6 & 6.9 & 52.6 &  D  \\
   \hline 
  \rule{0pt}{2ex}
   CO(3-2) ALMA project id: 2016.1.01535.S
 & J1430+1339 & 4.3 & 10.3 & 26.2 &  D   \\
   
  \hline 
  \rule{0pt}{2ex}
  CO(6-5) APEX id: E-0109.B-0710A-2022  & J0945+1737 & 10.2 & 82.1 & 128.7 & N   \\
   & J1010+0612 & 9.9 & 58.2 & 156.1 &  N   \\
   & J1100+0846 & 9.9 & 49.0 & 111.3 &  N   \\
   & J1430+1339 & 9.8 & 40.0 & 292.0 & L    \\
  \hline 
  \rule{0pt}{2ex}
  CO(7-6) APEX id: E-0109.B-0710A-2022  & J1108+0659 & 9.1 & 118.0 & 96.4 & N  \\
   & J1116+2200 & 8.8 & 124.9 & 87.5 &  N   \\
   & J1518+1403 & 8.8 & 68.3 & 245.2 &  N   \\
  \hline 
 \end{tabular}
\caption{Details of the CO observations including the CO line, telescope used and project id, the source name, the beam size, the rms, observing times ($\rm t_{obs}$) and whether we class a detection "D", low S/N detection "L" or a non-detection "N" based on criteria stated in Section~\ref{sec:spectra}.}
\label{tab:sample_obvs2}
\end{table*}

\begin{table*}
\centering
\begin{tabular}{ |c|c|c|c|c|c|c|c| } 
 \hline
 \rule{0pt}{2ex}
 Name & S/N & S$\rm _{CO(1-0)}$ & CO(1-0) $V_{50}$ & CO(1-0) $W_{80}$ & $\rm L'_{CO(1-0)}$ \\ 
   & &  (Jy \kms) & (\kms) & (\kms) & (1x10$^9$ $\times$ K \kms pc${^2}$)   \\
 \hline
 \rule{0pt}{2ex}
 J0909+1052 & -- & < 2.1 & -- & -- & < 2.8  \\
 J0945+1737 & 4.8 & 2.7 $\pm$ 0.7 & 243 $\pm$ 65 & 539 $\pm$ 167 & 2.1 $\pm$ 0.6 \\
 J0958+1439 & 8.6 & 2.6 $\pm$ 0.3 & -71 $\pm$ 13 & 215 $\pm$ 34 & 1.5 $\pm$ 0.2 \\
 J1000+1242 & 5.6 & 1.9 $\pm$ 0.6 & 259 $\pm$ 68 & 621 $\pm$ 246 & 1.8 $\pm$ 0.5 \\
 J1010+0612 & -- & -- & -- & -- & --  \\
 J1010+1413 & -- & -- & -- & -- & --  \\
 J1016+0028 (core) & 4.6 & 1.3 $\pm$ 0.4 & -116 $\pm$ 39 & 288 $\pm$ 101 & 0.8 $\pm$ 0.3 \\
 J1016+0028 North Lobe &  & 0.6 $\pm$ 0.1 & 611 $\pm$ 68 & 116 $\pm$ 23 & 0.4 $\pm$ 0.1 \\
 J1016+0028 South Lobe &  & 1.0 $\pm$ 0.2  & - 365 $\pm$ 16 & 126 $\pm$ 24 & 0.6 $\pm$ 0.1 \\
 J1055+1102 & 3.1 & 1.0 $\pm$ 0.4 & -6 $\pm$ 74 & 381 $\pm$ 189 & 1.0 $\pm$ 0.4 \\
 J1100+0846 (Total) & 7.6 & 3.7 $\pm$ 0.3 & 33 $\pm$ 42 & 378 $\pm$ 41 & 1.7 $\pm$ 0.3  \\
 J1100+0846 (red) & 6.2 & 2.3 $\pm$ 0.4 & 79 $\pm$ 16 & 142 $\pm$ 41 & 1.1 $\pm$ 0.3  \\
 J1100+0846 (blue) & 3.8 & 1.4 $\pm$ 0.4  & -197 $\pm$ 15 & 204 $\pm$ 45 & 0.7 $\pm$ 0.2  \\
 J1108+0659 (Total) & 32.2 & 8.4 $\pm$ 0.5  & 88 $\pm$ 5 & 354 $\pm$ 40 & 13.4 $\pm$ 3.1  \\
 J1108+0659 (core) & 22.1 & 4.7 $\pm$ 1.3 & 112 $\pm$ 5 & 66 $\pm$ 8 & 7.5 $\pm$ 2.0  \\ 
 J1108+0659 (blue wing) & 14.2 & 3.7 $\pm$ 1.4 & -64 $\pm$ 68 & 151 $\pm$ 40 & 5.9 $\pm$ 2.3 \\
 J1114+1939 & -- & < 2.0 & -- & -- & < 3.9  \\
 J1116+2200 & 12.4 & 7.4 $\pm$ 0.5 & 122 $\pm$ 17 & 578 $\pm$ 44 & 7.3 $\pm$ 0.5 \\ 
 J1222-0007 & -- & < 2.1 & -- & -- & < 3.0 \\
 J1316+1753 & -- & < 1.8 & -- & -- & < 2.4 \\
 J1356+1026 & -- & -- & -- & -- & --  \\
 J1430+1339 & -- & -- & -- & -- & --  \\
 J1518+1403 & 9.3 & 3.8 $\pm$ 0.5 & 47 $\pm$ 14 & 220 $\pm$ 36 & 1.2 $\pm$ 0.2 \\

 \hline
\end{tabular}
\caption{Properties of the CO(1-0) line emission observed with the ALMA ACA. We present the integrated S/N, the integrated flux from the single with Gaussian line fit, the velocity offset of the peak compared to $v$ = 0 \kms (defined from the SDSS redshift), line width ($W_{80})$ and the line luminosity. Upper limits are 3 $\sigma$ upper limits based on the median rms in the spectra and using the average line profile from other transitions of the same target, or from others with CO(1-0) detections. Those with no data shown do not have the required data to present.}
\label{fig:tableco10}
\end{table*}

\begin{table*}
\centering
\begin{tabular}{ |c|c|c|c|c|c|c|c| } 
 \hline
 \rule{0pt}{2ex}
 Name & S/N & S$\rm _{CO(2-1)}$ & CO(2-1) $V_{50}$ & CO(2-1) $W_{80}$ & $\rm L'_{CO(2-1)}$ \\ 
   & &  (Jy \kms) & (\kms) & (\kms) & (1x10$^9$ $\times$ K \kms pc${^2}$)   \\
 \hline
 \rule{0pt}{2ex}
 J0909+1052 & -- & < 34.1 & -- & -- & < 12.6  \\
 J0945+1737 & 5.1 & 11.6 $\pm$ 2.4 & 25 $\pm$ 49 & 528 $\pm$ 125 & 2.3 $\pm$ 0.5  \\
 J0958+1439 & -- & < 6.5 & -- & -- & < 1.0  \\
 J1000+1242 & 7.3 & 6.4 $\pm$ 1.2 & -64 $\pm$ 47 & 544 $\pm$ 119  &  1.7 $\pm$ 0.3  \\
 J1010+0612 & 4.2 & 12.3 $\pm$ 4.5 & 30 $\pm$ 74 & 453 $\pm$ 190 & 1.4 $\pm$ 0.5  \\
 J1010+1413 & 5.0 & 11.4 $\pm$ 2.4 & -10 $\pm$ 44 & 462 $\pm$ 112 & 5.5 $\pm$ 1.2 \\
 J1016+0028 (core) & 1.9 & < 6.6 & -- & - & < 1.1 \\
 J1055+1102 & -- & < 16.9 & -- & -- & < 5.9  \\
 J1100+0846 (Total) & 5.9 & 22.8 $\pm$ 3.4 & -18 $\pm$ 39 & 378 $\pm$ 41 & 2.7 $\pm$ 0.4  \\ 
 J1100+0846 (red) & 8.1 & 14.1 $\pm$ 2.6 & 79 $\pm$ 16 & 142 $\pm$ 41 & 1.7 $\pm$ 0.3 \\ 
 J1100+0846 (blue) & 5.0 & 8.8 $\pm$ 2.2  & -197 $\pm$ 15 & 204 $\pm$ 45 & 1.0 $\pm$ 0.3 \\ 
 J1108+0659 (Total) & 10.0 & 59.4 $\pm$ 15.6 & 9 $\pm$ 19 & 366 $\pm$ 112 & 24.0 $\pm$ 9.0  \\
 J1108+0659 (core) & 9.5 & 29.7 $\pm$ 13.2 & 35 $\pm$ 18 & 93 $\pm$ 26 & 12.0 $\pm$ 9.0  \\
 J1108+0659 (blue wing) & 6.4 & 29.8 $\pm$ 17.0 & -236 $\pm$ 185 & 277 $\pm$ 112 & 12.0 $\pm$ 7.0  \\ 
 J1114+1939 & 8.7 & 21.4 $\pm$ 6.6 & 280 $\pm$ 106 & 756 $\pm$ 270 & 10.3 $\pm$ 3.2  \\
 J1116+2200 & 6.3 & 26.4 $\pm$ 4.6 & 233 $\pm$ 41 & 517 $\pm$ 104 & 6.5 $\pm$ 1.1  \\
 J1222-0007 & -- & < 21.9 & -- & -- & < 9.7  \\
 J1316+1753 & 3.7 & 11.1 $\pm$ 3.6 & -370 $\pm$ 43 & 293 $\pm$ 110 & 3.0 $\pm$ 1.0 \\
 J1356+1026 & -- & < 30.9 & -- & -- & < 5.8  \\
 J1430+1339 & 5.7 & 21.2 $\pm$ 7.9 & 31 $\pm$ 106 & 631 $\pm$ 271 & 1.8 $\pm$ 0.7  \\
 J1518+1403 & 11.5 & 15.5 $\pm$ 1.8 & -14 $\pm$ 19 & 293 $\pm$ 49 & 3.6 $\pm$ 0.4 \\

 \hline
\end{tabular}
\caption{Properties of the CO(2-1) line emission observed with APEX. We present the integrated S/N, the integrated flux from the single with Gaussian line fit, the velocity offset of the peak compared to $v$ = 0 \kms (defined from the SDSS redshift), line width ($W_{80})$ and the line luminosity. Upper limits are 3$\sigma$ upper limits based on the median rms in the spectra and using the average line profile from other transitions of the same target, or from others with CO(2-1) detections. Those with no data shown do not have the required data to present.}
\label{fig:tableco21}
\end{table*}

\begin{table*}
\centering
\begin{tabular}{ |c|c|c|c|c|c|c| } 
 \hline
 \rule{0pt}{2ex}
 Name & S/N & S$\rm _{CO(3-2)}$ & CO(3-2) $V_{50}$  & CO(3-2) $W_{80}$ & $\rm L'_{CO(3-2)}$\\ 
   & &  (Jy km~$\rm s^{-1}$) & (km~$\rm s^{-1}$) & (km~$\rm s^{-1}$) & (1x10$^9$ $\times$ K km~$\rm s^{-1}$ pc${^2}$)  \\
 \hline
 \rule{0pt}{2ex}
 J0909+1052 & -- & < 22.5 & -- & -- & < 3.3  \\
 J0945+1737 & 2.7 & < 19.2 & -- & -- & < 1.7 \\
 J0958+1439 & 4.3 & 13.3 $\pm$ 2.7 & 188 $\pm$ 59 & 578 $\pm$ 137 & 1.5 $\pm$ 0.4 \\
 J1000+1242 & 3.1 & 8.6 $\pm$ 3.0 & -38 $\pm$ 81 & 576 $\pm$ 207 & 1.2 $\pm$ 0.4 \\
 J1010+0612 & 5.5 & 10.3 $\pm$ 3.1 & 179 $\pm$ 36 & 264 $\pm$ 94 & 0.5 $\pm$ 0.2 \\
 J1010+1413 & 12.1 & 37.5 $\pm$ 3.4 & -83 $\pm$ 30 & 673 $\pm$ 77 & 8.1 $\pm$ 0.7 \\
 J1016+0028 (core) & 2.1 & < 7.0 & -- & -- & < 0.5  \\
 J1055+1102 & -- & < 14.8 & -- & -- & < 1.7  \\
 J1100+0846 (Total) & 7.4 & 31.7 $\pm$ 3.7 & -21 $\pm$ 21 & 354 $\pm$ 36 & 1.7 $\pm$ 1.2  \\
 J1100+0846 (red) & 12.0 & 19.4 $\pm$ 2.7 & 58 $\pm$ 12 & 210 $\pm$ 36 & 1.0 $\pm$ 0.1  \\
 J1100+0846 (blue) & 7.6 & 12.3 $\pm$ 2.5  & -184 $\pm$ 13 & 167 $\pm$ 39 & 0.7 $\pm$ 0.1  \\
 J1108+0659 (Total) & 14.2 & 118.8 $\pm$ 20.5 & 55 $\pm$ 17 & 529 $\pm$ 141 & 21.0 $\pm$ 4.0 \\ 
 J1108+0659 (core) & 14.7 & 71.3 $\pm$ 13.9 & 142 $\pm$ 17 & 276 $\pm$ 41 & 12.7 $\pm$ 2.5  \\
 J1108+0659 (blue wing) & 8.3 & 47.6 $\pm$ 15.0 & -194 $\pm$ 53 & 417 $\pm$ 141 & 8.5 $\pm$ 2.7  \\
 J1114+1939 & -- & < 48.9 & -- & -- & < 6.9  \\
 J1116+2200 & 10.9 & 63.4 $\pm$ 5.7 & 53 $\pm$ 26 & 606 $\pm$ 65 & 6.7 $\pm$ 0.6  \\
 J1222-0007 & -- & < 26.7 & -- & -- & < 4.2  \\
 J1316+1753 & -- & < 11.8 & -- & -- & < 2.4  \\
 J1356+1026 & -- & < 14.5 & -- & -- & < 0.5  \\ 
 J1430+1339 & 9.3 & 17.5 $\pm$ 1.4 & 33 $\pm$ 19 & 532 $\pm$ 49 & 0.7 $\pm$ 0.1  \\
 J1518+1403 & 12.2 & 20.5 $\pm$ 2.1  & -34 $\pm$ 20 & 412 $\pm$ 50 & 2.1 $\pm$ 0.2  \\

 \hline
\end{tabular}
\caption{Properties of the CO(3-2) line emission observed with APEX. We present the integrated S/N, the integrated flux from the single with Gaussian line fit, the velocity offset of the peak compared to $v$~=~0~\kms (defined from the SDSS redshift), line width ($W_{80})$ and the line luminosity. Upper limits are 3$\sigma$ upper limits based on the median rms in the spectra and using the average line profile from other transitions of the same target, or from others with CO(3-2) detections. Those with no data shown do not have the required data to present.}
\label{fig:tableco32}
\end{table*}

\begin{table*}
\centering
\begin{tabular}{ |c|c|c|c|c|c|c|c| } 
 \hline
 \rule{0pt}{2ex}
 Name & S/N & S$\rm _{CO(6-5)}$  & CO(6-5) $V_{50}$ & CO(6-5) $W_{80}$ & $\rm L'_{CO(6-5)}$ \\ 
   & &  (Jy km~$\rm s^{-1}$) & (km~$\rm s^{-1}$) & (km~$\rm s^{-1}$) & (1x10$^9$ $\times$ K km~$\rm s^{-1}$ pc${^2}$)  \\
 \hline
 \rule{0pt}{2ex}
 J0945+1737 & -- & < 117.4 & -- & -- & < 2.6 \\
 J1010+0612 & 2.2 & < 59.2 & -- & -- & < 0.8\\
 J1100+0846 & 1.5 & < 47.5 & -- & -- & < 0.8 \\
 J1430+1339 & 6.7 & 42.9 $\pm$ 18.7 & 126 $\pm$ 121 & 669 $\pm$ 337 & 0.4 $\pm$ 0.2\\
 \hline
\end{tabular}
\caption{Properties of the CO(6-5) line emission observed with APEX. We present the integrated S/N, the integrated flux from the single with Gaussian line fit, the velocity offset of the peak compared to $v$~=~0~\kms (defined from the SDSS redshift), line width ($W_{80})$ and the line luminosity. Upper limits are 3$\sigma$ upper limits based on the median rms in the spectra and using the average line profile from other transitions of the same target, or from others with CO(6-5) detections. Those with no data shown do not have the required data to present.}
\label{fig:tableco65}
\end{table*}

\begin{table*}
\centering
\begin{tabular}{ |c|c|c|c|c|c|c|c| } 
 \hline
 \rule{0pt}{2ex}
 Name & S/N & S$\rm _{CO(7-6)}$ & CO(7-6) $V_{50}$  & CO(7-6) $W_{80}$ & $\rm L'_{CO(7-6)}$\\ 
   & & (Jy km~$\rm s^{-1}$) & (km~$\rm s^{-1}$) & (km~$\rm s^{-1}$) & (1x10$^9$ $\times$ K km~$\rm s^{-1}$ pc${^2}$)  \\
 \hline
 \rule{0pt}{2ex}
 J1108+0659 & -- & < 129.8  & -- & -- & < 4.2 \\
 J1116+2200 & -- & < 187.1  & --  & --  & < 3.8 \\
 J1518+1403 & -- & < 58.2 & --  & --  & < 1.5 \\
 \hline
\end{tabular}
\caption{Properties of the CO(7-6) line emission observed with APEX. We present the integrated S/N, the integrated flux from the single with Gaussian line fit, the velocity offset of the peak compared to $v$~=~0~\kms (defined from the SDSS redshift), line width ($W_{80})$ and the line luminosity. Upper limits are 3$\sigma$ upper limits based on the median rms in the spectra and using the average line profile from other transitions of the same target, or from others with or CO(7-6) detections. Those with no data shown do not have the required data to present.}
\label{fig:tableco76}
\end{table*}

\begin{table*}
\centering
\begin{tabular}{ |c|c|c|c|c|c|c|c| } 
 \hline
 \rule{0pt}{2ex}
 Name & Emission Line  & 30" $V_{50}$  & 30" $W_{80}$ & 3" $V_{50}$  & 3" $W_{80}$ & \# of components\\ 
   & &  (km~$\rm s^{-1}$) & (km~$\rm s^{-1}$) & (km~$\rm s^{-1}$)& (km~$\rm s^{-1}$) &  \\
 \hline
 \rule{0pt}{2ex}
 J0909+1052 & [O~{\sc III}]4959  & -34 $\pm$ 3 & 318 $\pm$ 8 & -38 $\pm$ 1 & 402 $\pm$ 3 & 1  \\
 J0945+1737 & [O~{\sc III}]5007   & 63 $\pm$ 2 & 734 $\pm$ 33  & 39 $\pm$ 1  & 798 $\pm$ 43 & 2 \\
 J0958+1439 & [O~{\sc III}]5007   & 16 $\pm$ 7 & 736 $\pm$ 44 & 18 $\pm$ 4 & 788 $\pm$ 25 & 2 \\
 J1000+1242 & [O~{\sc III}]5007  & -95 $\pm$ 4 & 673 $\pm$ 5 & -41 $\pm$ 6 & 808 $\pm$ 8 & 2  \\
 J1010+0612 & [O~{\sc III}]5007  & -17 $\pm$ 7 & 1216 $\pm$ 95 & -28 $\pm$ 14 & 1406 $\pm$ 55 & 2  \\
 J1010+1413 & [O~{\sc III}]5007  & -76 $\pm$ 7 & 1333 $\pm$ 30 & -14 $\pm$ 4 & 1440 $\pm$ 28  & 2  \\
 J1016+0028 & [O~{\sc III}]5007  & -130 $\pm$ 6 & 658 $\pm$ 8 & -41 $\pm$ 4 & 598 $\pm$ 20  & 3 \\
 J1055+1102 & [O~{\sc III}]5007 & -12 $\pm$ 1 & 473 $\pm$ 13 & -17 $\pm$ 1 & 473 $\pm$ 6  & 2  \\
 J1100+0846 & [O~{\sc III}]5007  & -5 $\pm$ 12 & 926 $\pm$ 45  & 0 $\pm$ 12 & 913 $\pm$ 49  & 2   \\
 J1108+0659 & $\rm H\beta$  &  40 $\pm$ 19 & 598 $\pm$ 36 & -47 $\pm$ 5 & 653 $\pm$ 11 & 2  \\
 J1114+1939 & [O~{\sc III}]5007 & 30 $\pm$ 4 & 616 $\pm$ 17  & 29 $\pm$ 2 & 648 $\pm$ 7  & 2  \\
 J1116+2200 & [O~{\sc III}]5007   & -35 $\pm$ 3 & 521 $\pm$ 43 & -57 $\pm$ 1 & 456 $\pm$ 18 & 2   \\
 J1222-0007 & $\rm H\beta$  & -79 $\pm$ 34 & 593 $\pm$ 51 & -54 $\pm$ 3 & 831 $\pm$ 12  & 3  \\
 J1316+1753 & [O~{\sc III}]5007   & -27 $\pm$ 4 & 1117 $\pm$ 56 & 29 $\pm$ 30 & 1152 $\pm$ 52  & 3   \\
 J1356+1026 & [O~{\sc III}]5007  & 75 $\pm$ 30 & 807 $\pm$ 188 & -10 $\pm$ 33 & 891 $\pm$ 111  & 3  \\
 J1430+1339 & [O~{\sc III}]5007  & 83 $\pm$ 15 & 661 $\pm$ 52  & 40 $\pm$ 14 & 796 $\pm$ 63  & 2   \\
 J1518+1403 & [O~{\sc III}]5007  & 0 $\pm$ 2 & 569 $\pm$ 64 & 1 $\pm$ 1 & 518 $\pm$ 36 & 2   \\
 \hline
\end{tabular}
\caption{Table presenting the MUSE line profile data, extracted from 30 and 3 arcsec apertures used as a comparison to the molecular gas presented in this work. Here we show the emission line used as a tracer for the ionised gas, the velocity offset ($V_{50}$) and line width ($W_{80}$). Also shown is the number of components present in the line (and therefore how many Gaussian components we fit to analyse the line profile).}
\label{fig:table_muse}
\end{table*}

\section{Spectra}\label{sec:spectra}

\begin{figure*}
    \centering
    \begin{minipage}{.32\textwidth}
        \centering
        \includegraphics[width=\linewidth, height=0.533\textheight]{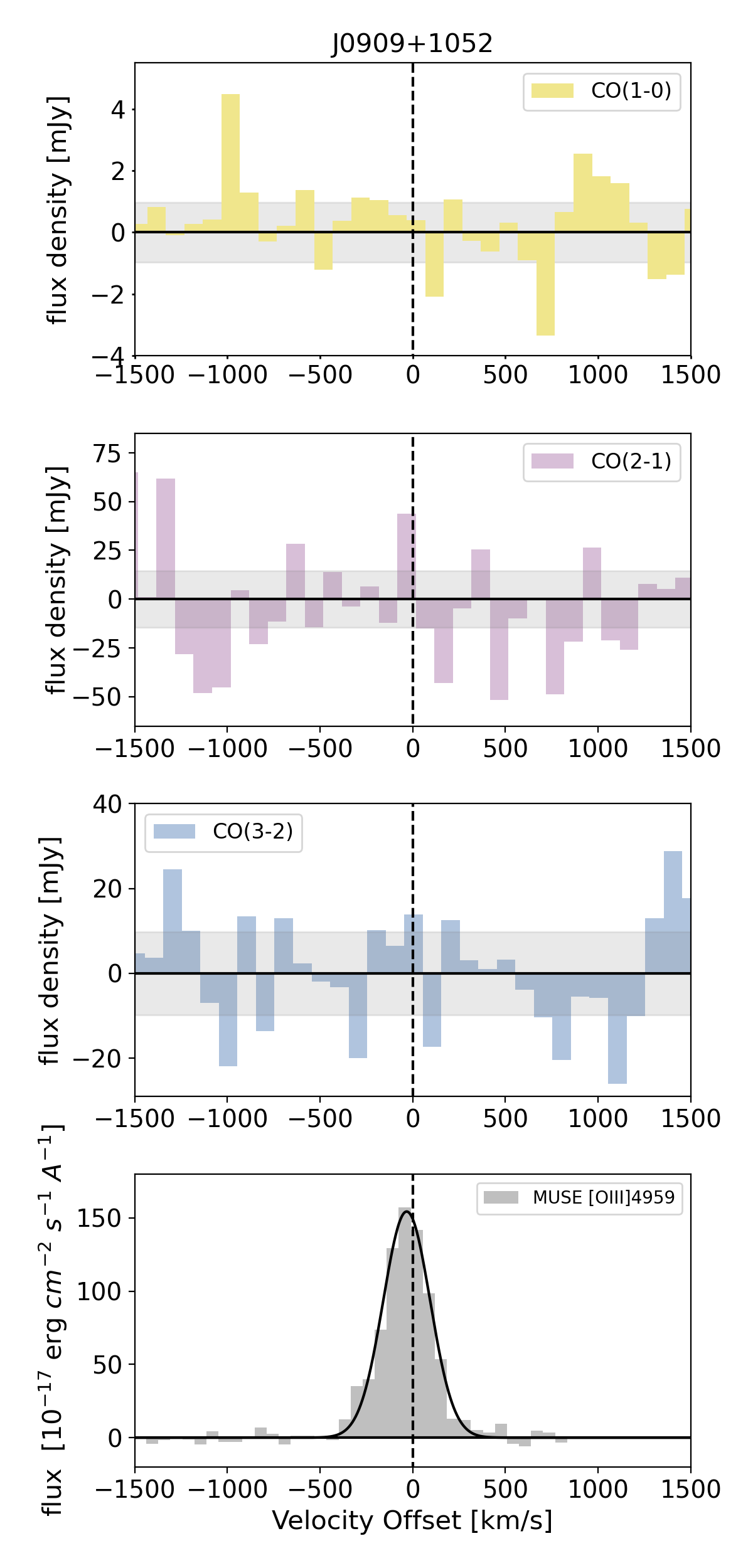}
        \caption{J0909+1052, non-detections across all three CO transitions. All shown with 100\kms bins. MUSE [O~{\sc III}] observations show a single with Gaussian.}
        \label{fig:J0909_spectra}
    \end{minipage}%
    \hspace{0.1cm}
    \begin{minipage}{0.32\textwidth}
        \centering
        \includegraphics[width=\linewidth, height=0.666\textheight]{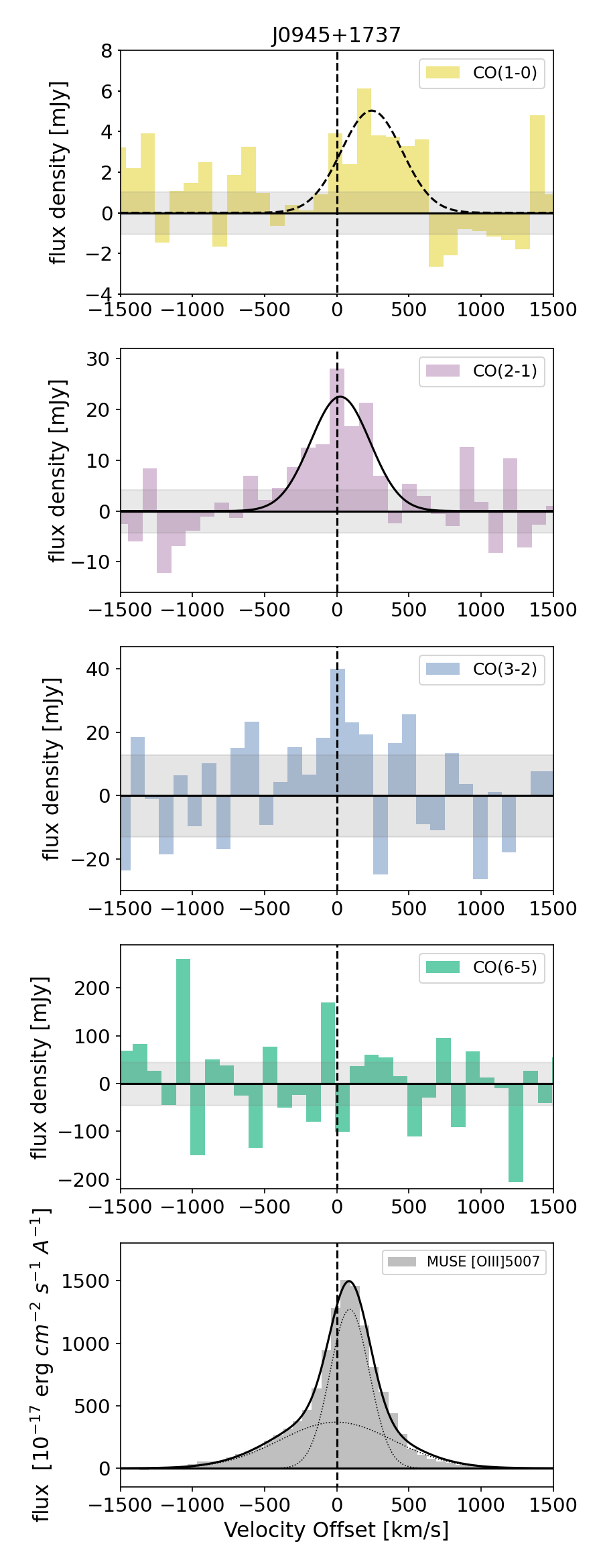}
        \caption{J0945+1737 spectra across four CO transitions. All with 100 \kms bins. Detection in CO(2-1) and low S/N detection in CO(1-0), non-detection in CO(3-2) (although tentative signs of a line at S/N = 2.7) or CO(6-5). MUSE [O~{\sc III}] observations show a two component fit with a narrow and broad component. }
        \label{fig:J0945_spectra}
    \end{minipage}
    \hspace{0.1cm}
    \begin{minipage}{0.32\textwidth}
        \centering
        \includegraphics[width=\linewidth, height=0.533\textheight]{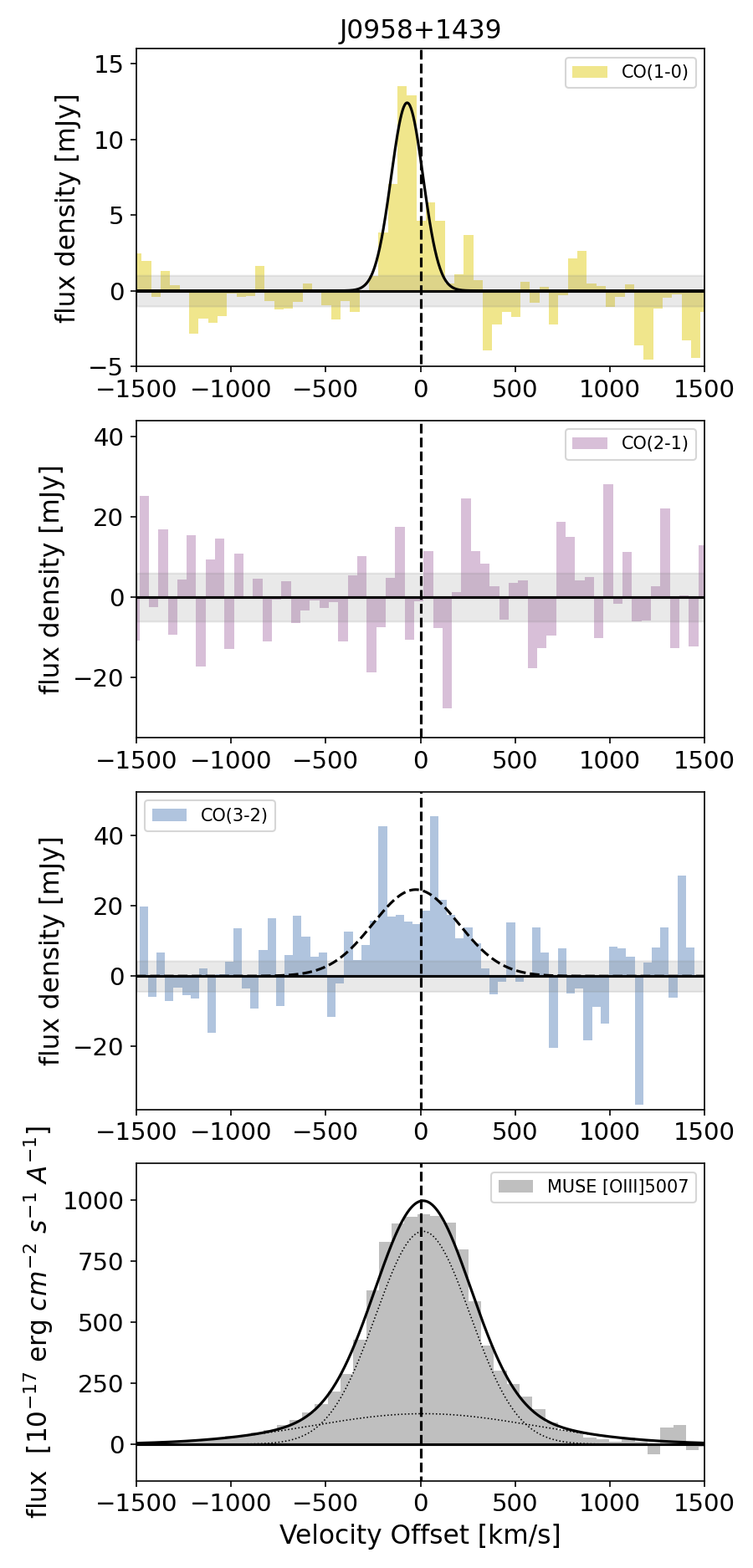}
        \caption{J0958+1439 detection in CO(1-0) and low S/N detection in CO(3-2). Non-detection in CO(2-1). MUSE [O~{\sc III}] observations show a two component fit with a narrow and broad component. }
        \label{fig:J0958_spectra}
    \end{minipage}
\end{figure*}

\begin{figure*}
    \centering
    \begin{minipage}{.32\textwidth}
        \centering
        \includegraphics[width=\linewidth, height=0.533\textheight]{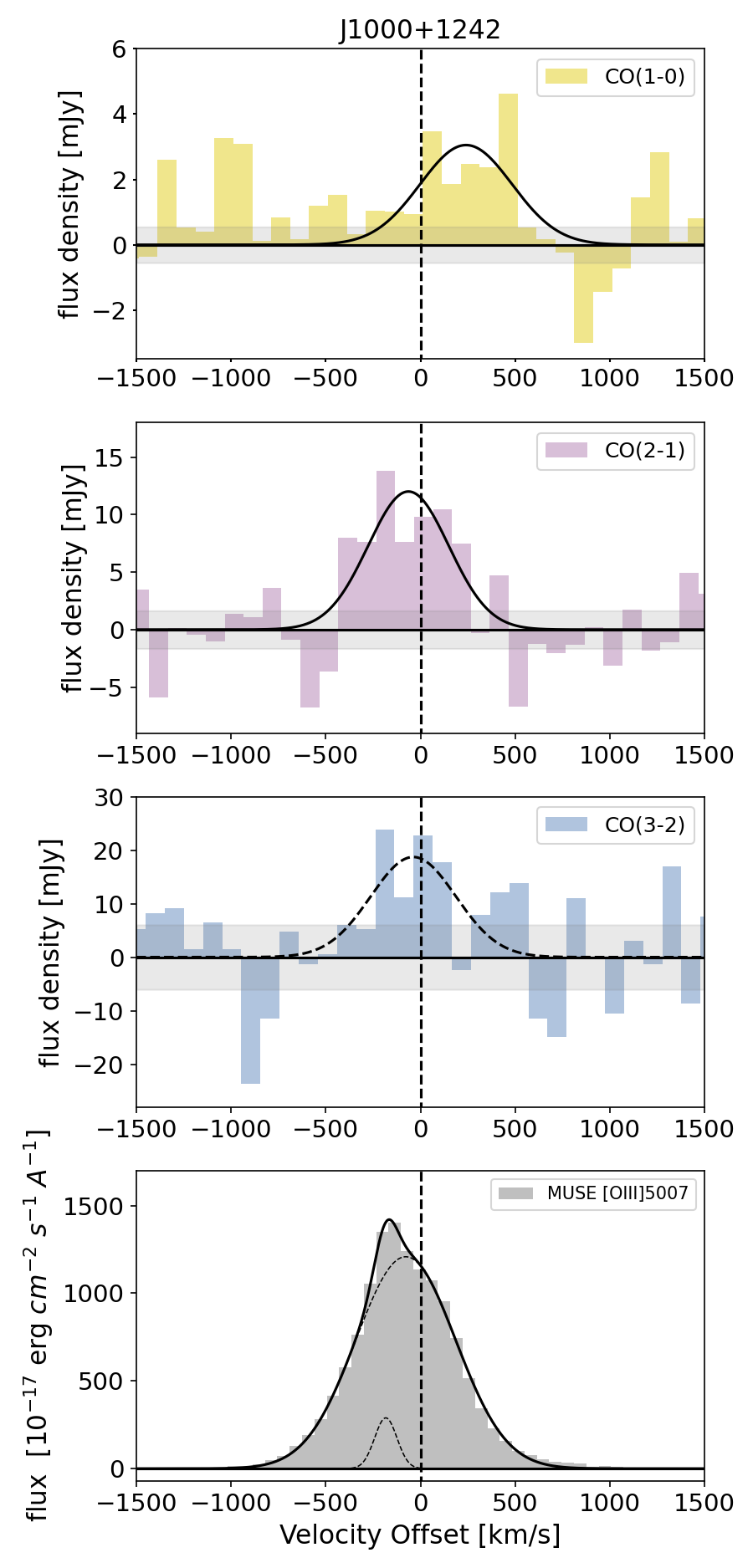}
        \caption{J1000+1242, detections in CO(1-0) and CO(2-1) with a low S/N detection in CO(3-2). MUSE [O~{\sc III}] observations show a broad single with Gaussian line profile.}
        \label{fig:J1000_spectra}
    \end{minipage}%
    \hspace{0.1cm}
    \begin{minipage}{0.32\textwidth}
        \centering
        \includegraphics[width=\linewidth, height=0.533\textheight]{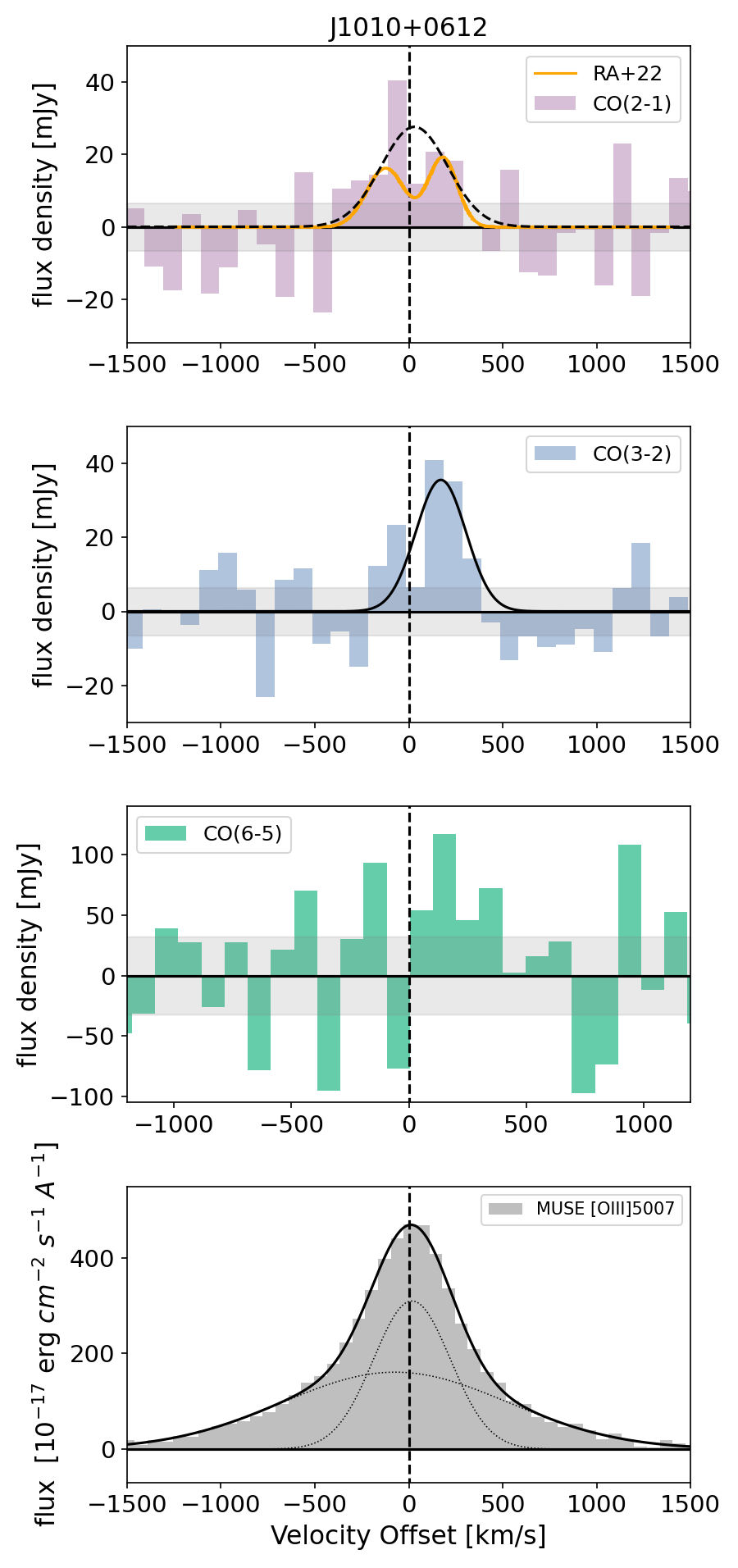}
        \caption{J1010+0612 spectra across three CO transitions, with no CO(1-0) observations available. All with 100 \kms bins. Detection in CO(3-2) and a low S/N detection in CO(2-1) which matches the two component line profile identified in \citealt{Almeida21} (spectra shown in orange, labelled RA+22) for which we see tentative signs here. MUSE [O~{\sc III}] observations show a two component fit with a narrow and broad component.}
        \label{fig:J1010+06_spectra}
    \end{minipage}
    \hspace{0.1cm}
    \begin{minipage}{0.32\textwidth}
        \centering
        \includegraphics[width=\linewidth, height=0.4\textheight]{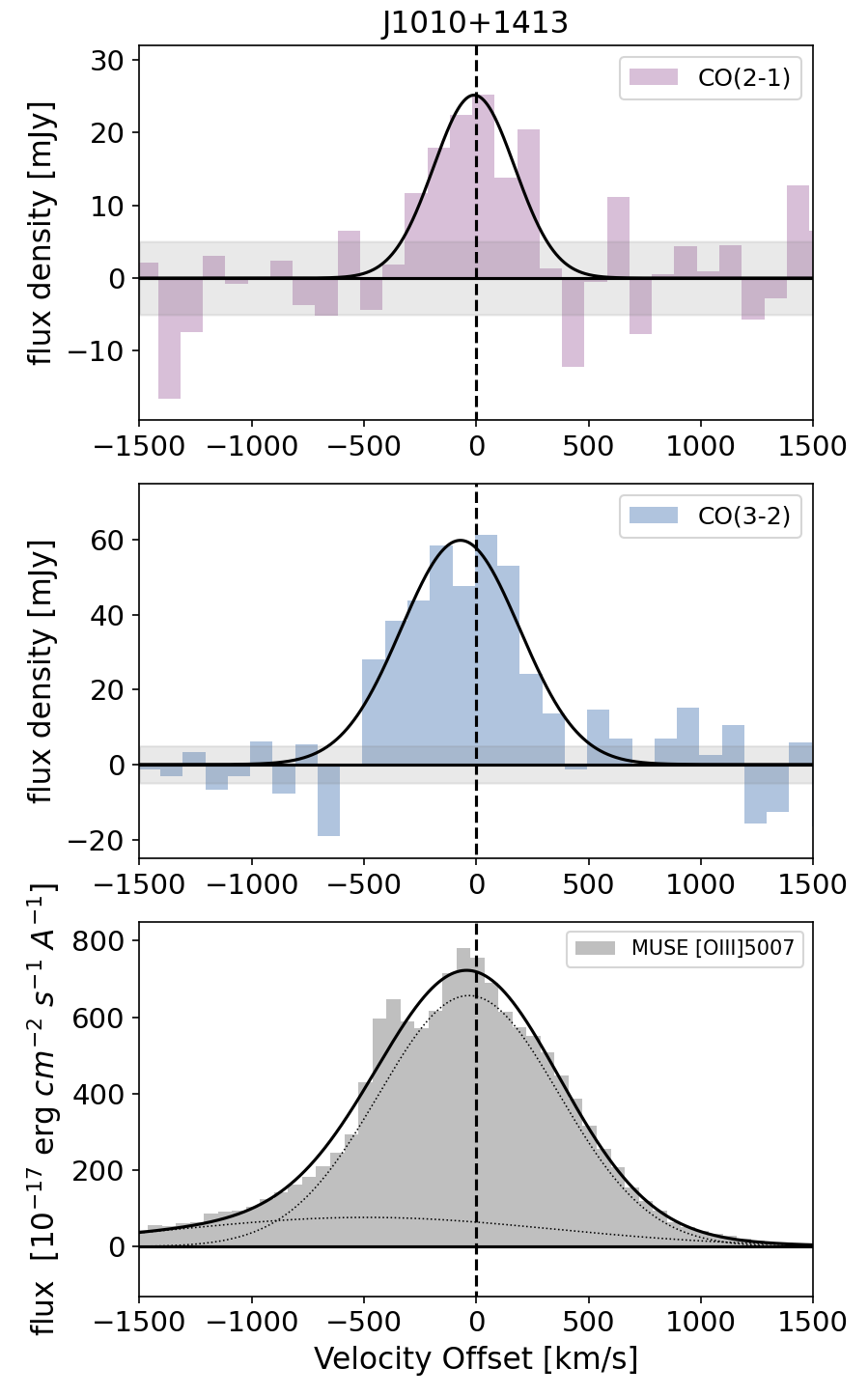}
        \caption{J1010+1413, detections in both CO(2-1) and CO(3-2). CO(1-0) not observed. MUSE [O~{\sc III}] observations show broad single with Gaussian line profile.}
        \label{fig:J1010+14_spectra}
    \end{minipage}
\end{figure*}

\begin{figure*}
    \centering
    \begin{minipage}{.32\textwidth}
        \centering
        \includegraphics[width=\linewidth, height=0.533\textheight]{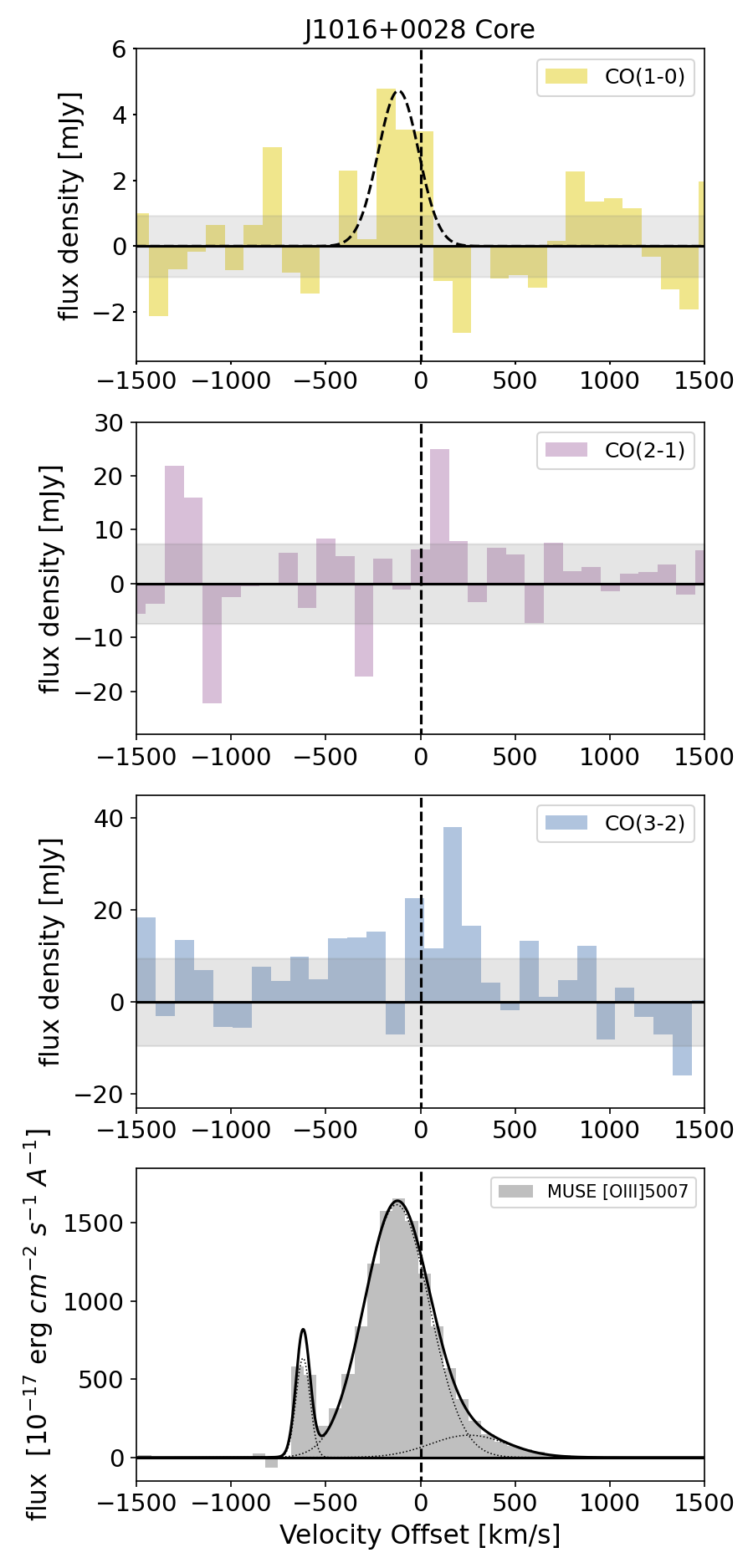}
        \caption{J1016+0028, low S/N detection in CO(1-0) and non detections in CO(2-1) and CO(3-2). MUSE [O~{\sc III}] observations show a two component line profile.}
        \label{fig:J1016_paper}
    \end{minipage}%
    \hspace{0.1cm}
    \begin{minipage}{0.32\textwidth}
        \centering
        \includegraphics[width=\linewidth, height=0.533\textheight]{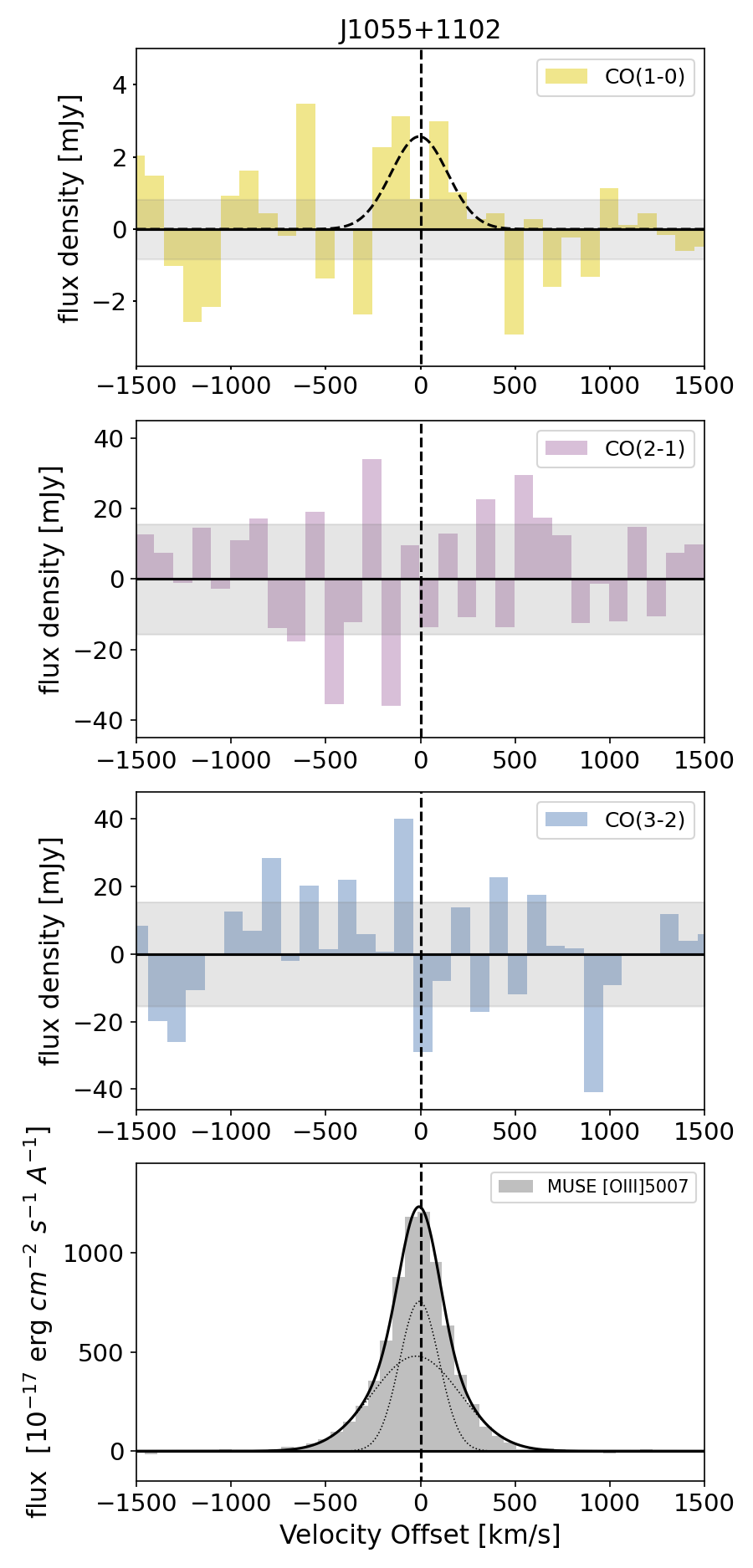}
        \caption{J1055+1102 only low S/N detection in CO(1-0), non detections in CO(2-1) and CO(3-2). MUSE [O~{\sc III}] observations show a two component fit with a narrow and broad component.}
        \label{fig:J1055_spectra}
    \end{minipage}
    \hspace{0.1cm}
    \begin{minipage}{0.32\textwidth}
        \centering
        \includegraphics[width=\linewidth, height=0.666\textheight]{J1100_spectra_paper_FINAL.png}
        \caption{J1100+0846 spectra across four transitions. CO(1-0) with 50 \kms bins and CO(2-1) and CO(3-2) with 25 \kms bins. Detection in all four transitions, all with same double peak profile. The same double peaked profile was found by \citealt{Almeida21} CO(2-1) observations (spectra shown in orange, labelled RA+22). MUSE [O~{\sc III}] observations show a two component fit with a narrow and broad component.}
        \label{fig:J1100_spectra}
    \end{minipage}
\end{figure*}

\begin{figure*}
    \centering
    \begin{minipage}{.32\textwidth}
        \centering
        \includegraphics[width=\linewidth, height=0.666\textheight]{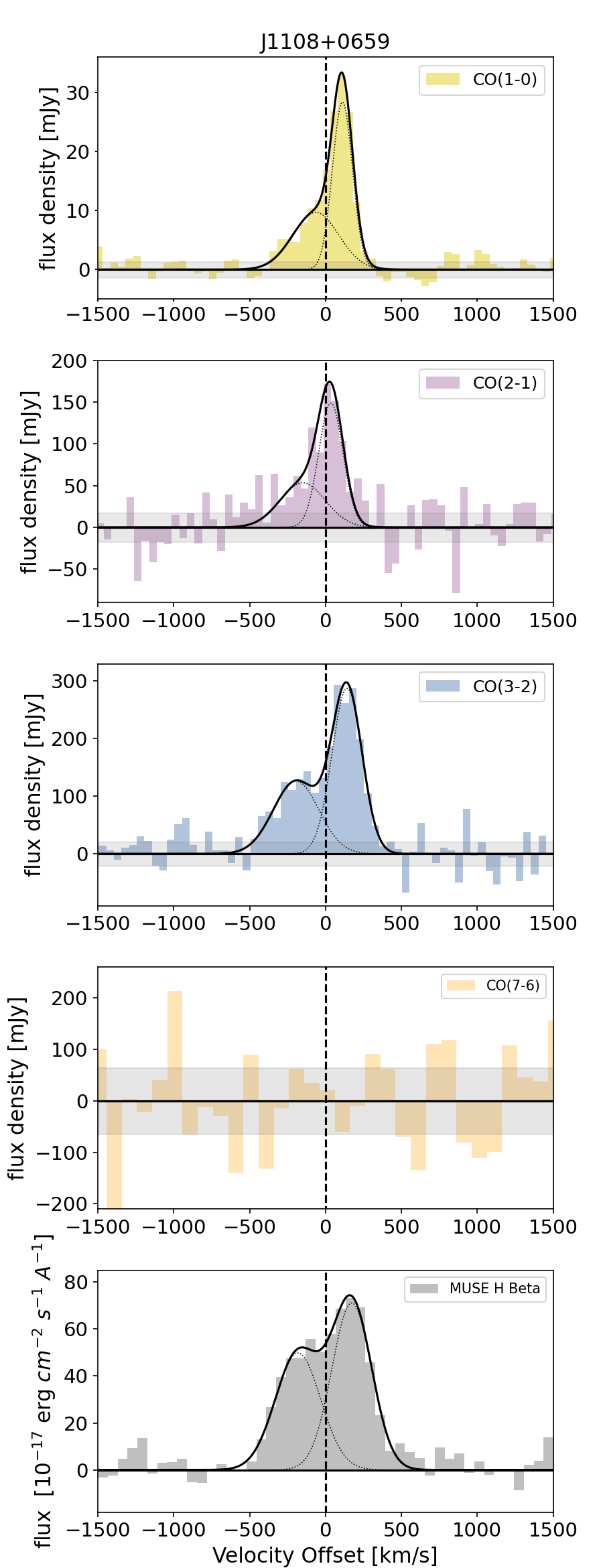}
        \caption{J1108+0659 spectra across four CO transitions. First three CO transitions plotted with 50 \kms bins and CO(7-6) plotted with 100 \kms bins due to a lack of detection. Blue wing in the first three CO transitions indicating a potential molecular outflow component. There is also potentially a small velocity shift in this outflow component across the first three CO transitions. We find a non-detection in CO(7-6). MUSE [O~{\sc III}] observations showing a similar blue wing to the CO data.}
        \label{fig:J1108_spectra}
    \end{minipage}%
    \hspace{0.1cm}
    \begin{minipage}{0.32\textwidth}
        \centering
        \includegraphics[width=\linewidth, height=0.533\textheight]{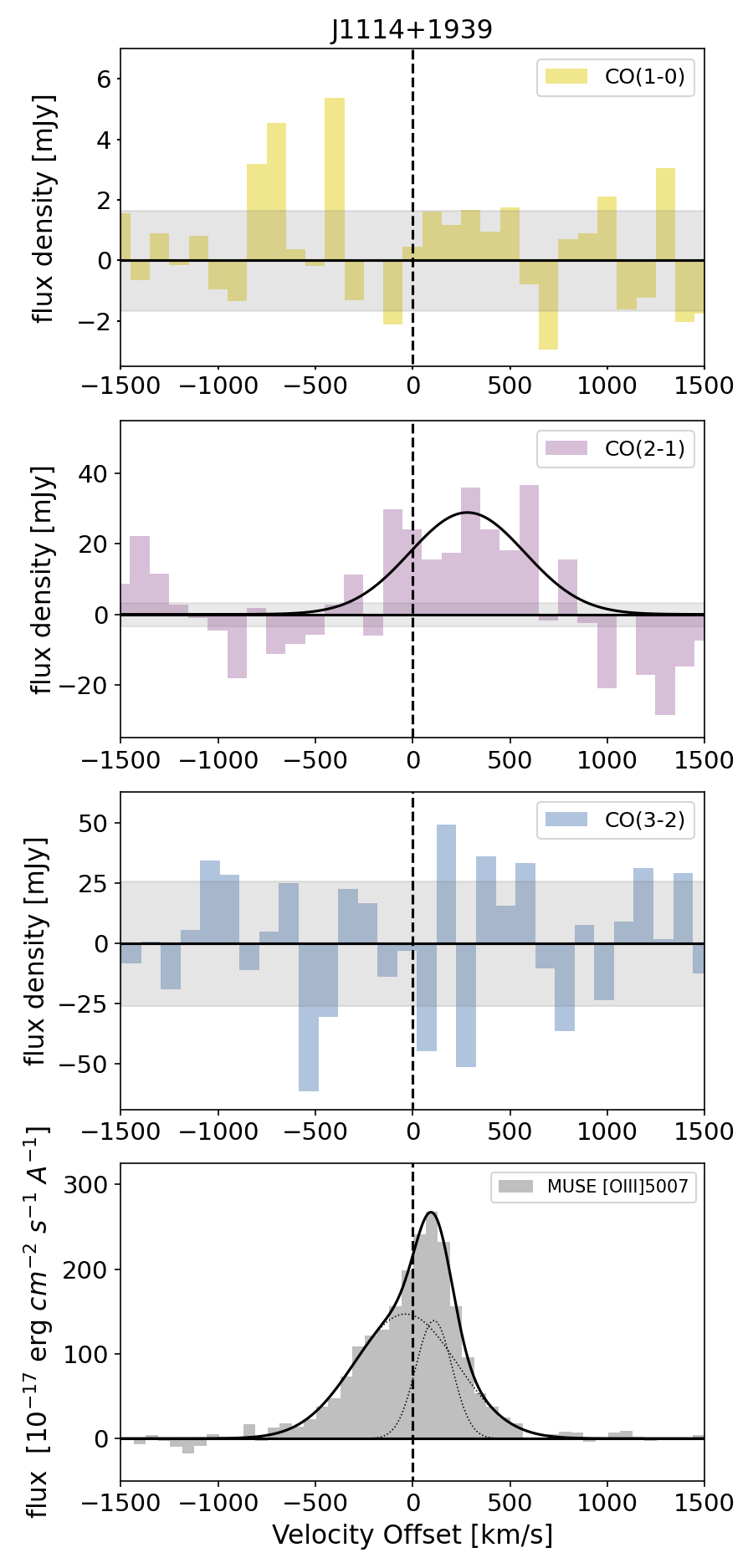}
        \caption{J1114+1939, only detection in CO(2-1). Non-detections in CO(1-0) and CO(3-2). MUSE [O~{\sc III}] observations show a two component fit with a narrow and broad component, with the broad component seemingly blueshifted potentially indicating an outflow.}
        \label{fig:J1114_spectra}
    \end{minipage}
    \hspace{0.1cm}
    \begin{minipage}{0.32\textwidth}
        \centering
        \includegraphics[width=\linewidth, height=0.666\textheight]{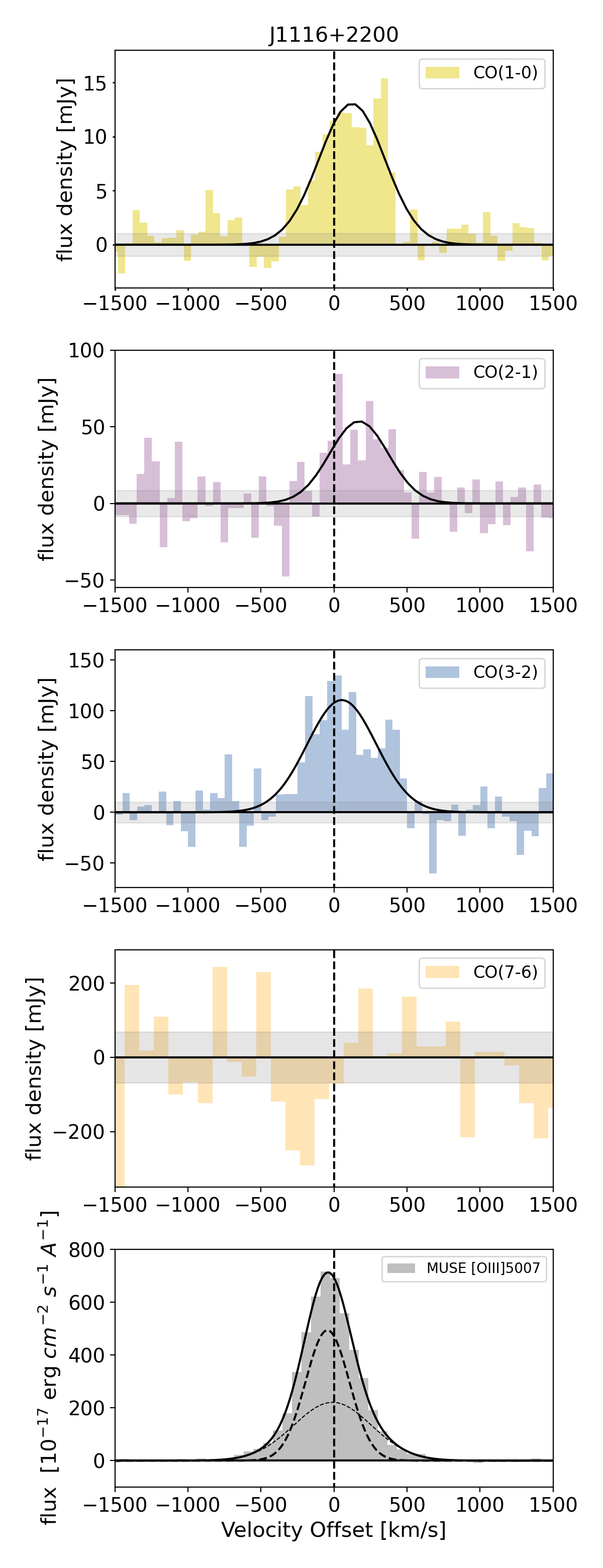}
        \caption{J1116+2200 spectra across four CO transitions. The first three transitions with 50 \kms bins and the CO(7-6) with 100 \kms bins. Detection of a broad line in the first three transitions. Non-detection in CO(7-6). MUSE [O~{\sc III}] observations show a single with Gaussian line profile.}
        \label{fig:J1116_spectra}
    \end{minipage}
\end{figure*}

\begin{figure*}
    \centering
    \begin{minipage}{.32\textwidth}
        \centering
        \includegraphics[width=\linewidth, height=0.533\textheight]{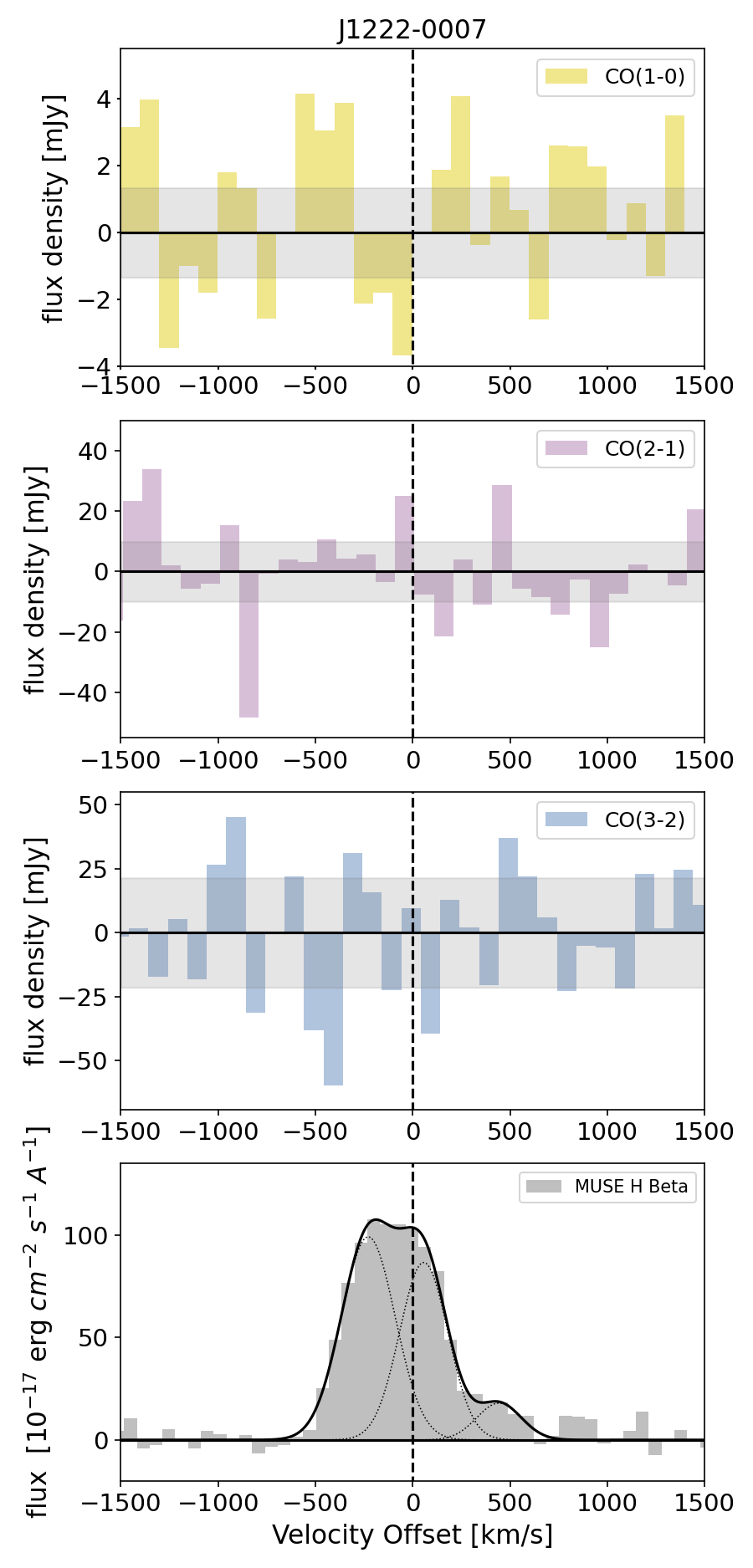}
        \caption{J1222-0007 non-detections in all three CO transitions. MUSE [O~{\sc III}] observations show a two component fit to the data.}
        \label{fig:J1222_spectra}
    \end{minipage}%
    \hspace{0.1cm}
    \begin{minipage}{0.32\textwidth}
        \centering
        \includegraphics[width=\linewidth, height=0.533\textheight]{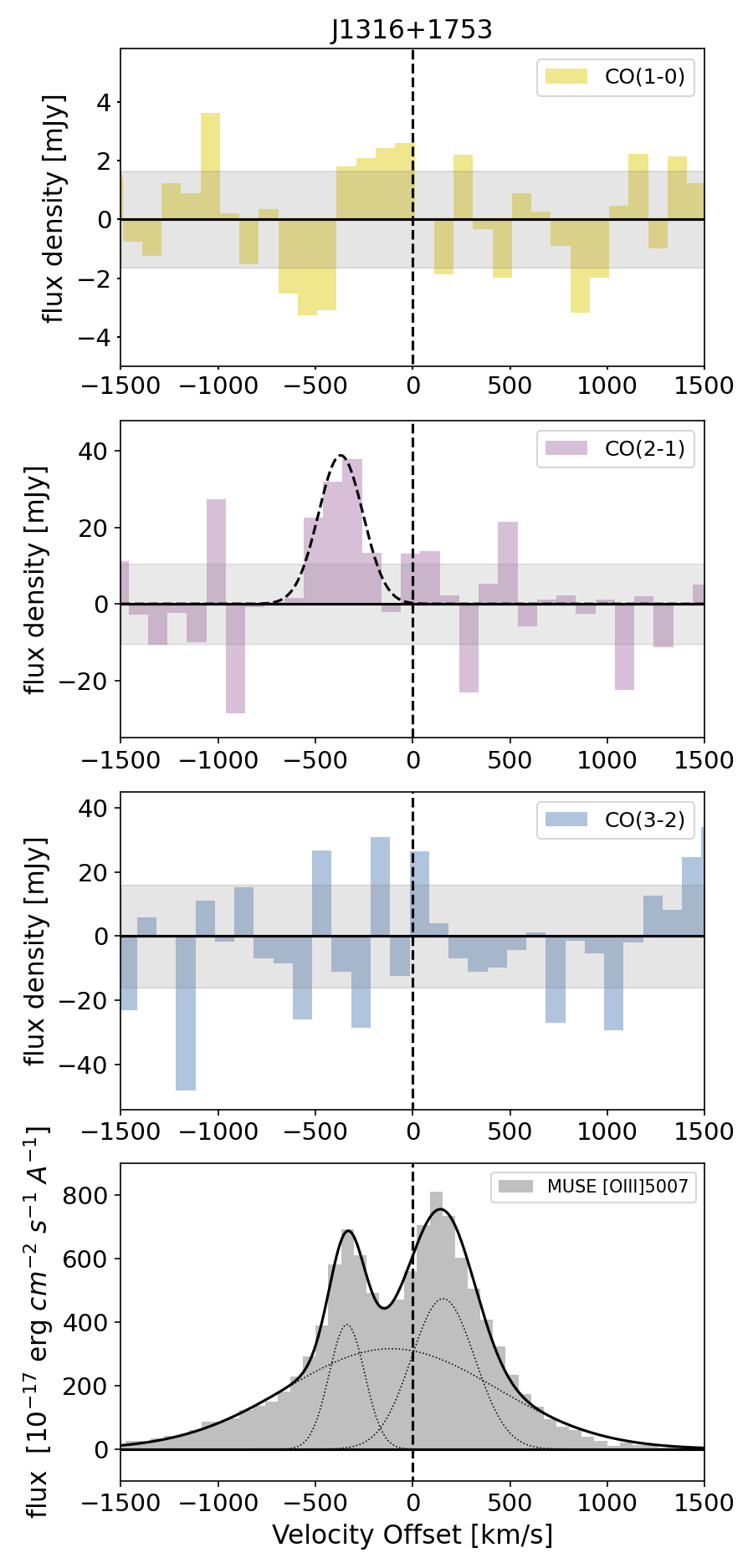}
        \caption{J1316+1753, low S/N detection in CO(2-1) and non-detection in CO(1-0) and CO(3-2). MUSE [O~{\sc III}] observations show a three component fit to the data. The low S/N detection in our CO(2-1) data seems to match the middle component present in the MUSE [O~{\sc III}] spectra.}
        \label{fig:J1316_spectra}
    \end{minipage}
    \hspace{0.1cm}
    \begin{minipage}{0.32\textwidth}
        \centering
        \includegraphics[width=\linewidth, height=0.4\textheight]{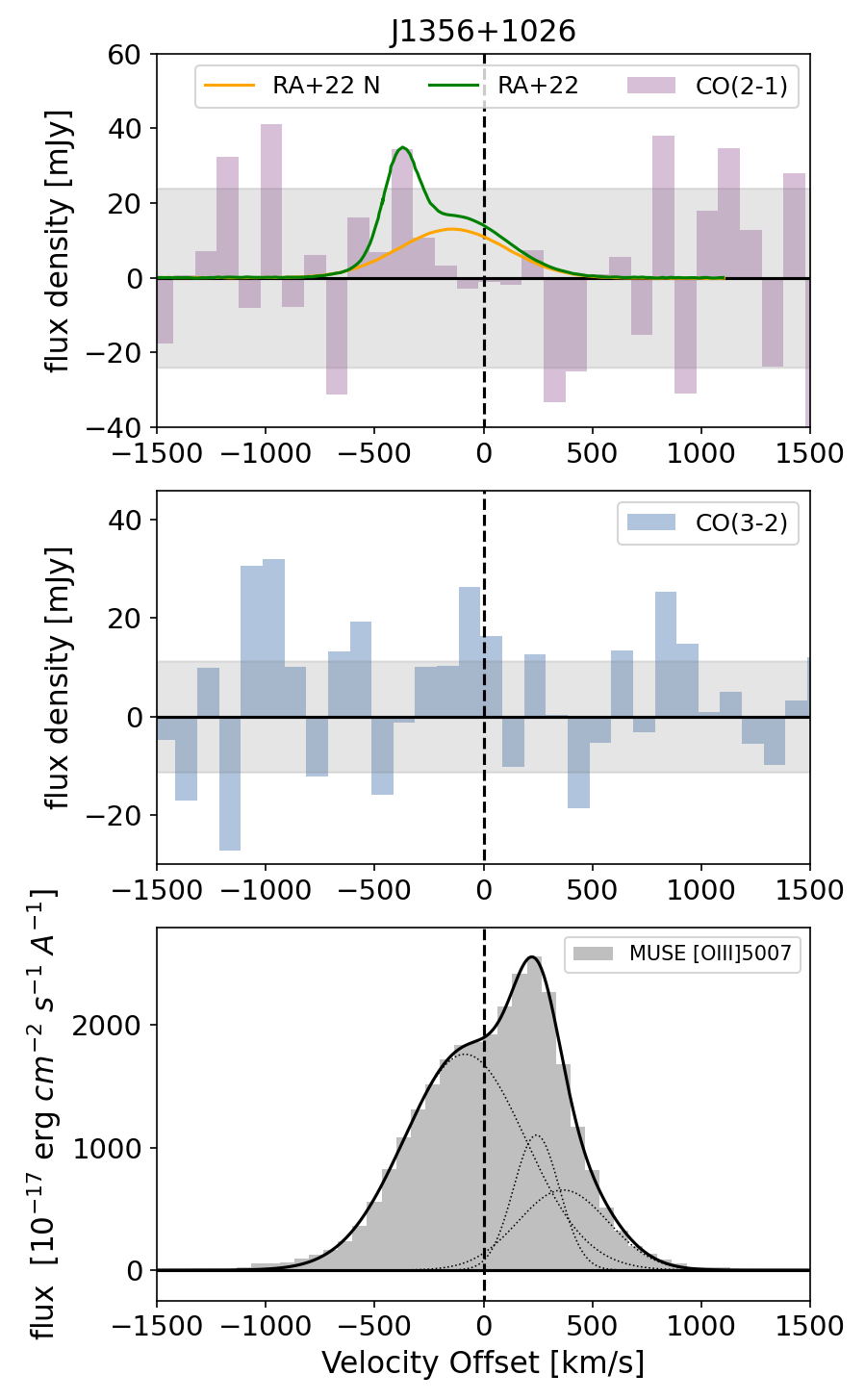}
        \caption{J1356 spectra across two CO transitions with no CO(1-0) data available. All set to 100 \kms bin widths. Non-detection in both transitions. Also shown is the detection from \citealt{Almeida21} detection in CO(2-1) (labelled RA+22 N in orange and RA+22 in green, where orange represents the nucleus only and green is the full line). We use the RA+22 data in our analysis as well lack a detection in CO(2-1)}. MUSE [O~{\sc III}] observations show a blue wing component.
        \label{fig:J1356_spectra}
    \end{minipage}
\end{figure*}

\begin{figure*}
    \centering
    \begin{minipage}{.48\textwidth}
        \centering
        \includegraphics[width=0.76\linewidth, height=0.6\textheight]{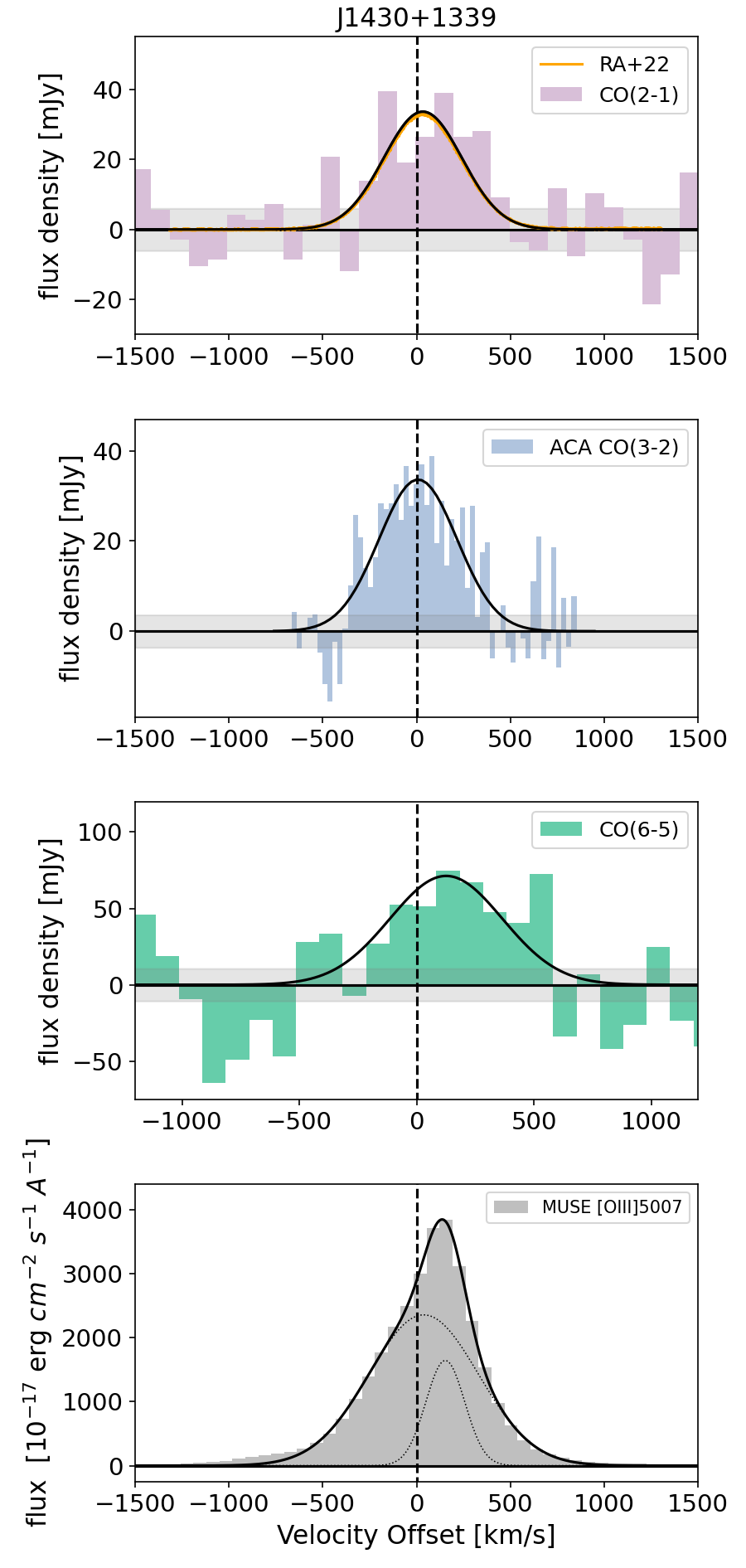}
        \caption{J1430+1339 spectra across three CO transitions, no observations in CO(1-0). CO(2-1) and CO(6-5) with 100 \kms bins, CO(3-2) with 26 \kms bins (ACA data). For CO(2-1) data from \citealt{Almeida21} was also available, shown here in orange (labelled RA+22) and showing an almost identical spectra to our APEX data. Detections across all CO transitions observed. MUSE [O~{\sc III}] observations show a two component fit including a blue wing.}
        \label{fig:J1430_spectra}
    \end{minipage}%
    \hspace{0.1cm}
    \begin{minipage}{0.48\textwidth}
        \centering
        \includegraphics[width=0.76\linewidth, height=0.68\textheight]{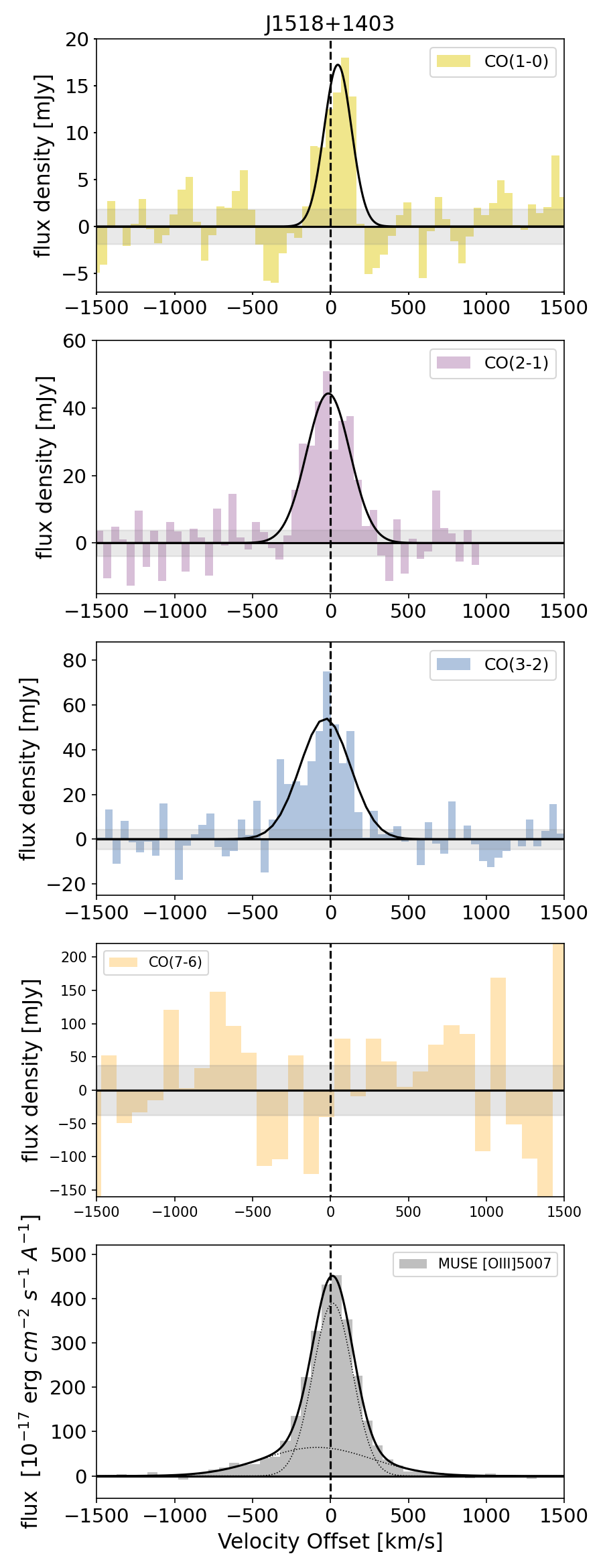}
        \caption{J1518+1403 spectra across four CO transitions. All with 50 \kms bins (except for CO(7-6) which is 100 \kms due to lack of a detection). Detections in both CO(1-0) and CO(3-2) and a low S/N detection in CO(2-1). MUSE [O~{\sc III}] observations show a two component fit with a narrow and broad component.} 
    \label{fig:J1518_spectra}
    \end{minipage}
\end{figure*}

Here we present all spectra for the targets in this paper including the multiple CO transition data and ionised gas observations from MUSE in Figures~\ref{fig:J0909_spectra} -- \ref{fig:J1518_spectra}. Top panels present the CO data available for that target and in every case we also present MUSE spectra for the [O III] line extracted from a 30 arcsec diameter aperture in the final panel. In all cases, solid black lines denote fits to the data. In some cases there are multiple components and as such, dotted black lines denote the different components that make the total fit to the spectra. In some CO(2-1) spectra we also show the fit from higher resolution ALMA observations \citep[solid orange line labelled RA+22,][]{Almeida21}. Shaded grey regions represent the 1$\sigma$ level. For further information about what is presented in the spectra plots (e.g. lines, fits to the spectra etc.) see Figure~3 in the main body of the paper.

\section{Analysis of the CO extent}\label{sec:CO_extent}

\begin{figure*}
\centering
        \includegraphics[width=0.95\linewidth]{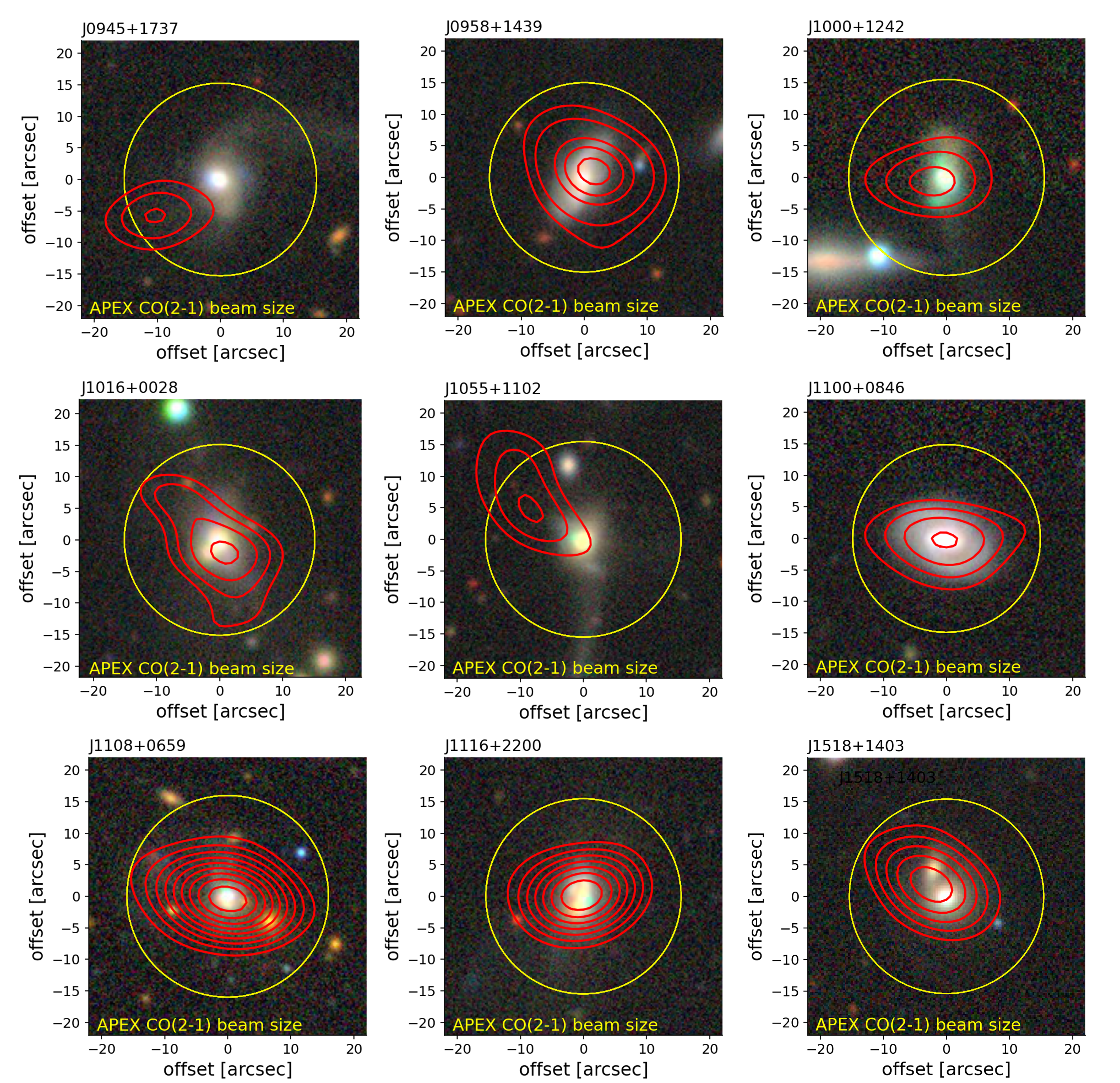}
        \captionsetup{justification=justified, labelfont=bf}
        \caption{Contour plots for the 9 targets with an integrated S/N $>$ 3 detection in the ACA CO(1-0) data, where red contours show the 2 $\sigma$ levels and upwards. The yellow circle denotes the beam size of the CO(2-1) observations, which is also the aperture used to extract the CO(1-0) spectra. The background images are rgb images from the DESI Legacy Imaging Survey in the (z, r, g) bands.} 
    \label{fig:Contour_plots}
\end{figure*}

In Figure~\ref{fig:Contour_plots} we present the contours of the ACA CO(1-0) data, plotted over optical images of the galaxies and the aperture in which the spectra were taken. Highlighting that the 30 arcsec aperture is a useful one to make sure we are obtaining accurate total flux values.

\bsp	
\label{lastpage}
\end{document}